\documentclass[a4paper,11pt]{report}
\usepackage[a4paper, total={6in, 8in}]{geometry}
\usepackage[toc,page]{appendix}
\usepackage[T1]{fontenc}
\usepackage[utf8]{inputenc}
\usepackage[english]{babel}
\usepackage{amsmath}
\usepackage{graphicx}
\usepackage{comment}
\usepackage{url}
\usepackage[table]{xcolor}
\usepackage{amssymb}
\usepackage{mathtools}
\usepackage[colorlinks=true,citecolor=violet,urlcolor=violet,linkcolor=blue]{hyperref} % https://latex.org/forum/viewtopic.php?t=16708
\usepackage{pdfpages}
\usepackage{cleveref}
\usepackage{courier}
\usepackage{amsthm}
\newtheorem{theorem}{Theorem}
\newtheorem{conjecture}{Conjecture}
\newtheorem{corollary}{Corollary}[theorem]

\newtheorem{proposition}[theorem]{Proposition}
\newtheorem{observation}[theorem]{Observation}
\numberwithin{equation}{section}

\definecolor{lightblue}{rgb}{0.8,0.8,1}

\makeatletter
\renewcommand*\cleardoublepage{\clearpage\if@twoside
  \ifodd\c@page \hbox{}\newpage\if@twocolumn\hbox{}%
  \newpage\fi\fi\fi}
\makeatother

\usepackage{titling}

\usepackage{eso-pic}
\usepackage[]{graphicx}
\AddToShipoutPictureBG*
  {%
    \put(\LenToUnit{.45\paperwidth}, \LenToUnit{.8\paperheight})
      {\includegraphics{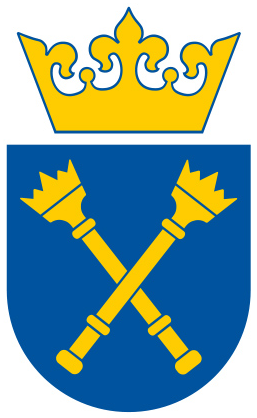}}%
  }

\def \paperslist {page.117}

\author{\bf \Large Wojciech Bruzda}

\title{
{\Huge Jagiellonian University in Krak\'{o}w}\\
\medskip
{Faculty of Physics, Astronomy\\ and Applied Computer Science}\\
\medskip
\medskip
\medskip
\medskip
\medskip
\medskip
\medskip
\medskip
\medskip
{\bf \Huge Structured Unitary Matrices\\and\\Quantum Entanglement\\}
\medskip
\medskip
\medskip
\medskip
\medskip
\medskip
\medskip
\medskip
\medskip
\medskip
\medskip
\medskip
\medskip
\medskip
\medskip
\medskip
{\small A doctoral dissertation}\\
\medskip
{\small written under the supervision of Prof. Karol \.{Z}yczkowski}
\medskip
\medskip
\medskip
\medskip
\medskip
\medskip
\medskip
\medskip
\medskip
\medskip
\medskip
\medskip
\medskip
\medskip
\medskip
\medskip
\medskip
\medskip
\medskip
}

\date{\small Krak\'{o}w, November 2021}

\begin{document}

%%%%%%%%%%%%%%%%%%%%%%%%%%%%%%%%%%%%%%%%%%%%%%%%%%%%%%%%%%%%%%%%%%%%

\maketitle

\newpage
\thispagestyle{empty}
~\newpage
\thispagestyle{empty}

{\Large Acknowledgments}

\medskip

I am grateful to my supervisor Karol \.{Z}yczkowski who guided me this far over all these years and supported me by all means.

\medskip

I acknowledge with gratitude the fruitful cooperation of the following people:

Daniel Alsina,
Ingemar Bengtsson,
Adam Burchardt,
Valerio Cappellini,
Åsa Ericsson,
Jonith Fischmann,
Shmuel Friedland,
Dardo Goyeneche,
Boris Khoruzhenko,
Arul Lakshminarayan,
Jan-Åke Larsson,
Grzegorz Rajchel-Mieldzio\'{c},
Suhail Ahmad Rather,
Marek Smaczy\'{n}ski,
Hans-J\"{u}rgen Sommers,
Wojciech Tadej,
Ond\v{r}ej Turek, all colleagues from Jagiellonian Quantum Information Team, and of course APPB Crew.

\medskip

I am indebted to Anna Szymusiak and Dardo Goyeneche who carefully read appropriate chapters
and whose criticism and remarks were instrumental for the final version of this work.

\medskip

Special thanks to:

Ania G{\l}owacka, pan Jacek, Sabrina, Zuzik, Miros{\l}aw ($\vec{F}=m\vec{g}$) Klewiec, Rockford,
Cycurek, $\#119$, unit $456$, and every sea $\square$ square.
%{\color{white}$\square$}\hfill$\square$

\medskip
\medskip
\medskip
\medskip
\medskip
\medskip
\medskip
\medskip
\medskip
\medskip
\medskip
\medskip
\medskip
\medskip
\medskip
\medskip
\medskip
\medskip
\medskip
\medskip
\medskip
\medskip
\medskip
\medskip
\medskip
\medskip
\medskip
\medskip
\medskip
\medskip
\medskip
\medskip
\medskip
\medskip
\medskip
\medskip
\medskip
\medskip
\medskip
\medskip
\medskip
\medskip
\medskip
\medskip
\medskip
\medskip
\medskip
\medskip
\medskip
\medskip
\medskip
\medskip
\medskip
\medskip
\medskip
\hfill for us

\hfill and them

\newpage
\thispagestyle{empty}
~\newpage
\thispagestyle{empty}

{\Large Abstract}

\medskip 

In this Thesis we explore the set of unitary matrices characterized by
a given structure in the context of their applications in the field of quantum information.
In the first part of the Thesis we focus on classification of special classes
of unitary matrices and possibility of introducing certain internal parameterizations.
Several new results and conjectures are discussed. In particular, we propose
the construction of families of isolated complex Hadamard matrices in infinitely many dimensions.
Second part of the Thesis is devoted to the concept of multipartite quantum entanglement.
We present a solution to the long-standing problem in theory of quantum information concerning
the status of absolutely maximally entangled states of four subsystems with six levels each.
Finally, we propose a fusion of two areas of research and we analyze the excess of a~matrix and the corresponding
Bell inequalities. This combination allows us to draw new conclusions related to quantum nonlocality.

\medskip 
\medskip 
\medskip 
\medskip 
\medskip 
\medskip

{\Large Streszczenie}

\medskip 

Dysertacja jest po\'{s}wi\k{e}cona badaniu macierzy unitarnych posiadaj\k{a}cych
pewn\k{a} zadan\k{a} struktur\k{e} w kontek\'{s}cie ich zastosowa\'{n} w teorii kwantowej informacji. %opisie fenomenu kwantowego spl\k{a}tania.
W pierwszej cz\k{e}\'{s}ci pracy skupiamy si\k{e} na problemie klasyfikacji specjalnych
klas macierzy unitarnych oraz mo\.{z}liwo\'{s}ci wprowadzenia pewnych wewn\k{e}trznych parametryzacji.
Przedstawiamy i~dyskutujemy szereg nowych rezultat\'{o}w oraz hipotez.
W szczeg\'{o}lno\'{s}ci omawiamy konstrukcj\k{e} rodzin izolowanych zespolonych macierzy Hadamarda w niesko\'{n}czenie wielu wymiarach.
Druga cz\k{e}\'{s}\'{c} pracy jest po\'{s}wi\k{e}cona wielo\-cz\k{a}stkowemu kwantowemu spl\k{a}taniu.
Prezentujemy d{\l}ugo oczekiwane rozwi\k{a}zanie jednego z problem\'{o}w w teorii kwantowej
informacji, kt\'{o}ry dotyczy istnienia stan\'{o}w absolutnie maksymalnie spl\k{a}tanych
dla czterech poduk{\l}ad\'{o}w o sze\'{s}ciu stopniach swobody.
Ostatni rozdzia{\l} pracy zawiera propozycj\k{e} zintegrowania dw\'{o}ch obszar\'{o}w bada\'{n}: ekscesu macierzowego oraz nier\'{o}wno\'{s}ci Bella.
Po{\l}\k{a}czenie tych dw\'{o}ch zagadnie\'{n} pozwala nam wyci\k{a}gn\k{a}\'{c} nowe wnioski dotycz\k{a}ce kwantowej nielokalno\'{s}ci.

\newpage
\thispagestyle{empty}
~\newpage
\thispagestyle{empty}
\clearpage
\pagenumbering{arabic} 

%%%%%%%%%%%%%%%%%%%%%%%%%%%%%%%%%%%%%%%%%%%%%%%%%%%%%%%%%%%%%%%%%%%%
\tableofcontents
%%%%%%%%%%%%%%%%%%%%%%%%%%%%%%%%%%%%%%%%%%%%%%%%%%%%%%%%%%%%%%%%%%%%
%\clearpage\null%\thispagestyle{empty}
%\clearpage\null%\thispagestyle{empty}
\chapter{Introduction}

Why do we do physics?

Nowadays it is a tricky question but
let us be so bold to suggest the following generic answer, which we believe is still valid to some extent, regardless the circumstances:
we do physics to ultimately understand the Universe.

We do not focus much on the philosophical aspects and the purpose of this all, but we are happy to know the mechanisms
explaining all processes at every scale; from elementary particles, through the classical world we live in, to the mechanics of
the most remote galaxies.
Probably, once we get the knowledge about everything that surrounds us and what is deep inside of our bodies,
we will eventually go back to the question concerning the objectives, or maybe the answer will come itself.
Either way, fortunately there is a long (very long indeed) way before us. 

Doing physics is a synergy of observation, intuition, imagination and mathematics.\footnote{Some say the patience and humility too.}
But in contrary to the last ingredient, which describes ideal concepts in its ideal realm, physicists notoriously
face the problem of permanent approximations and imprecision. 
At the subatomic scales, and in this Thesis we are implicitly interested exactly in subatomic scales, this
property of our reality is exhibited exceptionally harsh.
Let us take a very concrete example. Suppose we want to build (actually, we badly want to build!) a fully operative ``pocket''
quantum computer,
a successor of classical machines with computational power that is incomparably greater than the
one of the currently biggest superclusters.
Quantum mechanics, the mathematical framework that provides all necessary tools to operate at such low level,
imposes natural constraints and limits for the physical systems and their possible applications.
Roughly speaking, an action of a quantum computer can be described as a sequence of unitary operations (quantum gates)
applied to initial states (vectors in Hilbert space) which are subjected to measurement (outcomes of our calculations).
Somewhere in between the input and output, the unitary evolution should be 
maintained and supported by error correction procedures to minimize consequences of~immediate decoherence.
Process of decoherence, which is the interaction of quantum system with the environment causes
noise and the loss of interference, and the lack of interference in the quantum system makes it simply a classical one.
This is one of the biggest obstacles that we presently encounter.
If ever resolved, it will pave the way for further milestones on the way to quantum computing.

Quantum information theory (QIT)~\cite{Ho21} being a relatively young branch of contemporary physics, deals with the aforementioned
problems. At the heart of the QIT there is a notion of {\sl quantum information}, which can be understood
as a everything what we can infer from the quantum system described mathematically by a wave function.
As such, quantum information is the root of: entanglement, computation, randomness,
currency, metrology, cryptography, thermodynamics, Darwinism, teleportation, etc.,
where every item in the list deserves the prefix ``quantum''.

In the further perspective, apart from obviously attractive problem of quantum computers,
for which already awaits a bunch of problems to be solved,
we eagerly wish to address one additional issue concerning the quantum description of gravity, which is apparently not included in the
list above.
This is actually the last missing step in
understanding the properties of all four known fundamental interactions.
Meanwhile, short term expectations and the most imperative open problems in QIT have been
summarized in many lists. The priority of problems strongly depends on the community preparing such motivational aggregates.
However, the common idea remains
the same --- to gradually discover and understand the more areas of physics.
Here we recall two such lists.
``Five Open Problems in Theory of Quantum Information''~\cite{KCIKProblems},
and another one, taken from the Blog of Scott Aaronson, entitled
suggestively ``The Ten Most Annoying Questions in Quantum Computing''~\cite{TenAnnoyingProblems}.

Some problems from these lists like the existence of mutually unbiased bases,
absolutely maximally entangled states or special
families of quantum measurement, will be carefully (sometimes indirectly) explored in the chapters that follow.

%%%%%%%%%%%%%%%%%%%%%%%%%%%%%%%%%%%%%%%%%%%%%%%%%%%%%%%%%%%%%%%%%%%%%%%%%%%%%%%%
%\newpage
\section{Purpose and Structure of the Thesis}

This theoretically oriented dissertation contributes to few branches of QIT;
quantum computation, quantum measurement, quantum combinatorics and quantum nonlocality.
During the preparation of the Thesis some open problems have been partially or fully resolved.
Only the most interesting and, in large part, unpublished
results will be presented. %They led us to final outcomes presented in the papers, or preprints

Each chapter covers one specific topic and should be considered either as
a detailed supplement to the associated papers:
$[\hyperlink{\paperslist}{\rm A1}]$, % CHM
$[\hyperlink{\paperslist}{\rm A2}]$, % ISOLATED CHM
$[\hyperlink{\paperslist}{\rm A3}]$, % DELTA
$[\hyperlink{\paperslist}{\rm A4}]$, % AME
and
$[\hyperlink{\paperslist}{\rm A5}]$, % SIGMA (excess)
or a different view of the published (or to be published) results.
We focus only on the original contributions and, any time it is required to preserve
the continuity of exposition, we highlight references to our collaborators to emphasize their own achievements.

The logical composition of this Thesis can be divided onto two parts.
Chapters~\ref{chap:CHM} and~\ref{chap:DELTA} present the general approach to understanding the
structure of some classes of unitary matrices in arbitrary dimension.
While the remaining part with Chapters~\ref{chap:AME46} and~\ref{chap:EXCESS}
regards the idea of quantum entanglement
described in the language of matrix formalism.
This should comprehensively explain the concept, the structure and the title of the work.

Recall the schematic description of quantum computer mentioned at the beginning.
Three general ingredients involving quantum computation can be in some sense associated with the three first chapters:
\begin{itemize}
\item Chapter~\ref{chap:CHM} $\leftrightarrow$ quantum unitary evolution (quantum gates),
\item Chapter~\ref{chap:DELTA} $\leftrightarrow$ quantum measurement,
\item Chapter~\ref{chap:AME46} $\leftrightarrow$ quantum error correcting codes.
\end{itemize}
In each chapter we make (a successful) attempt to a better understanding of a given subject.

\medskip

The paper is organized as follows.
We begin with the problem of classification of {\sl complex Ha\-da\-mard matrices} (CHM).
Both, numerical and analytical aspects of searching for new examples of CHM are described.
We present a new family of CHM of size $8$ and continue the subject over the particular
subset of CHM --- isolated CHM, which leads to other previously unknown matrices of order $N>8$.
Two new constructions of isolated CHM are introduced,
and we propose two conjectures concerning existence of such objects in infinitely many dimensions.

Third chapter deals with the issue of extension of the notion of the defect of a unitary matrix adapted to the special
class of Hermitian unitary matrices corresponding to generalized quantum
measurements POVM ({\sl positive operator-valued measures}).
We provide case studies and concrete examples of geometrical structures being examined by this tool,
which allow us to draw conclusions about possibility of introduction of free parameters in
special classes of POVM including {\sl mutually unbiased bases} (MUB) and {\sl symmetric
informationally complete POVM} (SIC-POVM),
and also in other interesting objects used in the foundations of quantum mechanics and theory of quantum information.

In the second part of the Thesis we concentrate on the quantum entanglement.
Fourth chapter is devoted to the idea of multipartite entanglement and
the long-awaited status of existence of
{\sl absolutely maximally entangled} (AME) state of four quhexes, AME$(4,6)$. We provide elegant
analytical solution along with some off-topics related to this yet-to-be-fully-understood problem.
In particular, we elaborate on the problem of construction a real counterpart of such a complex state.

The last short chapter contains generalization of the notion of excess of a (real) Ha\-da\-mard matrix and
its applications in quantum nonlocality, bipartite entanglement
and special classes of Bell inequalities with the property of quantum advantage.

We summarize all addressed topics at the end.
Despite the fact that several issues have been ``disentangled'',
we stumbled upon many new unknown facts, which certainly require further investigation.
Hence, we also present a list of open questions along with a plan of the future research.

%%%%%%%%%%%%%%%%%%%%%%%%%%%%%%%%%%%%%%%%%%%%%%%%%%%%%%%%%%%%%%%%%%%%%%%%%%%%%%%%

%\newpage
\section{Notation Used}
\label{sec:notation}

We are going to employ standard mathematical notation with
minor exceptions where there is no general agreement upon the style or symbols.
We assume the set of natural numbers contains zero, $\mathbb{N}_0 = \{0\}\cup\mathbb{N}$. And every time
we use $p$ (not being a subscript, nor having any subscript), it should be viewed
as a prime number (Chapter~\ref{chap:CHM} and~\ref{chap:DELTA}). Bold symbols ${\bf p}$ and ${\bf q}$
in Chapter~\ref{chap:CHM} are reserved for polynomials.

We denote the set of permutation, orthogonal, unitary and general matrices of size $N$ by
$\mathbb{P}(N)$, $\mathbb{O}(N)$, $\mathbb{U}(N)$, and $\mathbb{F}^{N\times N}$, respectively,
where $N$ indicates the dimension and $\mathbb{F}$ is a field of real ($\mathbb{R}$) or complex
($\mathbb{C}$) numbers.
Other specific classes of matrices are defined in appropriate sections.
For instance, $\mathbb{H}(N)$ denotes the set of complex Hadamard matrices (CHM) of order $N$.

To be consistent with the notation used in publications, when traversing between chapters, we use
letters $N$ and $d$ to indicate dimension of the system, interchangeably. It should not cause too much confusion.
Special attention is only needed in Chapter~\ref{chap:DELTA}, where the defect of a unitary matrix (always in bold), ${\bf d}(U)$,
might clash with the dimension $d$.

In many cases (mainly in Chapter~\ref{chap:CHM} and~\ref{chap:AME46}), when presenting matrices
and associated operations, we adapt the notation
introduced in~\cite{TZ06}, developed in {\sl Catalogue of Complex Hadamard Matrices}~\cite{CHM_catalogue}
and widely accepted by community, e.g. operations ${\rm EXP(i M)}$ and $M_1 \circ M_2$, which stand for entrywise exponential
acting on $M$, and entrywise (Hadamard) product of two matrices $M_1$ and $M_2$, respectively.
To enhance clarity, all vanishing entries in $M$
are represented by black dots ($\bullet$) or completely omitted.
This form speeds up orientation over the structure of $M$ and highlights its most important features.
In general, we prefer concrete examples in place of abstract definitions for which
necessary resources are provided. 
Even, for further simplification, in few cases we reduce discussions to purely pictorial description.
However, occasionally we feel obliged to preserve rigorous mathematical formalism.

In the physics realm we switch to
bra-ket notation and we use different (but equivalent) forms of tensor multiplication
$|ab\rangle=|a,b\rangle=|a\rangle|b\rangle=|a\rangle\otimes |b\rangle$,
which are chosen at the convenience according to the situation.
We assume the Reader is familiar with
the fundamental concepts of quantum mechanics and basics of theory of quantum information, so
we intentionally omit some frequently used definitions and we avoid to explain the notions, % in theory of quantum information and quantum mechanics itself,
the meaning of which should be well known
or inferred directly from the context or, in the last resort, can be found in the attached papers.
We only provide detailed description of objects, which are crucial for our considerations.
In particular, we consistently adhere to the standard course of quantum mechanics and the generally
accepted postulates of this theory, without much deliberation about possible alternatives and different interpretations.
The sections containing main results are sometimes preceded by a short and popular
introduction to outline the physical background and motivation.
These descriptive fragments of the work by no means aspire to be an
exhaustive source of information on a given topic. We make it clear
by providing specific sources and the most relevant or up-to-date references. 

%%%%%%%%%%%%%%%%%%%%%%%%%%%%%%%%%%%%%%%%%%%%%%%%%%%%%%%%%%%%%%%%%%%%
\clearpage\null\thispagestyle{empty}
%\clearpage\null%\thispagestyle{empty}
\chapter{Complex Hadamard Matrices}
\label{chap:CHM}
Chapter based on~$[\hyperlink{\paperslist}{\rm A1}]$,~$[\hyperlink{\paperslist}{\rm A2}]$.
{\color{red}Update: Please, consider the second part of this chapter as a preliminary draft.
Many things and ideas have changed and been revised since November 2021. An improved version can be found at \href{https://arxiv.org/abs/2204.11727}{arXiv:2204.11727}.}
\medskip
\medskip
\medskip

\noindent A complex Hadamard matrix~\cite{TZ06} $H$ is a rescaled unitary matrix with unimodular entries
\begin{equation}
H H^{\dagger} = N \mathbb{I}_N  \quad : \quad H_{jk} = e^{i 2 \pi \varphi_{jk}} \quad \text{for some} \quad \varphi_{jk} \in [0, 1).
\end{equation}
Let $\mathbb{H}(N)$ be the set of all {\sl complex Hadamard matrices} (CHM) of size $N$.
If matrix $H$ has all its~entries in the form of
q$^\textrm{th}$ roots of unity, $\varphi_{jk} = m_{jk}/q$ for $m_{jk}\in\mathbb{N}$, we call it a {\sl Butson type}~\cite{Bu62, Bu63}.
Such matrices constitute a proper subset of CHM and will be denoted by
\begin{equation}
{\rm B}\mathbb{H}(N, q)\subsetneq \mathbb{H}(N).\label{BH_class}
\end{equation}
CHM is a natural extension of a real (with only $\mp 1$ entries) Hadamard matrix~\cite{Ha93}
and as such can be further generalized over quaternions~\cite{CD08} or general groups~\cite{Br88,Ha97}.
Note that real Hadamard matrices form a special case of Butson type matrices, ${\rm B}\mathbb{H}(N,2)$.
Here we focus only on the pure complex variant, while we return to classical real Hadamards in Chapter~\ref{chap:EXCESS}.

We also put here the definition of {\sl mutually unbiased bases} (MUB), which are tightly connected to CHM
and relevantly motivate the searching for new examples of CHM.
MUB will be recalled several times here, in Chapter~\ref{chap:DELTA}, and Chapter~\ref{chap:EXCESS}.
Let $\mathcal{B}_j=\big\{|\psi_1^{(j)}\rangle,...,|\psi_N^{(j)}\rangle\big\}$ for $j\in\{2,...,m\}$
be a collection of $2\leqslant m\leqslant N+1$ orthonormal bases in~$\mathbb{C}^N$.
We call them {\sl mutually unbiased} if the overlap
between each basis has constant magnitude,
\begin{equation}
\big|\langle\psi_a^{(j)}|\psi_b^{(k)} \rangle\big|^2=\delta_{jk}\delta_{ab}+\frac{1-\delta_{jk}}{N}\label{MUB_definition}
\end{equation}
for $a,b\in\{1,2,...,N\}$ and $j,k\in\{1,2,...,m\}$.
It is well known that the upper bound for MUB (complete set) in $\mathbb{C}^N$ is $N+1$, which is saturated for $N=p^l$ for prime $p$ and $l\geqslant 1$~\cite{Iv81, WF89}.
In general, the problem of existence of such bases (not necessarily a complete set) in non-prime-power dimensions is a major open problem,
especially for $N=6$~\cite{DEBZ10}.

Given two or more MUB; $\mathcal{B}_1$, $\mathcal{B}_2$, ... one such basis can be represented by $\mathbb{I}_N$.
It is then easy to infer that the remaining bases, when represented as matrices, comprise the set of different CHM.
We call such matrices {\sl mutually unbiased Hadamards} (MUH). Hence, searching for MUB of a given dimension is equivalent to searching
for a particular set of CHM~\cite{BBLTZ07}.

The role of complex Hadamard matrices in theoretical and experimental physics cannot be overestimated.
Let us briefly recall the most crucial examples of applications only in quantum information theory.
CHM are significant components behind the concept of {\sl dense coding} and {\sl quantum teleportation schemes}~\cite{We01}.
They form building blocks for {\sl nice error bases}~\cite{Kn96} and {\sl quantum designs}~\cite{Ha97,GS81}.
Constructions of {\sl unitary error bases} (UEB), {\sl bases of maximally entangled states} and {\sl unitary depolarizers}
directly exploit the existence of elements from $\mathbb{H}(N)$~\cite{We01, KR04, MV16}.
Optimal {\sl tomography of quantum states} is possible due to
measurement performed by means of MUB, which --- as we have just mentioned --- can be built out of CHM~\cite{Iv81, WF89}.
Additional applications of such matrices in pure mathematics can be found in the comprehensive monograph of Teo Banica~\cite{Ba21}.

There are many constructions of CHM. From simply rearranging of elements and/or expanding the size
of a given matrix~\cite{Wi44, Di04}, through
more advanced mathematical apparatus involving tiling abelian groups~\cite{MRS07}, 
graph theory~\cite{GS70, La16}, Golay sequences~\cite{LSO13}, and
orthogonal maximal abelian $*$-subalgebras~\cite{BN06}.
Further examples and references are summarized in the Catalogue of CHM~\cite{CHM_catalogue}.

It is easy to notice that $\forall\,N : \mathbb{H}(N)\neq\emptyset$ since such
an example is provided for any dimension $N$ by the Fourier matrix $F_N$,
\begin{equation}
F_N=\frac{1}{\sqrt{N}}\sum_{j,k=0}^{N-1}|j\rangle\langle k|\exp\Bigg\{\frac{2\pi ijk}{N}\Bigg\}\in{\rm B}\mathbb{H}(N,N)\subset\mathbb{H}(N).
\end{equation}
A bit more complicated example of a single parameter family of CHM of order $N=6$ reads
\begin{equation}
T_6^{(1)}(\gamma)=\left[\begin{array}{rrrrrr}
1 & 1   & 1   &  1 &  1 &  1\\
1 & a   & b   &  c &  a &  d\\
1 & b   & a   &  a &  c &  d\\
1 & b   & a   & -a & -1 & -d\\
1 & a   & b   & -1 & -a & -d\\
1 & c   & c   & -c & -c & -1
\end{array}\right],\label{matrix_T6}
\end{equation}
where an arbitrary phase $\gamma$ varies continuously, $\gamma\in[\frac{1}{2},\frac{3}{2}]$,
and the unimodular numbers are $a =\exp\{i\pi \gamma\}$, and
\begin{align}
b &=a^2,\\ 
c &= \frac{-2a - 1 - b - \sqrt{a^4+4a^3+2a^2+4a+1}}{2},\label{T6c}\\
d &= \frac{-2a - 1 - b + \sqrt{a^4+4a^3+2a^2+4a+1}}{2}.
\end{align}
Matrix $T_6^{(1)}$ is a slight modification of the first family in $\mathbb{H}(6)$
whose elements depend on each other in a nonlinear way.
It was found independently in $2006$~\cite{BBLTZ07}, and quickly shadowed by the equivalent and simpler form~\cite{BN06}.

\section{Classification and Invariants}

The main problem concerning CHM is their classification~\cite{TZ06, CHM_catalogue, SzPhD}.
Classifying all CHM of a given order, say $N=6$, would be helpful in tackling various problems in quantum information
theory, for instance the existence of MUB in this dimension.
There is no simple method that can tell if a given matrix represents a new or already known example of CHM.
It is mainly caused by additional degrees of freedom in $\mathbb{H}(N)$.
Two matrices $H_1$, $H_2\in\mathbb{H}(N)$ are called {\sl equivalent}~\cite{Ha96}, written
$H_1\simeq H_2$, if there
exist two diagonal unitary matrices
$D_1$, $D_2\in\mathbb{U}(N)$ and two permutation matrices
$P_1$, $P_2\in\mathbb{P}(N)$ such that
\begin{equation}
H_1 = D_1 P_1 H_2 P_2 D_2.\label{equivalence_relation}
\end{equation}

\begin{figure}[ht!]
\center
\includegraphics[width=3.5in]{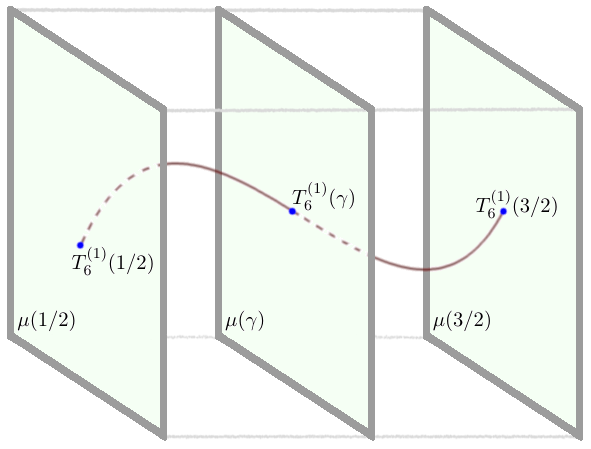}
\caption{Schematic illustration of the orbit of $T_6^{(1)}(\gamma)$~\eqref{matrix_T6} as a function of parameter
$\frac{1}{2}\leqslant \gamma\leqslant\frac{3}{2}$.
Given $\gamma$, each $\mu=\mu(\gamma)$ represents the equivalence space for $T_6^{(1)}(\gamma)$, that is the set of all possible products
$D_1P_1T_6^{(1)}(\gamma)P_2D_2$ for diagonal and permutation matrices defined in~\eqref{equivalence_relation}.
Equivalence space may form a multidimensional continuous manifold
% or can be just one particular class $[\mu(\gamma)]\in\mathbb{H}(N)/_{\simeq}$.
as above, where only three particular sections are depicted. For an isolated matrix there would be only one such space.}
\label{fig:T6_orbit}
\end{figure}

CHM come in a variety of forms.
We distinguish here three general classes: {\sl isolated points}, {\sl affine}
and {\sl nonaffine families} ({\sl orbits}). We say that $H$ is an {\sl isolated} matrix,
if its neighborhood\footnote{By a matrix neighborhood
we understand a natural generalization of a standard point-like neighborhood
extended over $\mathbb{C}^{N\times N}$ endowed with Euclidean topology.}
contains only equivalent matrices.
Otherwise, it is a part of a family of inequivalent matrices.
We decorate matrices with a superscript to indicate whether it is an isolated point, e.g. isolated matrix $S_6^{(0)}$ of order~six,
or it represents $\delta$-dimensional orbit (orbit depending on $\delta$ independent parameters), for instance,
a single parameter orbit $T_6^{(1)}$ of order~six too.
The character of variability of phases $\varphi_{jk}$ in $H_{jk} = e^{i 2 \pi \varphi_{jk}}$ as
functions of orbit parameters can be linear ({\sl affine}) or nonlinear ({\sl nonaffine}).
The latter case is represented by $T_6^{(1)}$~\eqref{matrix_T6}
and drawn schematically in Figure~\ref{fig:T6_orbit}.
Obviously, the dependence $c=c(a)=c(\gamma)$ in~\eqref{T6c}, where $\gamma$ is the only
orbit parameter, is highly nonlinear.
Plethora of examples of affine families can be found in~\cite{TZ06, CHM_catalogue}, and
the construction of a bit complicated nonaffine family will be presented in Section~\ref{sec:matrix_T8}.

%\newpage

We will mostly present matrices in the special {\sl dephased form} in which
first row and first column consists of ones.
Every CHM can be brought to its dephased form by multiplying by two diagonal unitary matrices.
This {\sl normalized} representation allows us to consider only permutation matrices
in the equivalence relation, and effectively work
with the $(N-1)$-dimensional {\sl core} of a given matrix.

Having equivalence relation we can introduce the notion of the quotient set
$\mathbb{H}(N)/_{\simeq}=[H_{(1)}]\cup [H_{(2)}]\cup ...\cup[H_{(j)}]\cup...$, so
the classification of CHM boils down
to determining all equivalence classes with respect to~$\simeq$.
Low dimensions, $N=2$, $3$, $4$, and $5$, have been completely described in~\cite{Ha96,Cr91} and the first open problem arises at $N=6$.
It is widely believed that the set of six-dimensional CHM consists of a single isolated
spectral matrix~\cite{Ta04} and $4$-parameter nonaffine family~\cite{Sz12} that connects all other $6$-dimensional examples,
\begin{equation}
\mathbb{H}(6)\stackrel{?}{=}S_6^{(0)}\cup \Big\{ G_6^{(4)}\Big\}.
\end{equation}

The diversity of forms for any $N\geqslant 6$ prevents us from
giving a single and simple answer to the question about the structure of $\mathbb{H}(N)$.
The task of checking whether a given matrix belongs to a known class is not easy indeed,
but there exist several methods~\cite{FG03, Ni06, Sz10} which can be used
to classify or rule out a matrix from a given subset of CHM.
We will briefly recall two of them: {\sl defect of a unitary matrix} and the set of {\sl Haagerup invariants}~\cite{Ha96}
as the main tools used in the rest of the chapter.

Defect of a matrix $H\in\mathbb{H}(N)$, denoted ${\bf d}(H)$,
was introduced and investigated in~\cite{TZ08, Ta18} as an algebraic tool
to solve the problem of possibility of deriving from a given matrix
a smooth family of inequivalent matrices.
Not only it can serve as a binary oracle that tells whether it is possible (or not) to stem
an orbit out of a given Hadamard matrix $H$,
but it can also assist in identifying and classifying the equivalence classes.
We skip here all mathematical formalism related to this concept
however, in Chapter~\ref{chap:DELTA} we will present the sketch of derivation of a similar tool in the context
of Hermitian unitary matrices.
Informally, we can treat ${\bf d}(H)$ as a non-negative number associated with a matrix $H$, which provides
the information about possible independent directions on the manifold of $\sqrt{N}\mathbb{U}(N)$ that $H$
can follow preserving the property of being CHM.
In other words, non-zero value of ${\bf d}(H)$ determines the upper bound of dimensionality of a family that
$H$ might (but not necessarily should) belong to.
In particular, we will extensively use the following fact (Lemma 3.3 in~\cite{TZ06}):
\begin{equation}
{\bf d}(H_N)=0 \Longrightarrow H_N=H_N^{(0)},\label{DEFECT_implication}
\end{equation}
so vanishing defect implies that $H_N$ cannot be part of any orbit of inequivalent matrices, and it is an isolated Hadamard matrix.
However, this is only one-way criterion and we know examples of isolated matrices for which defect does not vanish~\cite{MS19}.
Matrix $H$ which has non-zero defect and is a member of multidimensional orbit depending on
several parameters $\gamma_j$ for $1<j\leqslant {\bf d}(H)$ can have different defects for different values
of $\gamma_j$. This is simply explained by the fact that such hyperorbit might intersect other
orbits of different characteristics.

The Haagerup set~\cite{Ha96} of $H$, denoted by $\Lambda(H)$ is defined by
all possible products of quartets of matrix elements
\begin{equation}
\Lambda(H) = \Big\{H_{jk}H_{lm}H_{jm}^*H_{lk}^* : j,k,l,m\in\{1,...,N\}\Big\}\label{Lambda_H},
\end{equation}
where $^*$ is complex conjugate. Both, defect and Haagerup set are invariant with
respect to equivalence relation~\eqref{equivalence_relation}, which means that
${\bf d}(H) = {\bf d}(P_1 D_1 H D_2 P_2)$ and
$\Lambda(H) = \Lambda(P_1 D_1 H D_2 P_2)$, hence
we can formulate another helpful, criterion
\begin{equation}
\Lambda(H_1) \neq \Lambda(H_2) \Longrightarrow H_1 \not\simeq H_2.\label{LAMBDA_implication}
\end{equation}
Again, the converse is not true, and it is possible to
find two inequivalent matrices with identical sets $\Lambda$. 
Nevertheless, even though~\eqref{DEFECT_implication} and~\eqref{LAMBDA_implication} work only one way, they
will be extremely helpful in discriminating several cases of new matrices.

\medskip

We are ready to proceed to extend the set of known CHM.
As the work on dimension~$N=6$ can be roughly considered done and the next
$7$-dimensional set proved to be exceptionally hard,
we start with $N=8$. %introducing new complex Hadamard matrices in $\mathbb{H}(8)$.

\section{Extension of the Set $\mathbb{H}(8)$ of CHM of Order $8$}

As of 2018, the set of CHM of size eight contained the following elements:
$D_8^{(4)}$~\cite{Di09}, $S_8^{(4)}$~\cite{MRS07}, Fourier $F_8^{(5)}$, tensor product
of two Fouriers $F_2\otimes F_4\in{\rm B}\mathbb{H}(8,4)$~\cite{TZ06}, and
$V_8^{(0)}$~\cite{VE_private}.
Also a collection of Butson type matrices introduced in~\cite{OLS20} contribute to $\mathbb{H}(8)$.
It will be shown below that this list of matrices in $\mathbb{H}(8)$ can be extended.

\subsection{Random Walk over Core Phases}
\label{sec:RWCP}

Numerical algorithm used to search for new matrices $H\in\mathbb{H}(8)$ 
comes down to three simple stages:
\begin{enumerate}
\item draw phases $\varphi_{jk}$ of the core of $M=M(\varphi_{jk})$ at random, possibly fix some of them to impose a desired pattern,
\item gradually change the values of (unfixed) phases in order to minimize a target
function
\begin{equation}
\mathcal{Z}(M)=||MM^{\dagger}-N\mathbb{I}_N||_{\rm F},\label{ZIEL_function}
\end{equation}
where subscript ${\rm _F}$ denotes
Frobenius norm of a matrix and, here, $N=8$,
\item check the output for belonging to a given equivalence class and retrieve the analytical form.
\end{enumerate}
For $N\in\{6,7,8,9,10\}$ almost every initial (randomly chosen) matrix $M$ swiftly converges to $H$ 
such that $\mathcal{Z}(H)\approx 0$, so $H$ can be considered as a numerical approximation of CHM with arbitrarily high precision.
However, it is very common that in the last phase
we cannot simply discard or accept $H$ due to the lack of mathematical tools that can unequivocally
determine the status of the newly obtained matrix.
In general the most difficult task is to recover analytical values of the numerical result. % in the post-processing stage.
This can be slowly done in a systematic process of fixing phases and reiterating the optimization procedure until
we begin to observe more easy-to-guess numbers. With this method we can quite efficiently determine
individual matrices. If we anticipate that matrix may belong to a family (especially, a nonaffine family),
we should fix phases in a clever way, customized to a particular
problem, so that subsequent optimizations ought to return some apparent patterns.
Eventually, mixing all together,
and solving orthogonality constraints for $H$, we can get ultimate functional dependencies among the
entries of the matrix.

\subsection{New Nonaffine Family $T_8^{(1)}$}
\label{sec:matrix_T8}

During numerical generation of matrices we noted that in the sequence of many outputs there is
one, $T_8$, such that its defect amounts to $3$.
This specific value had not been observed before for any matrix in $\mathbb{H}(8)$, and
according to the aforementioned fact that defect is invariant with respect to permutations,
it was a clear signature that $T_8$ may represent a new class of equivalence.
Then, after several attempts,
we were able to extrapolate analytically some numerically expressed relations in
the matrix core and reconstruct
entire character of the matrix $T_8$ admitting the form of a nonaffine family:
\begin{equation}
T_8^{(1)}(\gamma)={\rm EXP}\left(i\frac{\pi}{10} R_8\right)
\circ\left[\begin{array}{rrrrrrrr}
1 & 1 & 1 & 1 & 1 & 1 & 1 & 1\\
1 & x & y & z & -\frac{y z}{x} & 1 & 1 & 1 \\
1 & -\frac{u}{y} & \frac{u}{x} & -\frac{u z}{x y} & -\frac{u z}{x^2} & 1 & 1 & 1\\ 
1 & y & x & -\frac{y z}{x} & z & 1 & 1 & 1\\ 
1 & \frac{u}{x} & -\frac{u}{y} & -\frac{u z}{x^2} & -\frac{u z}{ x y} & 1 & 1 & 1\\ 
1 & -i\frac{x}{z} & i\frac{x}{z} & -i\frac{z}{x} & i\frac{z}{x} & 1 & 1 & 1\\ 
1 & u & -u & -\frac{u z^2}{x^2} & \frac{u z^2}{x^2} & 1 & 1 & 1\\ 
1 & i\frac{u z}{x} & i\frac{u z}{x} & -i\frac{u z}{x} & -i\frac{u z}{x} & 1 & 1 &1 \\ 
\end{array}\right]\label{T8_matrix}
\end{equation}
where the matrix of phases reads ($\bullet=0$)
\begin{equation}
R_8=\left[\begin{array}{rrrrrrrr}
 \bullet & \bullet & \bullet &   \bullet &   \bullet &   \bullet &   \bullet &   \bullet\\
 \bullet & \bullet & \bullet &   \bullet &   \bullet &   3 & 15 & 18\\
 \bullet & 8 & 8 &   8 &   8 & 13 & 15 &   8\\
 \bullet & 2 & 2 &   2 &   2 & 17 &   5 &   2\\
 \bullet & \bullet & \bullet &   \bullet &   \bullet &   7 &   5 & 12\\
 \bullet & \bullet & \bullet &   \bullet &   \bullet &   \bullet & 10 & 10\\
 \bullet & \bullet & \bullet &   \bullet &   \bullet & 10 &   \bullet & 10\\
 \bullet & \bullet & \bullet &   \bullet &   \bullet & 10 & 10 &   \bullet\\
 \end{array}\right],
\end{equation}
and depends on a single parameter $\gamma\in[\frac{2}{5},\frac{4}{5}]\cup[1,\frac{7}{5}]$ hidden in $x=e^{i\pi\gamma}$.
The range of~the orbit parameter $\gamma$, being the sum of two intervals,
is given only as the approximation. Accurate limits require complex calculations, but
the presented scope is absolutely sufficient to ensure unimodularity of entries of $T_8^{(1)}$.
For the same reason, formulas for $y=y(x)$, $z=z(x,y)$, and $u=u(x,y,z)$ %(while $x=e^{i\pi\gamma}$)
which are overly complicated, should not be reproduced here in full.
Their explicit forms do not add anything to the further analysis, perhaps except the fact how
rich the structure of $\mathbb{H}(N)$ is as $N\to\infty$.
Detailed description of $T_8^{(1)}$ can be found in~$[\hyperlink{\paperslist}{\rm A1}]$.

Meanwhile, let us observe that for particular values of $\gamma$,
two different Butson type matrices
can be extracted from the orbit $T_8^{(1)}(\gamma)$.
For $\gamma\in\big\{\frac{1}{2}, 1, \frac{13}{10}\big\}$ we obtain three matrices equivalent to
\begin{equation}
B_{8a}={\rm EXP}\left(i\frac{\pi}{10}\left[\begin{array}{rrrrrrrr}
\bullet &   \bullet &   \bullet &   \bullet &   \bullet &   \bullet &   \bullet &   \bullet\\
\bullet &   5 & 10 & 13 &   8 &   3 & 15 & 18\\
\bullet & 10 &   5 & 18 &   3 & 13 & 15 &   8\\
\bullet & 12 &   7 & 10 & 15 & 17 &   5 &   2\\
\bullet & 17 &   2 & 15 & 10 &   7 &   5 & 12\\
\bullet &   7 & 17 &   3 & 13 &   \bullet & 10 & 10\\
\bullet &   2 & 12 &   8 & 18 & 10 &   \bullet & 10\\
\bullet & 15 & 15 &   5 &   5 & 10 & 10 &   \bullet
\end{array}\right]\right)\in{\rm B}\mathbb{H}(8,20),
\end{equation}
and another Butson type matrix for $\gamma=\frac{4}{5}$
\begin{equation}
B_{8b}={\rm EXP}\left(i\frac{\pi}{10}\left[\begin{array}{rrrrrrrr}
\bullet &   \bullet &   \bullet &   \bullet &   \bullet &   \bullet &   \bullet &   \bullet\\
\bullet &  8 & 10 & 13 &  5 &  3 & 15 & 18\\
\bullet & 18 & 10 &  3 &  5 & 13 & 15 &   8\\
\bullet & 12 & 10 &  7 & 15 & 17 &   5 &   2\\
\bullet &  2 & 10 & 17 & 15 &  7 &   5 & 12\\
\bullet & 10 &  \bullet &  \bullet & 10 &   \bullet & 10 & 10\\
\bullet & 10 &  \bullet & 10 &  \bullet & 10 &   \bullet & 10\\
\bullet &  \bullet &  \bullet & 10 & 10 & 10 & 10 &   \bullet
\end{array}\right]\right)\in{\rm B}\mathbb{H}(8,20).
\end{equation}
Different defects ${\bf d}(B_{8a})=7$ and ${\bf d}(B_{8b})=11$ prove inequivalence of the two
representatives, and value of $q=20$~\eqref{BH_class} confirms the novelty of the finding,
since at the time of publication of~$[\hyperlink{\paperslist}{\rm A1}]$,
such matrices did not appear in the literature nor in online catalogs~\cite{CHM_catalogue, BH_home}.

The defect ${\bf d}\big(T_8^{(1)}\big)=3$ theoretically allows $T_8^{(1)}$ to be a part of at most three-dimensional family,
and during the laborious numerical analysis, we encountered strong numerical evidence supporting such possibility;
$\big\{T_8^{(1)}\big\} \subset \big\{T_8^{(3)}\big\}$.
However, the complexity of the internal structure prevents us from determining (or even from guessing) the explicit
formulas of such a family. At this stage we should consider Eq.~\eqref{T8_matrix} as the final form.

Dimension $N=8=2^3$, being a square of prime, facilitates several constructions~\cite{Ch02},
implications and applications mentioned at the beginning of this section.
Nevertheless, still, apart from a one step further in full classification of CHM, we can use
the matrix $T_8^{(1)}$ as an ingredient to build new pairs of MUB, UEB, quantum gates,
or just special orthogonal bases in $\mathbb{C}^8$.

In the next section, we present another method to obtain CHM numerically,
and later we will (re)use it as a starting point for % to
a more general construction leading to special families of CHM.

\section{Extension of the Set $\mathbb{H}(9)$ of CHM of Order $9$}

Similarly to $\mathbb{H}(8)$, all known CHM of order $N=9$ forming $\mathbb{H}(9)$ can be listed as follows:
$F_3\otimes F_3$ and $F_9^{(4)}$~\cite{TZ06}, $S_9^{(0)}$~\cite{MW12},
$B_9^{(0)}$~\cite{BN06},
$N_9^{(0)}$~\cite{BN06}, and $K_9^{(2)}$~\cite{Ka16}.
Again, a number of Butson type matrices~\cite{OLS20} must be added to complete the picture.

\subsection{Sinkhorn Algorithm Revisited}
\label{sec:Sinkhorn_AA}

Before we continue the subject of CHM, we shall recall another numerical recipe
proposed by Richard Sinkhorn in 1964
that was originally designed to generate (random) bistochastic matrices~\cite{Si64, SK67}.

Suppose we have a mathematical object (e.g. matrix) which is characterized by at least two properties $\mathcal{P}_j$ ($j\geqslant 1$).
They need not be complementary, nor disjoint. For example $\mathbb{H}(N)$
can be equivalently redefined as intersection of two sets (properties): rescaled unitary matrices and unimodular matrices,
\begin{equation}
\mathbb{H}(N) = \sqrt{N}\mathbb{U}(N) \cap \mathbb{T}(N),
\end{equation}
where $\mathbb{T}(N)$ denotes an $N$-dimensional torus.

To get a matrix $M^{*}$ having all desired properties, $M^{*}\in\mathcal{P}_1\cap\mathcal{P}_2\cap...$
we can try to perform alternating mappings (projections) $\pi_j$ onto $\mathcal{P}_j$, applied to a given initial argument $M$, so that
$\pi_j(M)\in\mathcal{P}_j$.
Provided that such iterative procedure exhibits contractive behavior, it should also be convergent
to an element in a non-empty subset $\mathcal{P}$, common for all $\mathcal{P}_j$,
\begin{equation}
\lim_{k\to\infty}\Big(\prod_{j=1}^k\pi_j\Big)(M)=M^{*}\in\mathcal{P}\subset\bigcap_{j}\mathcal{P}_j,
\end{equation}
where the product of $\pi_j$'s above should be understood in terms of map compositions.

In the particular case of CHM, the algorithm takes the following form:
\begin{enumerate}
\item draw a complex matrix $M\in\left(\mathbb{C}\setminus\{(0,0)\}\right)^{N\times N}$ at random (we exclude zeros to avoid problems in the next step),
\item normalize (unimodularize) each entry of $M$ so $M\to M'=\pi_1(M)\in\mathbb{T}^N$ such that $M'_{jk}=M_{jk}/|M_{jk}|$,
\item perform polar decomposition of $M'=U\sqrt{M'^{\dagger}M'}$ to obtain (nearest) unitary matrix
$U=M'/\sqrt{M'{^\dagger}M'}$, so $M'\to M''=\pi_2(M')$ such that $M''=M'/\sqrt{M'^{\dagger}M'}$,
\item repeat steps 2. and 3. until $\mathcal{Z}(\pi_2(\pi_1(...(\pi_1(M))...)))\approx 0$ up to
a given precision (where the target function $\mathcal{Z}$ is defined in~\eqref{ZIEL_function}).
\end{enumerate}
The convergence of the alternating procedure is assured in the case of convexity
of all components $\mathcal{P}_j$~\cite{BB96}.
Despite the fact that none of subsets determining $\mathbb{H}(N)$ is convex,
the above method applied to an initial matrix $M\in\left(\mathbb{C}\setminus\{(0,0)\}\right)^{N\times N}$
can effectively produce very accurate approximations of CHM from $\mathbb{H}(N)$.
We are not going to investigate mathematical details of this algorithm,
instead, we will use it as a reliable tool that proved its fitness in many numerical simulations.
The advantage of this technique over the previous one is the speed and performance.
However, one pays the price of loosing full control over the matrix structure, 
as it is no longer possible to easily fix given entries of $M$ imposing a concrete appearance.
Nevertheless, as we will see, this method can produce matrices with a surprisingly high degree of symmetry.
Original variant of this algorithm was successfully used in construction
of bistochastic matrices~\cite{CSBZ09}.
Slightly modified version will provide the main result in Chapter~\ref{chap:AME46}.
In the next section we present a $9$-dimensional CHM obtained in this way, the form
of which will trigger additional observations and further results.

\subsection{A Novel Isolated Matrix $T_9^{(0)}$}
\label{sec:matrix_T9}

We present another CHM obtained during numerical studies with the help
of Sinkhorn algorithm described in previous section.\footnote{No phases were fixed, all entries were initially drawn at random according to normal distribution and such initial point was supplied
to the algorithm.}
Originally, after dephasing,\footnote{Usually matrices produced in this way
are not in the dephased form. We call them {\sl undephased}.} which sets all entries of the first row and the first column to one,
the matrix emerged in the form of
\begin{equation}
T_9=\left[
\begin{array}{lllllllll}
1 & 1& 1& 1& 1& 1& 1& 1& 1\\
1 & a & d & a^{*} &  c^{*} &  b^{*} & c & b & d^{*}\\
1 & b & c & b^{*} & a & d^{*} & a^{*} & d & c^{*}\\
1 & c & b^{*} & c^{*} & d & a & d^{*} & a^{*} & b\\
1 & b^{*} & c^{*} & b & a^{*} & d & a & d^{*} & c\\
1 & d & a^{*} & d^{*} & b & c^{*} &  b^{*} & c & a\\
1 & a^{*} & d^{*} & a & c & b & c^{*} & b^{*} & d\\
1 & c^{*} & b & c & d^{*} & a^{*} & d & a & b^{*}\\
1 & d^{*} & a & d & b^{*} & c & b & c^{*} & a^{*}
\end{array}\right]\label{T9}
\end{equation}
\begin{align}
\text{with}:\qquad a&\approx -0.3396 + 0.9406i, \qquad b\approx -0.9635 + 0.2676i,\label{quadruplet_ab}\\
c&\approx -0.0365 + 0.9993i, \qquad d\approx +0.8396 + 0.5432i,\label{quadruplet_cd}
\end{align}
and $x^{*}=\frac{1}{x}$ denoting complex conjugate of unimodular numbers.
As in the previous case, attention was focused on this particular example because of the individual features
that distinguish it from other $9$-dimensional candidates. Namely, its defect vanishes and rough 
analysis shows that it does not belong to the Butson type class~\eqref{BH_class}.

There are only four different parameters $a$, $b$, $c$, and $d$ in $T_9$, which is very
favorable circumstance (in comparison with a core of $64$ different values that in general would
make it hardly possible to deal with). 
To express these parameters analytically, one can try to solve the unitarity constraints for $T_9$.
They form a system of five nonlinear equations with four variables
\begin{equation}
\left\{
\begin{array}{r}
1+a+\dfrac{1}{a}+b+\dfrac{1}{b}+c+\dfrac{1}{c}+d+\dfrac{1}{d}=0,\\
\\
1+a^2+\dfrac{1}{a^2}+b^2+\dfrac{1}{b^2}+c^2+\dfrac{1}{c^2}+d^2+\dfrac{1}{d^2}=0,\\
\\
1+\dfrac{a}{b}+\dfrac{b}{a}+\dfrac{b}{d}+\dfrac{d}{b}+\dfrac{c}{d}+\dfrac{d}{c}+ac+\dfrac{1}{ac}=0,\\
\\
1+\dfrac{a}{c}+\dfrac{c}{a}+ab+\dfrac{1}{ab}+cd+\dfrac{1}{cd}+bd+\dfrac{1}{bd}=0,\\
\\
1+\dfrac{a}{d}+\dfrac{d}{a}+\dfrac{b}{c}+\dfrac{c}{b}+bc+\dfrac{1}{bc}+ad+\dfrac{1}{ad}=0.
\end{array}\label{UC_T9}
\right.
\end{equation}
There are many solutions of the above system, including those which
do not meet the condition of unimodularity.
Supported by numerical software, we find that $128$ possible quadruplets $(a,b,c,d)$
fulfilling~\eqref{UC_T9}
can be divided onto three distinct classes. Class a) solutions that correspond to $T_9$, class b)
can be used to build a Butson type matrix of size $9$, and class c)
to be discarded as they do not provide CHM.
We should not continue the thread about the Butson solution, because in this very case it
does not reveal anything new. Let us focus on the class a) --- solution for $T_9$.

\begin{figure}[ht!]
\center
\includegraphics[width=5in]{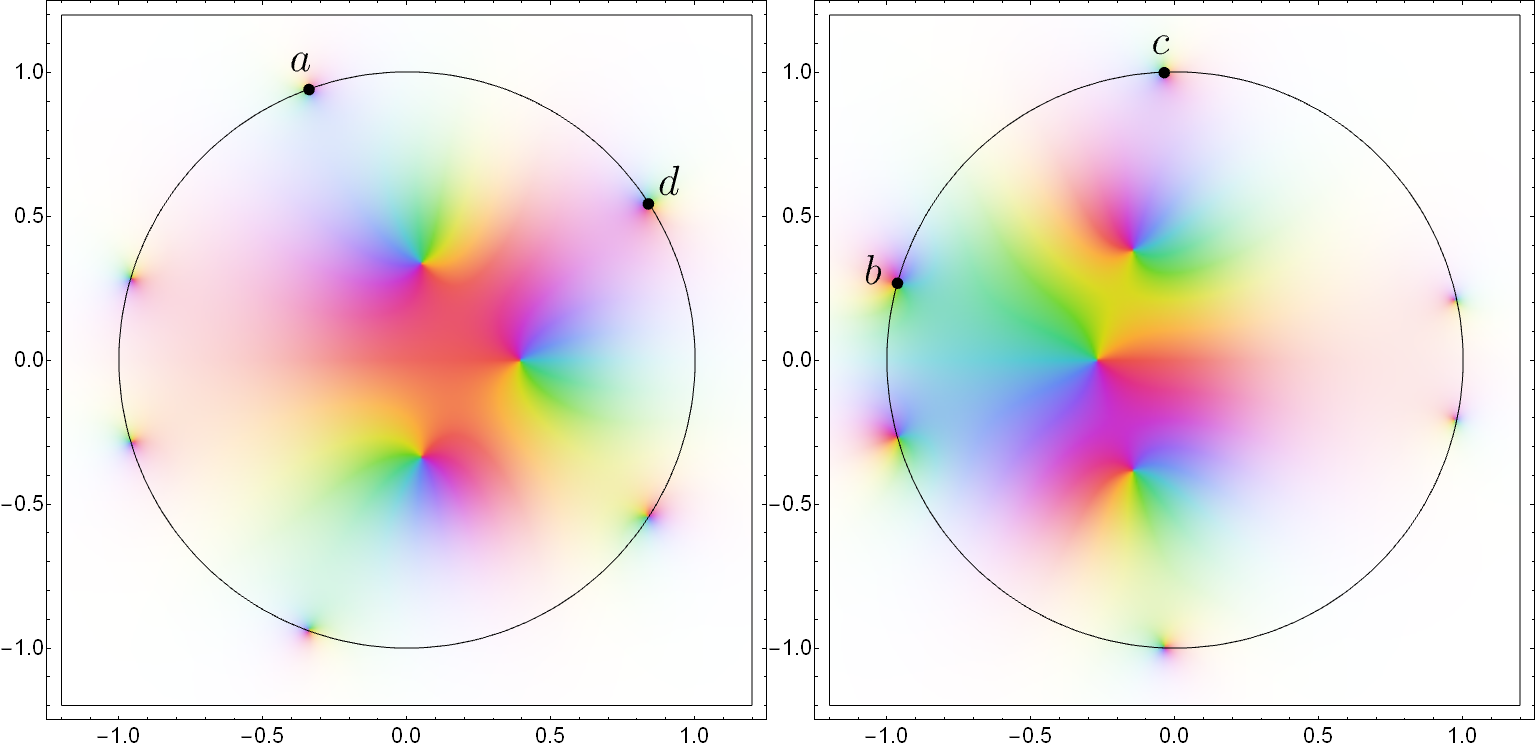}
\caption{Complex roots of two palindromic polynomials ${\bf p}_1$ and ${\bf p}_2$
associated with unitarity constraints~\eqref{UC_T9}
for $T_9$. Six of them (excluding complex
conjugates) from
the unit circle contain four parameters $a$, $b$, $c$, and $d$~\eqref{quadruplet_ab}--\eqref{quadruplet_cd}
that characterize $T_9$.}
\label{fig:palindromic_roots}
\end{figure}

After rearranging, reducing and simplifying~\eqref{UC_T9}, we observe that numbers in~\eqref{quadruplet_ab} and~\eqref{quadruplet_cd}
can be calculated as roots of two monic palindromic polynomials\footnote{Monic polynomial ${\bf p}$ of degree $n$ has leading coefficient $c_n=1$. For brevity,
we write only coefficients of ${\bf p}$ (in ascending order).
For example, the standard polynomial notation ${\bf p}(x)=1+2x+3x^2+2x^3+x^4$ corresponds to $(1,2,\underline{3},2,1)$, where we emphasized
the middle factor to ease confirmation that they really form a palindromic sequence: $c_j=c_{n-j}$ for $j\in\{0,...,n\}$.} or degree $12$ (see Figure~\ref{fig:palindromic_roots})
\begin{align}
{\bf p}_1(x)&=(1, -3, 9, -16, -12, 6, \underline{3}, 6, -12, -16, 9, -3, 1),\\
{\bf p}_2(x)&=(1, 6, 15, 26, 3, -24, \underline{-27}, -24, 3, 26, 15, 6, 1).
\end{align}
We exploit the well known fact from algebra of palindromic polynomials which says that
every such ${\bf p}(x)$ of even degree $2k$ can be expressed as
\begin{equation}
{\bf p}(x)=x^{k}{\bf q}\big(x+1/x\big),\label{palindromic_reduction}
\end{equation}
where ${\bf q}$ is another polynomial (in variable $y=x+1/x$) of degree $k$~(cf. Section C.2 in \cite{SzPhD}).
Sometimes, this can shorten the form of ${\bf p}(x)$ significantly.
Indeed, in our case the problem can be reduced to 
calculating the roots of polynomials of $3^{\rm rd}$ degree
and, eventually, we arrive at\footnote{The Author is indebted to Oliver Reardon-Smith for pointing out an error in earlier calculations.}
\begin{align}
a&=\sqrt{\gamma_1^2-1}-\gamma_1,\quad b=\sqrt{\gamma_2^2-1}-\gamma_2,\\
c&=\sqrt{\gamma_3^2-1}+\gamma_3,\quad d=\sqrt{\gamma_4^2-1}+\gamma_4,
\end{align}
where
\begin{align}
\gamma_1&=\frac{1}{4}\sqrt{1-\zeta_2-\zeta_2^*}-\frac{1}{4},\qquad \gamma_2=\frac{\sqrt{2}}{2}\sqrt{1-\zeta_1-\zeta_1^*}+\frac{1}{2},\\
\gamma_4&=\frac{1}{4}\sqrt{1-\zeta_2-\zeta_2^*}+\frac{1}{4},\qquad \gamma_3=\frac{\sqrt{2}}{2}\sqrt{1-\zeta_1-\zeta_1^*}-\frac{1}{2},
\end{align}
and
\begin{align}
\zeta_1&=\frac{\omega}{2^{4/3}}\left(43-3i\sqrt{771}\right)^{1/3},\\
\zeta_2&=2^{5/3}\omega^2\left(43+3i\sqrt{771}\right)^{1/3} \quad \text{with} \quad  \omega=\exp\Big\{i \pi \frac{5}{3}\Big\}.
\end{align}
Analytic expressions for $a$, $b$, $c$ and $d$ match perfectly numerical values in~\eqref{quadruplet_ab} and~\eqref{quadruplet_cd}.

Now we can formally confirm that $T_9\in\mathbb{H}(9)$ and its defect ${\bf d}(T_9)=0$, so $T_9=T_9^{(0)}$ is an isolated matrix.
Obviously $T_9\not\in{\rm B}\mathbb{H}(9,q)$ for any $q$.
Since the only known isolated matrices
in $\mathbb{H}(9)$ are
$N_9^{(0)}$ whose entries belong to $\mathbb{Q}\left[\frac{1}{4}\left(1-i\sqrt{15}\right)\right]$,
$S_9^{(0)}\in{\rm B}\mathbb{H}(9,6)$, and several other Butson type examples,
we ultimately prove that the matrix $T_9^{(0)}$ is the next new representative of $9$-dimensional set of CHM.

\section{Families of (Isolated) Complex Hadamard Matrices}
\label{sec:CHM_LS}

Up to this moment we were generating CHM numerically and then, in the post-processing phase,
they underwent tedious analysis to be finally presented in an analytic form.
In this section, we propose a more systematic construction 
which might lead to infinite families of CHM with additional property of being isolated.

Let us recall the definition of an object that will be
intensively studied in Chapter~\ref{chap:AME46}.
A {\sl Latin square} (LS) of size $N$ is an array of $N\times N$ elements (numbers or symbols)
such that in every row and in every column there is no repetition of two identical components.
Two Latin squares are said to be {\sl orthogonal} (OLS) if, when superimposed, they contain distinct
(ordered) pairs of symbols~\cite{CD01}. The example below explains why OLS are also called Graeco-Latin squares
\begin{equation}
\left[\begin{array}{ccc}
a & b & c\\
c & a & b\\
b & c & a
\end{array}\right]\
\cup \
\left[\begin{array}{ccc}
\alpha & \beta & \gamma\\
\beta & \gamma & \alpha\\
\gamma & \alpha & \beta
\end{array}\right] \
= \
\left[\begin{array}{ccc}
a\alpha & b\beta & c\gamma\\
c\beta & a\gamma & b\alpha\\
b\gamma & c\alpha & a\beta
\end{array}\right].
\end{equation}

There are known connections that employ both of LS and (complex) Hadamard matrices and combine
them together in order to build more advanced structures. One of
these is a method of building MUB
in square dimensions~\cite{WB05}.
Also, they are tightly connected via the construction of
orthonormal bases of unitary operators, for which
the existence of Hadamard matrix and LS is required~\cite{We01, MV16}.
Also, the existence of special classes of (real) Hadamard matrices can be related to orthogonal LS~\cite{Ve82}.
However, it is not known to us, whether there exists a systematic method of constructing
complex Hadamard matrices that explicitly makes use of the idea of LS.
In the recent paper~\cite{Al18} contributing to the field of statistical analysis,
the author introduces the term ``Latin Hadamard Matrix'',
but beside a familiar name, it is by no means related to our current problem.

Our proposition directly involves the concept of LS (or even
more general combinatorial structures).
As we have seen, to build a CHM one can a priori impose certain initial configuration on entries in
a matrix (like in Eq.~\eqref{T9}),
and solve the associated system of nonlinear equations representing unitarity constraints.
However, not every configuration is valid,
as there might exist some patterns for which the only solution is the empty set.
This technique, already presented in the previous section,
gave rise to a new matrix $T_9^{(0)}\in\mathbb{H}(9)$. We will show that
particularly chosen configuration of entries
leads to several other previously unknown CHM of order $N\geqslant 9$,
and may set the ground for a more general conjecture.

\subsection{Family of Complex Hadamard Matrices $L_N$}
\label{sec:family_LN}

We will follow a heuristic reasoning. Consider again the very first form of $T_9^{(0)}$~\eqref{T9}.
Permuting rows and columns allows us to present it equivalently as
\begin{equation}
T_9^{(0)}\simeq
\left[
\begin{array}{l|ll|ll|ll|ll}
1 & 1 &1 &1 &1 &1 &1 &1 &1\\
\hline
1 & a & a^{*}&  b & b^{*} & c & c^{*} & d & d^{*}\\
1 & a^{*} & a & b^{*} & b & c^{*} & c & d^{*} & d\\
\hline
1 & b & b^{*} & d & d^{*} & a^{*} & a & c & c^{*}\\
1 & b^{*} & b & d^{*} & d & a & a^{*} & c^{*} & c\\
\hline
1 & c & c^{*} & a^{*} & a & d^{*} & d & b^{*} & b\\
1 & c^{*} & c & a & a^{*} & d & d^{*} & b & b^{*}\\
\hline
1 & d & d^{*} & c & c^{*} & b^{*} & b & a^{*} & a\\
1 & d^{*} & d & c^{*} & c & b & b^{*} & a & a^{*}
\end{array}\right]\label{T9_LS_form},
\end{equation}
where grid lines additionally accent the internal structure.
Suddenly, the core of the matrix resembles symmetric LS with additional complex conjugates ($^*$) in few places,
\begin{equation}
\text{core}\Big(T_9^{(0)}\Big)=\left[
\begin{array}{llll}
\fbox{\texttt{A}} & \fbox{\texttt{B}} & \fbox{\texttt{C}} & \fbox{\texttt{D}}\\
\fbox{\texttt{B}} & \fbox{\texttt{D}} & \fbox{\texttt{A}}^* & \fbox{\texttt{C}}\\
\fbox{\texttt{C}} & \fbox{\texttt{A}}^* & \fbox{\texttt{D}}^* & \fbox{\texttt{B}}^*\\
\fbox{\texttt{D}} & \fbox{\texttt{C}} & \fbox{\texttt{B}}^* & \fbox{\texttt{A}}^*
\end{array}\right],
\end{equation}
where every $\fbox{\texttt{X}}$ corresponds to a $2\times 2$ matrix
\begin{equation}
\fbox{\texttt{X}}=\left[\begin{array}{ll}
x & x^*\\
x^* & x
\end{array}\right]\label{2x2_X}
\end{equation}
for $(x,\texttt{X})\in\{(a,\texttt{A}),(b,\texttt{B}),(c,\texttt{C}),(d,\texttt{D})\}$.
This observation may gently suggest a deeper relationship between the CHM and LS.
Let us tentatively propose a formal construction
and then test it for several dimensions.

Define
\begin{equation}
L_N=\left[\begin{array}{ll}
1 & 1^{1\times (N-1)}\\
1^{(N-1)\times 1} & \text{core}(L_N)
\end{array}\right]\in\mathbb{C}^{N\times N},\label{LS_pattern}
\end{equation}
where $1^{a\times b}$ denotes vectors of ones of size $a\times b$
and
\begin{equation}
\text{core}(L_N)=\text{circ}\left(
\left[\begin{array}{ll}
c_0 & c_0^*\\
c_0^* & c_0
\end{array}\right],
\left[\begin{array}{ll}
c_1 & c_1^*\\
c_1^* & c_1
\end{array}\right],
...,
\left[\begin{array}{ll}
c_{n-1} & c_{n-1}^*\\
c_{n-1}^* & c_{n-1}
\end{array}\right]
\right) \ : \ c_j\in\mathbb{C}
\end{equation}
forms a submatrix being
a $(2\times 2)$-block circulant Latin square\footnote{Note that we do not introduce complex conjugates
in blocks, as it was in the case of $T_9^{(0)}$~\eqref{T9_LS_form}.} of size $N-1$ with $n=(N-3)/2$. By construction $N$ must be odd.
The question is: For what values of $N$, the matrix $L_N$ can be a member of the set $\mathbb{H}(N)$ of complex Hadamard matrices?

Given $N$, we can immediately build $L_N$, write unitarity constraints and try to solve them.
It turned out that this particular construction stemming from the circulant LS
provides CHM only under one additional restriction.
Namely, we can find a solution only in every second odd dimension, $N=3+4k$ for $k\in\mathbb{N}_0$.
Due to circulant symmetries, the total number of possible equations forming a
nonlinear system associated with unitarity constraints can be reduced from $N(N-1)/2$ to only $2+3k$ such equations with $k\geqslant 0$.
This significant simplification allows us to solve such systems for $N$ ranging from $11$ to $127$ with a very high precision,
and initial analysis shows that in each case the defect of $L_N$ vanishes and $L_N\not\in{\rm B}\mathbb{H}(N,q)$ for any $1<q<2^{16}$.

Obviously for $N=3$ we recover the well known Fourier matrix $F_3^{(0)}$ as well as $F_7^{(0)}$ for $k=1$ and $N=7$.
But for $k\geqslant 2$ which implies $N\geqslant 11$, all constructed matrices are believed to be new as the %, because of their properties.
fact of being isolated non-Butson-like matrices automatically excludes them from many known families.
Additionally, computer assisted proof confirms that the cardinalities of the Haagerup invariants~\eqref{Lambda_H} are unique.
For $N=11$ we have
\begin{align}
\#\Lambda\left(C_{11A}^{(0)}\right) = \#\Lambda\left(C_{11B}^{(0)}\right) &= 5,\\
\#\Lambda\left(N_{11A}^{(0)}\right) = \#\Lambda\left(N_{11B}^{(0)}\right) = \#\Lambda\left(N_{11C}^{(0)}\right) &= 10,\\
\#\Lambda\left(F_{11}^{(0)}\right) &= 11,\\
\#\Lambda\left(Q_{11A}^{(0)}\right) = \#\Lambda\left(Q_{11B}^{(0)}\right) &= 63,\\
\#\Lambda\left(L_{11}^{(0)}\right) &= {\bf 191},
\end{align}
and similarly for $N=15$
\begin{align}
\#\Lambda\left(A_{15A}^{(0)}\right) = \#\Lambda\left(A_{15B}^{(0)}\right) &= 5,\\
\#\Lambda\left(A_{15G}^{(0)}\right) = \#\Lambda\left(A_{15H}^{(0)}\right) &= 38,\\
\#\Lambda\left(L_{15}^{(0)}\right) &= {\bf 461}.
\end{align}
There is no need to check invariants in higher dimensions because,
to the best of our knowledge, no isolated matrices
of this size (apart of Butson matrices~\cite{BH_home, MW12}) were ever presented in the literature.
All these observations allow us to formulate
\begin{conjecture}
For any $k\in\mathbb{N}_0$ and $N=3+4k$, there exists an isolated matrix
$L_N^{(0)}\in\mathbb{H}(N)$ which is not of the Butson type.\label{LN_conjecture}
\end{conjecture}

Before we comment the possible implications of this conjecture,
in the next section we shall introduce even simpler method producing a sequence of matrices with identical properties.

\subsection{Family of Complex Hadamard Matrices $V_N$}
\label{sec:family_VN}

Instead of struggling with a circulant core, let us simply analyze entire circulant matrix
\begin{equation}
V_N=\text{circ}\left[c_0,c_1,c_2,...,c_{N-1}\right]\in\mathbb{C}^{N\times N} \quad : \quad c_j\in\mathbb{C}.
\end{equation}
Unusual undephased representation of $V_N$ provides an extremely nice and compact form
of the unitarity constraints:
\begin{equation}
\sum_{j=0}^{N-1}\frac{c_j}{c_{(j+k)\bmod N}}=0 \qquad \text{for} \qquad k\in\{1,2,...,N-1\}\label{UC_LS_undephased}.
\end{equation}
This time the above $N-1$ nonlinear equations for $c_j\in\mathbb{C}$ can be solved
in any dimension $6\leqslant N\leqslant 64$
resulting in a sequence of isolated and non-Butson matrices.
We intentionally discarded all solutions reproducing Butson type matrices as it is very unlikely
to find something new in this way.
Also, we ignored the possibility of getting something new for $9\leqslant N\leqslant 16$ as these cases
have been studied for a long time.
Instead we focused on checking how
far we can go with $N$ still getting matrices with the given properties, to draw
\begin{conjecture}
For any $N\geqslant 6$ the solution of Eq.~\eqref{UC_LS_undephased} contains at least one isolated CHM matrix $V_N^{(0)}$ which
is not of the Butson type.\label{VN_conjecture}
\end{conjecture}
While for the previous sequence of $L_N$ numerical simulations
suggest that each (proper) solution leads only to an isolated matrix,
for $V_N$ the defect is not strictly restricted to zero. We observed such a situation already for $N=8$, and
also for some slightly higher orders. However, the bigger $N$, the less chance of hitting any potential
representative of a multidimensional family.
A curiosity concerning $N=8$ will be explained in the last section of this chapter.

\section{Comment on the Structure of the Set $\mathbb{H}(8)$}

In light of the facts from the previous sections, we end this chapter with two short remarks concerning Elser's results from $2011$.
Veit Elser found the matrix $V_8^{(0)}$
quite accidentally as a by-product of
computations with a completely different purpose~\cite{VE_private}.
It has been identified as an isolated point, put into the Catalogue of CHM~\cite{CHM_catalogue} and apparently forgotten for years.
Its nice undephased structure reads
\begin{equation}
V_8^{(0)}=\left[
\begin{array}{rrrrrrrr}
-1 & -1 &  b &  b &  c &  c &  a &  a \\
-1 &  b & -1 &  c &  b &  a &  c & -a \\
 b & -1 &  c & -1 &  a &  b & -a &  c \\
 b &  c & -1 &  a & -1 & -a &  b & -c \\
 c &  b &  a & -1 & -a & -1 & -c &  b \\
 c &  a &  b & -a & -1 & -c & -1 & -b \\
 a &  c & -a &  b & -c & -1 & -b & -1 \\
 a & -a &  c & -c &  b & -b & -1 &  1 \\
\end{array}\right].
\end{equation}
The shape of the characteristic bands of numbers revolving symmetrically around the diagonals
remains the mystery of the Sinkhorn's algorithm that was used to find this matrix.
The only difference in the Elser's version of the routine was that
he used QR decomposition instead
of polar decomposition, to get a projection onto the set $\mathbb{U}(8)$.

Firstly, let us quickly resolve one issue concerning $V_8^{(0)}$ and recover its pure analytic form.
Unitarity constraints for $V_8^{(0)}$
\begin{equation}
\left\{\begin{aligned}
-\frac{1}{b} - b + \frac{b}{c} + \frac{c}{b} + \frac{c}{a} + \frac{a}{c} & = 0, \\
-\frac{1}{b} -b - \frac{1}{c} -c   - \frac{c}{a} - \frac{a}{c}+ \frac{b}{a} + \frac{a}{b} & = 0, \\
 \frac{1}{c} + c + \frac{1}{a} + a+ \frac{b}{a} + \frac{a}{b} & = 0,
\end{aligned}\right.
\end{equation}
can be rearranged as in the case of $T_9^{(0)}$ and presented in the form of a monic palindromic polynomial ${\bf p}$,
so the set of roots of ${\bf p}$ contains parameters $a$, $b$, and $c$.
Seemingly, the less parameters the better, since computational complexity
grows rapidly with dimension $N\gg 1$. But in this case,
with only three parameters and very special symmetries among the entries of $V_8^{(0)}$,
the polynomial ${\bf p}$ turns out to be
more complicated than the one for $T_9^{(0)}$, and its degree is as much as $16$. Thus, to save space,
we show only the first half of its $17$ coefficients
\begin{equation}
{\bf p}(x)=(1,16,64,16,-332,-1040,-1984,-2832,\underline{-3194},...)
\end{equation}
Fortunately, using~\eqref{palindromic_reduction}, it can be simply reduced to an octic polynomial ${\bf q}(x)$,
%which, however, no longer reflects palindromic properties
\begin{equation}
{\bf q}(x)=(-16,256,-96,-896,-696,-96,56,16,1),
\end{equation}
which can be further factorized into
\begin{align}
{\bf q}(x)=&\left(x^4+8x^3+\alpha(0)x^2+\beta(0) x+\gamma(0)\right)\times\nonumber\\
\times&\left(x^4+8x^3+\alpha(1)x^2+\beta(1) x+\gamma(1)\right)
\end{align}
where
\begin{align}
\alpha(\nu)&=-4+2(-1)^{\nu}\sqrt{116-2\sqrt{2}},\\
\beta(\nu)&=-16+8(-1)^{\nu}\sqrt{10-\sqrt{2}},\\
\gamma(\nu)&=-4\sqrt{2}+4-4(-1)^{\nu}\sqrt{4-2\sqrt{2}},
\end{align}
and $\nu\in\{0,1\}$.
Applying well known formulas for roots of a quatric polynomial and taking
into account all changes of variables,
we can obtain analytically all the triplets
that fully determine $V_8^{(0)}$. Due to the complicated character of the final results we do not present them here.
An example solution reads\footnote{Details can be found in the Catalogue of CHM~\cite{CHM_catalogue}.}
\begin{equation}
a\approx -0.6509\mp 0.7592i,\quad b\approx -0.7799\pm 0.6258i,\quad c\approx +0.6183\mp 0.7859i.
\end{equation}
Hence, $V_8^{(0)}$ can be formally accepted as the (isolated) member of $\mathbb{H}(8)$.
The matrix $V_8^{(0)}$ with non-commensurable phases cannot belong to the Butson class.

Secondly, the sequence $V_N$ presented in the previous section\footnote{Coincidence of the family name $V_N$ and the matrix $V_8^{(0)}$ is intentional.}
tightly relates $V_8^{(0)}$ with other $8$-dimensional matrices.
Surprisingly, the unitarity constraints~\eqref{UC_LS_undephased} for $N=8$
include the solution for matrices equivalent to
$V_8^{(0)}$, $A_8^{(0)}$, possibly $S_8^{(4)}$, and perhaps some other elements of $\mathbb{H}(8)$.
Before $A_8^{(0)}$ was rediscovered as the solution of the system~\eqref{UC_LS_undephased},
it was found numerically~$[\hyperlink{\paperslist}{\rm A1}]$
using a random walk procedure over core phases (see Section~\ref{sec:RWCP}) with additional
symmetry constraints, $A_8^{(0)}=A_8^{(0)\rm T}$.
Now we see that this symmetry is just a special case of a more general structure that covers also more complicated
patterns of $V_8^{(0)}$.
In general, one can readily check, that the core of any matrix from the family $V_N$ is symmetric along the antidiagonal.

Shortly, the two matrices $V_8^{(0)}$ and $A_8^{(0)}$ surely
originate from the same general circulant seed, ${\rm circ[c_0,c_1,c_2,c_3,c_4,c_5,c_6,c_7]}$.
Actually, these are their equivalent forms.
Other $8$-dim\-ensional representatives of $\mathbb{H}(8)$ are still waiting for a confirmation
whether they belong to this super-class of CHM.
This previously unknown observation might be a clue for a more detailed analytic research.

%Much stronger symmetries are visible in $V_8^{(0)}$ that originally emerged out of the alternative numerical studies.
%Temporarily unresolved problem is whether the shape of the characteristic bands of numbers revolving symmetrically around the diagonals,
%can be used to build similar matrices in even dimensions $N=4, 6, 10, 12, ...$, which are
%equivalent to elements of the sequence $V_N$.

\medskip

We have reached the point that slowly exceeds the scope of this Thesis, so we stop here further considerations
concerning the contents of the set of complex Hadamard matrices.
All these insights along with the result from Section~\ref{sec:matrix_T8} 
contribute to a more accurate description of $\mathbb{H}(N)$.
We encourage those interested in the topic to follow up the preprint~$[\hyperlink{\paperslist}{\rm A2}]$ that will appear soon.

\section{Summary}

In this chapter we have extended two sets of complex Hadamard matrices, which
can be used for diverse purposes in constructing novel objects relevant for quantum theory. We introduced
the nonaffine family $T_8^{(1)}$ and the isolated point $T_9^{(0)}$. We also
discussed some other details concerning the analytic description of set $\mathbb{H}(8)$.

We asked if there was any deeper relationship between complex Hadamard matrices and Latin squares,
and it turns out that in fact it is possible to identify some extremely simple connections between the two, that lead to previously
unknown observations and, above all, to discovery of new examples of CHM.

Based on the numerical evidence we put forward two conjectures concerning existence
of a sequence of isolated CHM in infinitely many dimensions.
In each conjecture we consider only two possible patterns of CHM, both related to some extent to the notion of LS.
Certainly, this does not exhaust the whole range of the other possibilities, in particular those which,
as shown in the initial case of $T_9^{(0)}$~\eqref{T9}, can be enriched
with additional, carefully selected, complex conjugates or even more advanced features (symmetries).
For a moment we cannot see how they could be possibly classified into any systematic or coherent way.
Hence, we are not going to explore other configurations at least until they prove to exhibit some particularly
desirable physical (or mathematical) properties.

There are roughly three systematic methods of construction of isolated CHM.
The first one is based purely on numerical approach.
Second one follows from the theorem, which says that for any prime dimension $p$ the Fourier matrix is
isolated~\cite{TZ06},
\begin{equation}
N=p\Longrightarrow F_p=F_p^{(0)}.\label{Fp_isolated}
\end{equation}
The third one has its source in the construction described in Theorem 3 in~\cite{MW12}.
If these conjectures were proven, a new way to get an infinite sequence of isolated CHM would be found.
But even if they fail to be true, we are left with two manageable constructions that so far generated several dozen examples
of CHM in a number of dimensions $N\geqslant 8$. One could ask here what is the physical motivation
for searching isolated matrices and why we put so much effort to classify them. 
Again, one possible answer is the straightforward connection of CHM with MUB. %the following. If we would like to solve the problem of MUB (say for $N=6$ or $N=10$)
It is much easier to examine multidimensional families and check if their representatives form possible sets of MUB.
But what if the complete set of MUB in $\mathbb{C}^6$ or $\mathbb{C}^{10}$ is built of (yet unknown) isolated CHM?\footnote{Actually, it is very unlikely
that for $N=6$ new isolated matrices will be discovered, but for $N=10$ our knowledge is more limited.}
Another reason to study isolated structures will be explained in the next chapter.

Looking at the final forms for $T_8^{(1)}$, $V_8^{(0)}$, and $T_9^{(0)}$ it seems that analytic approach to solve the system
of equations corresponding to unitarity constraints
for $L_N$ and $V_N$ for any $N$ might be out of reach. On the other hand, it does not require
much effort to obtain highly accurate numerical approximations. The aforementioned sequences
$\{L_N\}_{N=3+4k}$ for $k\geqslant 0$ and $\{V_N\}_{N\geqslant 6}$ supporting both conjectures
were %home-made using mildly outdated hardware over the period of three days
calculated quite quickly without much pressure on optimization in terms of speed, we rather wanted good quality solutions.
Mathematical computation software was chosen to be Matlab/Octave serving appropriately tuned
function \texttt{fsolve}, which proved to be more than enough for our purpose. 
We are aware of many other specialized methods designed to solve such problems, but (upon examination)
none of them can provide immediate results at the analytic level.

Presented facts may find applications in several branches of modern physics, including
creation of new pairs of MUB, being building blocks for unitary error bases (UEB)~\cite{MV16}
or used in entanglement detection~\cite{SHBAH12,HL14}.
As it has been recently shown, even incomplete (unextendible) set of MUB can be more effective to detect entanglement~\cite{HMBRBC21}.
We will return to the problem of unextendible sets of MUB in Chapter~\ref{chap:EXCESS}, where
we also present the relation of CHM to some classes of Bell inequalities, while the phenomenon of entanglement will be
treated in Chapter~\ref{chap:AME46}.

\medskip

Let us conclude this chapter with a list of some open questions:
\begin{enumerate}
\item Are Conjecture~\ref{LN_conjecture} and~\ref{VN_conjecture} true?
Especially, how to prove that the construction of $L_N$ lead only to zero-defect matrices?
\item What about different patterns~\eqref{LS_pattern} which might provide a solution? How to classify them?
Is the presented connection between CHM and Latin squares accidental or fundamental?

We shall mention here that apart from $T_8^{(0)}$, $T_9^{(0)}$, $\{L_N\}$, and $\{V_N\}$ we found several other matrices
of order $N\in\{11,12,13,14,15,16\}$.
They all can be found online~\cite{GITHUB} and look promising.
Currently they await to be discarded or accepted as valid members of the set $\mathbb{H}(N)$ of inequivalent complex Hadamard matrices.

\item Are there any other genuine numerical methods to generate CHM?
Inspired by~\cite{SM19} we made some attempts to employ machine learning and artificial intelligence techniques
to estimate possibilities of finding new classes of CHM. But currently such research is at the very preliminary stage.

\end{enumerate}  

%\bigskip

\begin{figure}[ht!]
\center
\includegraphics[width=0.5in]{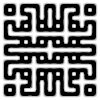}
\end{figure}

%\bigskip

In the next chapter we are going to explore more advanced structures.
Based on the presented notion of defect of a unitary matrix, we will propose another analytical tool to examine
possibility of introduction of free parameters in several objects representing quantum measurement operators.

%%%%%%%%%%%%%%%%%%%%%%%%%%%%%%%%%%%%%%%%%%%%%%%%%%%%%%%%%%%%%%%%%%%%
%\clearpage\null%\thispagestyle{empty}
%\clearpage\null%\thispagestyle{empty}
\chapter{Restricted Defect of a Unitary Matrix}
\label{chap:DELTA}
Chapter based on~$[\hyperlink{\paperslist}{\rm A3}]$.
\medskip
\medskip
\medskip

\noindent In our approach to explore quantum mechanics various classes of matrices are the main framework for many concepts.
Matrix algebra provides a comprehensive mathematical description of physical systems and dynamics at the subatomic level.
Having chosen convenient reference frame (a basis), they
represent operators, states or just special purpose bases in Hilbert spaces being a background for our considerations.
We have already investigated very particular class of complex Hadamard matrices.
Every such matrix rescaled by the square root of its size\footnote{Caution. In Chapter~\ref{chap:CHM}
we used the notion of defect usually denoted by ${\bf d}(H)$. In the current
chapter letter $d$ is reserved for a (local) dimension and wherever it is needed, the defect of a unitary matrix is always denoted by ${\bf d}(U)$.}
automatically becomes unitary, 
$\frac{1}{\sqrt{d}}\mathbb{H}(d)\subsetneq\mathbb{U}(d)$.
We have shown that these matrices might admit special parameterization forming families depending
on free parameters.

Unitarity is the first fundamental property of quantum dynamics.
Evolution of quantum states represented by density matrices $\rho$ of size $d$ is governed by unitary matrices,
$\rho\to U\rho U^{\dagger}$, $U\in\mathbb{U}(d)$.
Another mathematical attribute pertaining to the quantum world is hermiticity.
All operators corresponding to physical observables must be Hermitian (self-adjoint) to guarantee
the fact that eigenvalues representing possible measurement outcomes are real.
In theory of quantum information we encounter many structures that are Hermitian by construction or by demand.
In some cases they allow to introduce free parameters as it is for CHM.
The question is when and how many degrees of freedom can we expect.

In this chapter, we will answer that question by adapting the idea of the defect of a unitary
matrix~\cite{TZ08, Ta18} used in Chapter~\ref{chap:CHM} to a 
very particular subset of $\mathbb{U}(d)$, namely unitary and Hermitian matrices having constant diagonal.
Due to its character, it should be called {\sl restricted defect} (denoted $\Delta$), and it will serve as a suitable tool
to determine the maximal number of free parameters allowed by a given structure.
As a consequence, we will calculate $\Delta$ for a wide range of special operators, identify
some families and classify isolated objects.

\section{Quantum Measurement}
\label{sec:QM_GM}

Measurement is a primordial concept in physics.
No matter how attractive theory to model our reality we propose, it is completely useless unless
we can confirm and reproduce its expectations in a series of independent experiments,
and every experiment must involve at least one measurement of a physical quantity.
The measurement process has destructive nature, as by definition it is an interaction
of an apparatus with the physical system. Such interaction might invoke some changes, either in the system
or in the measuring tool (or both), which are represented by real numbers out of which we may draw appropriate conclusions.
In the classical world, we do not observe this invasive property,
since the analysis of the system does not influence on it at all or the impact is so tiny that it is immeasurable and negligible.
However, at the mesoscopic or subatomic level, if the act of measurement can destroy the system and reflects probabilistic character,
this has much more profound consequences making it an irreversible step versus
fully reversible unitary evolution.\footnote{We express the hope that the Reader is aware
of the fact that such a naively presented essence of quantum measurement hides much more content and nuances.
For instance, see~\cite{CNB19} and references therein.}

In the axiomatic approach to quantum mechanics~\cite{BZ17}, classical
probability vectors are replaced by states $|\psi\rangle$
represented by normalized complex vectors from Hilbert space $\mathcal{H}_d$ of a given dimension $d$.
Such a state vector $|\psi\rangle$, called a {\sl pure state}, carries all information about the isolated (closed) system.
In general, for open systems (interacting with an environment) we should consider
a statistical ensemble $\{(|\psi_j\rangle, p_j)\}$, where $p_j$ is the probability
that the system is in the state $|\psi_j\rangle$.
This leads to a formalism of {\sl density matrices} (operators) defined as
\begin{equation}
\rho=\sum_{j}p_j|\psi_j\rangle\langle\psi_j|,
\end{equation}
so that $\rho$ is Hermitian ($\rho=\rho^{\dagger}$), positive-semidefinite ($\rho\geqslant 0$) and normalized (${\rm Tr}(\rho)=1$)
operator acting on $\mathcal{H}_d$.

To obtain the information about the quantum system, one must perform a measurement.
The most general form of the quantum measurement is described by a set of Kraus operators $K_j$,
such that the following resolution of identity holds
\begin{equation}
\sum_j K_j^{\dagger}K_j=\mathbb{I}_d.
\end{equation}
Such a form can be explained by the fact that given a density matrix $\rho$,
all the information that is contained in $\rho$ can be extracted by means of
the expectation values, $q_j={\rm Tr}(O_j\rho)$, for some operators $O_j$.
If we require $q_j$ to be probabilities, i.e. $q_j\geqslant 0$ and $\sum_jq_j=1$, then
it implies that $O_j\geqslant 0$ and $\sum_jO_j=\mathbb{I}_d$.
Hence, the probability of $j^{\rm th}$ measurement outcome is
\begin{equation}
p_j={\rm Tr}\big(K_j\rho K_j^{\dagger}\big)
\end{equation}
and the system collapses to the state
\begin{equation}
\rho'=\frac{K_j\rho K_j^{\dagger}}{p_j}.
\end{equation}

In this work, we consider a particular scenario called {\sl positive-operator valued measure} (POVM)
restricted to the variant with rank-$1$ projectors.
Thus, we define POVM as a set of $N$ semi-definite normalized operators $\Gamma_j$ of size $d$ % subnormalized
such that
\begin{equation}
\sum_{j=0}^{N-1}\Gamma_j=\mathbb{I}_d,
\end{equation}
where each $\Gamma_j$ is proportional to a rank-$1$ projector $\Pi_j$ and corresponds to the $j^{\rm th}$ measurement outcome.
Projectors $\Pi_j$ satisfy the geometrical relation
\begin{equation}
{\rm Tr}\big(\Pi_j\Pi_k\big)=S_{jk}\label{geometrical_relation}
\end{equation}
for a symmetric matrix $S=S^{\rm T}\in\mathbb{C}^{N\times N}$.

Let us put formal definitions of some distinguished subclasses of POVM
having prescribed symmetry given by the real symmetric matrix $S$.
A matrix
\begin{equation}
S = \mathbb{I}_N + \frac{N - d}{d(N-1)}\left(\mathbb{J}_N-\mathbb{I}_N\right)\label{S_ETF}
\end{equation} 
corresponds to {\sl equiangular tight frames} (ETF) composed of $N$ vectors each of size $d$~\cite{STDH07}.
The symbol $\mathbb{J}_N$ denotes the flat matrix of size $N$ with all entries equal to $1$.
For $N=d^2$, one obtains an important subclass of ETF called {\sl symmetric, informationally complete POVM} (SIC-POVM)~\cite{RKSC04}
which, for $d=8$, includes a very particular example called the {\sl Hoggar lines}~\cite{Ho98}.
In Chapter~\ref{chap:CHM}, we frequently referred to {\sl mutually unbiased bases} (MUB) --- see definition in Eq.~\eqref{MUB_definition}.
Associated symmetric matrix~\eqref{geometrical_relation} reads
\begin{equation}
S = \mathbb{I}_{dm}+\frac{1}{d}\left(\mathbb{J}_m-\mathbb{I}_m\right)\otimes \mathbb{J}_d.\label{S_MUB}
\end{equation}
These geometrical structures, visualized in the hierarchical scheme
\begin{equation}
\text{POVM} \supset \left\{\begin{array}{l}\text{MUB}\\\text{ETF}\supset\text{SIC-POVM},\end{array}\right. 
\end{equation}
have important applications in quantum physics.
For example, ETF serve in coding theory~\cite{CK03} and error correction~\cite{HP04}.
SIC-POVM and MUB are widely used in quantum tomography i.e. reconstruction of density matrix of an unknown system~\cite{Iv81, AS10},
as they maximize the information gained during measurement while minimizing influence of statistical errors.
They also define entropic uncertainty relations~\cite{Ra13},
find applications in detecting entanglement~\cite{SHBAH12}, bound entanglement~\cite{HL14}, and locking
classical information in quantum states~\cite{BW07}.

Having presented above categories one might rightly suppose they will be subjected
to examination in view of possibility of introduction of free parameters.
Indeed, our current objective can be briefly summarized in the following question:
\begin{align}
&\textit{Given generalized quantum measurement of predefined geometrical structure,}\nonumber\\
&\textit{is it possible to extend it to a smooth family of POVM?}\label{POVM_question}
\end{align}

For $d>3$, the presence of free parameters in SIC-POVM is an unresolved task. 
On the other hand, some families of MUB in low dimensions are known~\cite{GG15}.
Construction of such families can find pragmatic applications providing
flexibility when designing experimental implementations of measurement settings.
Free parameters can be tuned to produce customized machinery for concrete purposes.
For instance, taking the well known one-parameter family of SIC-POVM in $d=3$~\cite{RKSC04},
only one particular representative maximizes the informational power,
that is, the classical capacity of a quantum-classical channel
generated by the SIC-POVM~\cite{ASz14}.
It is also worth to mention another class of measurements not
belonging to any family. Such isolated objects might play important role
when there is a need to have a unique solution which optimizes a certain problem,
like a unique maximal violation of a Bell inequality, which is 
a crucial component in procedures of self-testing~\cite{MY04}.

The general question about the existence of SIC-POVM and MUB is a completely different story.
We will not deal with this subject here and only mention the two biggest problems.
In his thesis~\cite{Za99}, Gerhard Zauner posed a conjecture that in every dimension $d\geqslant 2$, there exist
a special vector called {\sl fiducial} one, which is a kind of a generator for SIC-POVM (we will later see one such vector for $d=3$).
Since currently only finitely many SIC-POVM are known analytically (this includes aforementioned
one-parameter family in $d=3$),
there are many attempts to tackle the problem numerically~\cite{SG10}.
Mutually unbiased bases, first introduced in~\cite{Sch60} have been investigated since early 1980s~\cite{Iv81}.
Alike SIC-POVM, MUB are also far from being completely classified.
Despite many efforts, the seemingly simple six-dimensional case contributes to the both lists
of the most wanted open problems~\cite{KCIKProblems, TenAnnoyingProblems}.

\newpage
\section{Gram Matrix}

Before we attempt to give an answer to~\eqref{POVM_question}, we need two things for completeness.
First, we recall the notion of the Gram matrix.
Consider $N$ vectors $|\varphi_j\rangle\in\mathbb{C}^d$. The Gram matrix $G(|\varphi_1\rangle,...,|\varphi_N\rangle)$
associated with $\{|\varphi_j\rangle\}$
is given by 
\begin{equation}
G_{jk}=\langle\varphi_j|\varphi_k\rangle\label{Gram_matrix}
\end{equation}
for $j,k\in\{1,2,...,N\}.$
One can show~\cite{TDHS05} that $G^2=GN/d$
is the necessary and sufficient condition
for a Gram matrix $G$ to represent a POVM with $N$ vectors in $\mathbb{C}^d$ for $d\leqslant N$.
Then, from the fact that ${\rm Tr}(G)=N$ and
\begin{equation}
{\rm eig}(G)=\Big\{\underbrace{\frac{N}{d},...,\frac{N}{d}}_{d},\underbrace{0,...,0}_{N-d}\Big\},
\end{equation}
we infer that $U=\mathbb{I}_N-2dG/N$ is unitary, $U\in\mathbb{U}(N)$.
Thus the existence of a POVM with a symmetry imposed by matrix $S$ given in~\eqref{geometrical_relation}
can be considered in terms of a Hermitian unitary
matrix $|U_{jk}|=2d\sqrt{S_{jk}}/N$ having positive constant diagonal, $U_{jj} = 1 - 2d/N$.

As an example~\cite{GT17}, the Gram matrix of a generic ETF corresponding to~\eqref{S_ETF} has the form
\begin{equation}
G_{\rm ETF}=\left[\begin{array}{cccc}
1 &      \alpha e^{i\beta{12}} & \cdots & \alpha e^{i\beta_{1N}}\\
                    * & 1 & \cdots & \alpha e^{i\beta_{2N}}\\
               \vdots & \vdots                &        & \vdots\\
                    * &                     * & \cdots & 1
\end{array}\right] \ : \ \mathcal{\alpha}=\sqrt{\frac{N-d}{d(N-1)}} \ , \ \ \ \ \beta_{jk}\in[0,2\pi).
\end{equation}
The symbols $*$ below the diagonal denote complex conjugated entries. %; $G_{\rm ETF}=G_{\rm ETF}^{*}$.

The last missing link is a counterpart of the defect of a unitary matrix tailored to Hermitian
unitary matrices with constant diagonal. We will derive it in the next section.

\section{Restricted Defect $\Delta$}
\label{sec:DELTA_derivation}

Based on the idea of the defect of a unitary matrix $U$ (Chapter~\ref{chap:CHM}), we will present a heuristic derivation of a similar
formula for the restricted defect applicable to a Hermitian unitary matrix with constant
diagonal. In mathematical language, restricted defect is defined in full analogy to its unitary equivalent
as a dimension of a real space of possible directions
that preserve unitarity, hermiticity, diagonal entries and the moduli of all matrix elements
in the first order or perturbations.
We shall not strictly adhere to mathematical rigor to make the description
more accessible and easier to understand. Formal derivation details including
the special treatment of the case of matrices with zero entries\footnote{We thank Wojciech Tadej for his patient
explanations of mathematical details of multidimensional tangent spaces and manifolds.}
can be found in~\cite{Ta18}.

Let $d>1$. Consider a unitary matrix
\begin{equation}
U=\left[\begin{array}{lllll}
            c &  U_{12} &   U_{13} & \dots & U_{1d} \\
U_{12}^* &            c &   U_{23} & \dots & U_{2d} \\
U_{13}^* &     U_{23}^* &        c & \dots & U_{2d} \\
  \vdots &       \vdots &   \vdots &       & \vdots \\
U_{1d}^* &     U_{2d}^* & U_{3d}^* & \dots &      c \\
\end{array}\right],
\end{equation}
such that (by construction)
\begin{enumerate}
\item $U$ is unitary and Hermitian; $UU^{\dagger}=U^2\propto\mathbb{I}_d$ (we relax strict normalization, hence only proportionality is required),
\item $U$ has constant diagonal; $c\in\mathbb{R}$,
\item $|U_{jk}|$ is constant for $j\neq k$ (not necessarily unimodular).
\end{enumerate}
We look for the most general antisymmetric perturbation matrix $R$ such that
\begin{equation}
U(R)=U\circ {\rm EXP}(it R) \quad : \quad t\in\mathbb{R}\label{UitR}
\end{equation}
fulfills the three conditions above.\footnote{Entrywise product of two matrices is denoted by $\circ$, while
entrywise exponent of a matrix $A$ reads $[{\rm EXP}(iA)]_{jk}=e^{iA_{jk}}$, see~\cite{TZ06}.}
In other words, we examine all possible (independent) ``directions'', we can follow so that
the properties of $U$ are not violated.
For example, for $d=3$ relevant calculations lead to
\begin{align}
\big(U(R)\big)^2&=
\left[\begin{array}{lll}
c & U_{12}e^{itR_{12}} & U_{13}e^{itR_{13}}\\
U_{12}^*e^{-itR_{12}} & c & U_{23}e^{itR_{13}}\\
U_{13}^*e^{-itR_{12}} & U_{23}^*e^{-itR_{13}} & c\\
\end{array}\right]^2=\\
&=
\left[\begin{array}{lll}
c^2 + |U_{12}|^2 + |U_{13}|^2 & \ast & \ast\\
2cU_{12}e^{-itR_{12}} + U_{23}U_{13}^*e^{it(R_{23}-R_{13})} & \text{constant} & \ast\\
... & ... & \text{constant}
\end{array}\right]\propto\mathbb{I}_d.\label{URUR}
\end{align}
For brevity we do not fill the places over the diagonal, and the diagonal itself
is obviously constant. Also, in~\eqref{URUR} we present only one product
in the second entry in the first column
since the other off-diagonal elements take exactly the same form, differing only in indices.
One readily recognizes the general pattern
\begin{equation}
2cU_{jk}^*e^{-iR_{jk}}+\sum_{l\neq j, k}U_{kl}U_{jl}^*e^{i(R_{kl}-R_{jl})}=0\label{2cUUURR}
\end{equation}
for $1\leqslant j<k\leqslant d$.
The system of nonlinear equations~\eqref{2cUUURR} contains full information about
the conditions that the perturbation matrix $R$ must meet in order for~\eqref{UitR}
to represent the matrix with all the required properties.
However, the solution of such a system of equations is generally intractable.

A dummy real parameter $t\in\mathbb{R}$, which has been artificially introduced in~\eqref{UitR},
now allows us to perform linearization of~\eqref{2cUUURR}. We write
\begin{equation}
\lim_{t\to 0}\frac{d}{dt}\Big(U(R)\Big)^2=[0,...,0]^{\rm T}\in\mathbb{R}^d\label{lim_d_dt}
\end{equation}
which gives us explicit first order approximation on conditions for~\eqref{UitR}.
This eventually implies the system of linear equations for real variables $R_{jk}$,
\begin{equation}
-2cU_{kj}R_{jk}+\sum_{l\neq j, k}U_{kl}U_{lj}(R_{kl}-R_{jl})=0\label{2cUUURRL}
\end{equation}
for $1\leqslant j<k\leqslant d$ and $1\leqslant l\leqslant d$.
Note, that $U_{j...}^*=U_{... j}$.
System~\eqref{2cUUURRL} consists of $\tau=d(d-1)/2$ linear equations but
usually it is not of full rank. Let $r$ denote the rank of the matrix $\mathcal{R}$
associated with the system~\eqref{2cUUURRL}.

Equivalence relation $U \simeq D U D^{*}$ for
$D={\rm diag}[1, e^{i\beta_2},e^{i\beta_2},\dots,e^{i\beta_d}]$
with appropriately chosen $\beta_j$, $j\in\{2,...,d\}$,
preserves the diagonal of $U$ and absorbs $f=d-1$ phases (this corresponds to $1^{\rm st}$ row/column of $U$).
Moreover, zeros that might appear in $U$ are obviously not affected by the relation $\simeq$
and further reduce the number of phases (free parameters).
Let the number of zeros above the main diagonal in the matrix $U$ be denoted by $z$.

Putting all together, we are in a position to pose a definition of 
the {\sl restricted defect} $\Delta$ of the Hermitian unitary matrix $U$ of size $d$. It reads
\begin{align}
\Delta(U)=& \ \tau-f-z-r\label{DELTA_definition}.
\end{align}

Originally, the defect ${\bf d}(U)$ of the unitary matrix $U$
was used to determine an upper bound on the number
of free parameters allowed by a complex Hadamard matrix~\cite{TZ06}.
Looking at the definition of $\Delta$~\eqref{DELTA_definition} we recognize
a close relation between $\Delta(U)$ and ${\bf d}(U)$ which
has been adapted to the set of matrices of a special structure.
By construction, in the particular case of a Hermitian unitary matrix $U\in\mathbb{U}(d)$ with constant diagonal
without off-diagonal zeros ($z=0$), the following bounds hold
\begin{equation}
0\leqslant \Delta(U)\leqslant {\bf d}(U)\leqslant (d-1)^2.
\end{equation}
All properties of the unitary defect are naturally transferred to $\Delta$.
In both cases, defect is invariant with respect to the equivalence
relation which for $\Delta$ takes the simpler form (to preserve constant diagonal and hermiticity) of
\begin{equation}
\Delta(U) = \Delta\big(PDUD^{*}P^{\rm T}\big),
\end{equation}
for $D={\rm diag}[1,e^{i\beta_2},...,e^{i\beta_d}]$ and $P\in\mathbb{P}(d)$.
And, its vanishing value implies
the impossibility of introduction of free parameters, hence
it detects isolated matrices.
In the case of the restricted defect, this fact has further consequences.

It should be adverted that the way of deriving the Hermitian defect of a matrix $U$, as well as its standard unitary defect,
provides an explicit construction of affine families stemming from a given matrix $U$. 
We are not going to construct such families here
and we refer the Reader to~$[\hyperlink{\paperslist}{\rm A3}]$ or~\cite{TZ08} for more information.

\subsection{Elementary Examples}

To gain some practice, let us discuss some examples of simple matrices of size $d=3, 4, 5$, and $6$.
Take a matrix $U$ of the form
\begin{equation}
U=\left[\begin{array}{rrr}
-1/2 & 1 & 1\\
1 & -1/2 & 1\\
1 & 1 & -1/2\end{array}\right].
\end{equation}
Matrix $U$ perturbed by an affine entrywise rotation by phase $t$ reads
\begin{equation}
U(R)=\left[\begin{array}{rrr}
-1/2 & 1 & 1\\
1 & -1/2 & 1\\
1 & 1 & -1/2\end{array}\right]\circ {\rm EXP}(itR).
\end{equation}
Quick arithmetic,
\begin{equation}
\frac{d}{dt}\Big(U^2(R)\Big)\Bigg|_{t=0}=
\left[
\begin{array}{c}
0\\
0\\
0
\end{array}
\right]\Longleftrightarrow\mathcal{R}=
%\left\{
\begin{cases}
& R_{12}+R_{23}-R_{13} =0\\
& R_{13}-R_{12}-R_{23} =0\\
& R_{23}+R_{12}-R_{13} =0
\end{cases},\label{DELTA_example_N3}
%\right.
\end{equation}
shows that all equations in~\eqref{DELTA_example_N3} are linearly dependent so $r=1$
and the restricted defect given by Eq.~\eqref{DELTA_definition} reads $\Delta(U)=0$.

Similar calculations for $U={\rm circ}[-1, 1, 1, 1]$ of size $d=4$
lead to a system of $d(d-1)/2=6$ linear equations whose matrix form is given by a matrix $\mathcal{R}$ of order six
\begin{equation}
\mathcal{R}=\left[\begin{array}{rrrrrr}
 2& -1 & -1 & 1 & 1 & 0 \\
-1&  2 & -1 &-1 & 0 & 1 \\
-1& -1 &  2 & 0 &-1 &-1 \\
 1& -1 &  0 & 2 &-1 & 1 \\
 1&  0 & -1 &-1 & 2 &-1 \\
 0&  1 & -1 & 1 &-1 & 2 \\
\end{array}\right].
\end{equation}
Hence $r=r(\mathcal{R})=3$, again $z=0$, so $\Delta(U)=0$ as well.
The same vanishing value for the restricted defect $\Delta$ we obtain for another circulant matrix $U={\rm circ}[-\frac{3}{2}, 1, 1, 1, 1]$.

A slightly more complicated example involves MUB for a single-qubit system,
|$\varphi_j\rangle=|j\rangle$ for $j=0,1$ and $|\psi_{\mp}\rangle=\frac{1}{\sqrt{2}}\big(|0\rangle\mp|1\rangle\big)$.
The Gram matrix associated with the set of two unbiased bases reads
\begin{equation}
G_{\rm MUB}=\frac{1}{2}\left[\begin{array}{rrrr}
2 & 0 & 1 & 1\\
0 & 2 & 1 &-1\\
1 & 1 & 2 & 0\\
1 &-1 & 0 & 2
\end{array}\right].
\end{equation}
In general we can introduce four parameters, so the Gram matrix takes the form of
\begin{equation}
G_{\rm MUB}=\frac{1}{2}\left[\begin{array}{rrrr}
2 & 0 & e^{itR_{13}} & e^{itR_{14}}\\
0 & 2 & e^{itR_{23}} &-e^{itR_{24}}\\
e^{-itR_{13}} & e^{-itR_{23}} & 2 & 0\\
e^{-itR_{14}} &-e^{-itR_{24}} & 0 & 2
\end{array}\right].
\end{equation}
After absorbing three phases by acting with $D={\rm diag}[1,e^{-R_{23}},e^{itR_{13}},e^{itR_{14}}]$
on $G_{\rm MUB}$,
we arrive at the unitary matrix
\begin{align}
U(R)&=\mathbb{I}_4-\frac{2\cdot 2}{4}G_{\rm MUB}\\
&=\frac{1}{2}\left[\begin{array}{rrrr}
0 & 0 & 1 & 1\\
0 & 0 & 1 &-e^{itR_{24}}\\
1 & 1 & 0 & 0\\
1 &-e^{-itR_{24}} & 0 & 0
\end{array}\right].
\end{align}
It is easy to notice that the rank of the associated system of equations for~\eqref{2cUUURRL} is $r=1$, and $z=2$,
therefore $\Delta=0$.
This result is in accordance with the well known fact that the pair of MUB
in $\mathbb{C}^2$ is unique up to a global rotation~\cite{BWB10}.

Lastly, consider a one-parameter family of Hermitian conference\footnote{Complex {\sl conference matrix} $C$ of size $d$
satisfies two conditions: $|C_{jk}|=1-\delta_{jk}$ and $CC^{\dagger}=(d-1)\mathbb{I}_d$.
} matrices~\cite{ETa14}
\begin{equation}
C_6(b) = \left[\begin{array}{rrrrrr}
0 & 1 & 1 & 1 & 1 & 1 \\
1 & 0 &-1 &  e^{ib} & 1 & -e^{ib} \\
1 &-1 & 0 & -e^{ib} & 1 &  e^{ib} \\
1 & e^{-ib} & -e^{-ib} & 0 & -1 & 1 \\
1 & 1 & 1 & -1 & 0 & -1 \\
1 & -e^{-ib} & e^{-ib} & 1 & -1 & 0
\end{array}\right] \ \ : \ \  b \in [0,2\pi).
\end{equation}
For a generic value of $b\in[0,2\pi)$, the restricted defect $\Delta(C_6(b))=1$, while $\Delta(C_6(0))=\Delta(C_6(\pi))=4$.
This fact implies that the family of conference matrices $C_6(b)$ is one-dimensional and one cannot
extend it to the second dimension.

\newpage

\section{Isolated Mutually Unbiased Bases (MUB) and Symmetric, Informationally Complete Positive-Operator Valued Measures (SIC-POVM)}

Our first main results in this chapter are summarized in the two following statements:

\begin{proposition}
Maximal sets of $d+1$ MUB for $d=4$, $8$, $9$, and $16$ are isolated.
\end{proposition}

\begin{proposition}
SIC-POVM in dimension $4\leqslant d\leqslant 16$ are isolated.
\end{proposition}

Let us briefly comment these observations.
For a maximal set of $m=d+1$ MUB in dimension $d$, one can prove that the number of zeros in the upper triangular part
of the associated Gram matrix $G$ is $z=\frac{1}{2}d(d^2-1)$, hence the total number
of parameters in $G$ is $\tau-f-z=\frac{1}{2}d^4+\frac{1}{2}d^3-d^2-d+1$.
For $d=4$, $8$, $9$, and $16$ the rank of the corresponding system of linear equations~\eqref{2cUUURRL}
reads: $141$, $2233$, $3556$ and $34545$, respectively --- See Table~\ref{tab:DELTA_MUB}.
We took maximal sets of MUB from~\cite{WF89, Ch02, SK14}.
\begin{table}[ht!]
\center
\begin{tabular}{ l | l | l | l }
 $d$ & $\tau-f-z$ & $r$ & $\Delta=\tau-f-z-r$ \\
\hline
 $4$ & $141$ & $141$ & $0$\\  
 $8$ & $2233$ & $2233$ & $0$\\  
 $9$ & $3556$ & $3556$ & $0$\\  
 $16$ & $34545$ & $34545$ & $0$\\  
\end{tabular}
\caption{Values of the restricted defect for maximal set of $d+1$ MUB in dimension $d=4$, $8$, $9$, and $16$.
As $\Delta=0$, each set of analyzed MUB is isolated.}
\label{tab:DELTA_MUB}
\end{table}
%For prime power we can squares of primes we can build maximal set of MUB~\cite

Another observation concerns certain incomplete sets of MUB in dimensions $2\leqslant d\leqslant 9$.
The restricted defects $\Delta$ calculated for sets of $2\leqslant m\leqslant d+1$ MUB are
presented in Table~\ref{TABLE_MUB}.
Most of all, question marks in Table~\ref{TABLE_MUB} for $d=6$ denote the current status of the six-dimensional MUB problem,
while $\diamond$ points to the case of three such bases in $\mathbb{C}^6$, which is still obscured and not fully understood issue~\cite{Go15}.

\begin{table}[ht!]
\center
\begin{tabular}{ l | r | l | l | l | l | l | l | l }
$m\setminus d$ & 2 & 3 & 4 & 5 & 6 & 7 & 8 & 9\\ 
\hline
\hline
2 & $\Delta=0$ & 0 & 3 & 0 & 4 & 0 & 21 & 16\\
\hline
3 & 0 & 0 & 3 & 0 & $\diamond$ & 0 & 27 & 20\\
\hline
4 &   & 0 & 0 & 0 & ? & 0 & 19 & 32\\
\hline
5 &   &   & 0 & 0 & ? & 0 & 7  & 0 \\
\hline
6 &   &   &   & 0 & ? & 0 & 0  & 0 \\
\hline
7 &   &   &   &   & ? & 0 & 0  & 0 \\
\hline
8 &   &   &   &   &   & 0 & 0  & 0 \\
\hline
9 &   &   &   &   &   &   & 0  & 0 \\
\hline
10&   &   &   &   &   &   &    & 0
\end{tabular}
\caption{Values of the restricted defect for subsets of MUB in dimension $2\leqslant d\leqslant 9$. The results do not depend on the choice
of $m$ subsets of MUB out of the full set of $d+1$ MUB.}
\label{TABLE_MUB}
\end{table}

One sees that for prime dimensions, $d=p=2$, $3$, $5$, $7$, the Hermitian defect vanishes for every $2\leqslant m\leqslant d+1$.
This can be explained by the fact that every CHM contributing to a particular set of MUB is equivalent to
the isolated Fourier $F_d=F_p^{(0)}$, see~\eqref{Fp_isolated}.
Note that for $m=2$ (pair of MUB), $\Delta$ coincides with the standard unitary defect.
That is because the Gram matrix associated with two MUB $\mathcal{B}_1$ and $\mathcal{B}_2$,
\begin{equation}
G_{\rm MUB}=\left[\begin{array}{cc}
\mathbb{I}_d & H\\
H^{\dagger} & \mathbb{I}_d
\end{array}\right] \quad:\quad H = \mathcal{B}_1^{\dagger}\mathcal{B}_2,
\end{equation} 
allows one to introduce exactly as many free parameters as it is required for $H$ itself.
Here, each basis $\mathcal{B}_j$ represents a CHM.
This clarifies the first row ($m=2$) in Table~\ref{TABLE_MUB} --- for non-prime dimensions
there are non-zero values of $\Delta$. However, since we know that in every low dimension (excluding $d=4$),
there exists an isolated CHM (and probably for infinitely many dimensions greater than $4$ --- see Chapter~\ref{chap:CHM}),
we can always construct exceptional pairs of MUB, e.g.
\begin{equation}
\Bigg\{\mathbb{I}_6, \frac{1}{\sqrt{6}}S_6^{(0)}\Bigg\}, \Bigg\{\mathbb{I}_7, \frac{1}{\sqrt{7}}F_7^{(0)}\Bigg\}, \Bigg\{\mathbb{I}_8, \frac{1}{\sqrt{8}}A_8^{(0)}\Bigg\}, \Bigg\{\mathbb{I}_9, \frac{1}{\sqrt{9}}N_9^{(0)}\Bigg\}, ..., \Bigg\{\mathbb{I}_d, \frac{1}{\sqrt{d}}V_d^{(0)}\Bigg\}, ...  
\end{equation}
such that $\Delta$ vanishes.
For $d=8$ and $9$, we examined subsets of MUB taken from maximal sets~\cite{WF89, SK14}.
For $m\geqslant 6$ and $m\geqslant 5$ respectively, these subsets are isolated too.

The last observation regarding MUB explains the coincidence between the value of the restricted defect $\Delta$
for $d$ and $d+1$ sets, as maximal set of $d+1$ MUB is explicitly determined by the first $d$ elements.
It is not the case for $2\leqslant m<d$, where $\Delta$ might depend on the choice of a given subset.
Table~\ref{TABLE_MUB} contains values of Hermitian defect consistent for every subset of MUB.

%\medskip

The case of SIC-POVM also reflects isolated character, at least in low dimensions.
For a qubit ($d=2$) and qutrit systems ($d=3$)
the existence of SIC-POVM is well known.
In the former case, we have a unique (up to a global rotation) solution in form of a tetrahedron
inscribed into the Bloch sphere, while in the latter case there
exists a one-parameter family~\cite{Sz14}.
As we have already mentioned, for $d>4$ the problem of SIC families remains open.
We solved associated system of linear equations built from analytical~\cite{Za99, Gr09, Ap05} and numerical
formulas~\cite{SG10}, and computer simulations indicate impossibility
of introduction of free parameters into all SIC-POVM solutions for $4\leqslant d\leqslant 16$.
For $d=8$ this includes a particular example of Hoggar lines~\cite{Ho98}, being
a very special class of supersymmetric SIC-POVM that finds applications
in the theory of quantum-classical channels~\cite{SS16}.

\subsection{Isolated Kochen--Specker Sets}

In Chapter~\ref{chap:EXCESS} of this Thesis we are going to describe some problems
concerning foundations of quantum mechanics, local hidden variables theories and Bell inequalities.
Here we preliminarily mention a single aspect of this subject.
In 1967 Simon Kochen and Ernst Specker~\cite{KS67} proved that a deterministic local hidden variable theory cannot exist,
unless we make some additional assumptions.
This result accompanies the work of John Bell
and implies that a determinism in quantum mechanics leads to a conclusion that
physical observables can be understood only in the context of a given measurement,
while the context stands for the whole idea describing an event.
In other words, local hidden variable theories, which are non-contextual are incompatible with quantum mechanics
and should be discarded. Detailed review of this topic can be found in the recent paper~\cite{BCGKL21}.
%the outcomes of a measurement depend on the way how the measurement is being performed.

The original reasoning presented in~\cite{KS67} is based on $N=117$ vectors (KS vectors) in dimension $d=3$.
Then it can be shown that a system prepared in a state $\rho$ together with KS vectors $\{\phi_j\}_{j=1,...,N}$,
out of which we can prepare a number of orthonormal bases, cannot provide
$N$ deterministic probabilities $p_j={\rm Tr}\big(\rho|\phi_j\rangle\langle\phi_j|\big)\in\{0,1\}$ and,
in consequence, any predefined values of observables must depend on the measurement's context.
Later, another inequivalent KS sets were introduced:
$N=13$ vectors in $d=3$~\cite{YO12}, $N=18$, $d=4$~\cite{CEG96}, and $N=21$, $d=6$~\cite{LBPC14}.
Observe that they all form POVM, hence proceeding as in the previous cases we can build associated Gram matrix,
and our mechanism of restricted defect can be applied to check
the possibility of introduction of free variables. % into the KS vectors.
Appropriate values are tabulated below,

\begin{table}[ht!]
\center
\begin{tabular}{ll | lll | l | l}
$N$ & $d$ & $z$ & $\tau$ & $r$ & $\Delta$ & \# free parameters\\
\hline
13 & 3 & 24  &  78  & 66  & 12 & 0\\
18 & 4 & 63  &  90  & 83  & 7  & 0\\
21 & 6 & 105 &  105 & 103 & 2  & 0
\end{tabular}
\caption{Values of the restricted defect for the three KS sets. Despite the fact that $\Delta\neq 0$, each set is isolated.}
\label{tab:DELTA_KS}
\end{table}

\noindent One immediately notes that in no case the restricted defect $\Delta$ vanishes, which does not automatically excludes these sets
from the isolated domain. However, 
all possible free parameters allowed by a positive restricted defect are absorbed
as global phases of the KS vectors.
Hence, our reasoning implies that all presented sets of KS vectors are isolated.

\section{Free Parameters in MUB, Equiangular Tight Frames and SIC-POVM}

Let us return for a moment to Table~\ref{TABLE_MUB}, which contains upper bounds for the maximal number
of free parameters allowed by all possible subsets of $2\leqslant m\leqslant d+1$ MUB in dimension $2\leqslant d\leqslant 9$.
Now, let us start the analysis beginning from the third row, corresponding to $m=3$.
The restricted defect $\Delta$ for the triplets in dimension $d=4$ is three, which perfectly matches
the maximal number of free parameters~\cite{BWB10}. The possibility of existing of
multidimensional families is also allowed in $d=8$ for $m=2$, $3$, $4$, $5$ and $d=9$ for $m=2$, $3$, $4$.
In particular for $(d,m)=(8,5)$ several such families do exist~\cite{GG15}.

%\medskip

Equiangular tight frames ETF$(d,N)$ composed of $N=k^2$ vectors, each of dimension $d=k(k-1)/2$,
are associated with the Hermitian Fourier matrices
\begin{equation}
[F_N]_{ab}=\exp\Bigg\{\frac{2\pi i}{k}\left({\rm mod}(a,k)\Bigg\lfloor\frac{b}{k}\Bigg\rfloor-{\rm mod}(b,k)\Bigg\lfloor\frac{a}{k}\Bigg\rfloor\right)\Bigg\}\label{HFourier},
\end{equation}
where ${\rm mod}(\cdot,k)$ denotes remainder of the integer division $\cdot/k$.
For prime values of $k=p$ our results suggest that the following statement might be true.
\begin{conjecture}
Given a Fourier matrix $F_{p^2}$ defined in~\eqref{HFourier} for prime $p\geqslant 2$, its Hermitian unitary defect reads
\begin{equation}
\Delta(F_{p^2}) = \frac{1}{2}(p-2)(p-1)(p+1).\label{DHF_formula}
\end{equation}
\end{conjecture}
\noindent Formula~\eqref{DHF_formula} matches (for primes) all values of $\Delta$ presented in the table
\begin{equation}
\begin{tabular}{r|l|l|l|l|l|l|l|l|l|l|l|l|l}
$k$ & 2 & 3 & 4 & 5 & 6 & 7 & 8 & 9 & 10 & 11 & 12 & 13 & 14\\
\hline
$\Delta(F_{k^2}) $ & 0 & 4 & 21 & 36 & 112 & 120 & 273 & 352 & 576 & 540 & 1237 & 924 & 1632
\end{tabular},
\end{equation}
where calculated values of restricted defect $\Delta$ of Hermitian Fourier matrices for $2\leqslant k\leqslant 14$ are presented.
The only case of $k=2$ represents isolated regular three-dimensional simplex ETF$(3,4)$.
For $k=3$, we have $\Delta=4$, however only two-dimensional orbit exist~\cite{RKSC04}, while 
for $k=4$ several six-dimensional families of ETF$(6,16)$ can be found~\cite{GT17}.
Other cases of $k\geqslant 5$ allow us to introduce several free parameters.
Currently we are not able to provide a single formula for $\Delta=\Delta(k)$ nor to prove equation~\eqref{DHF_formula}.

%\medskip

Finally, we switch to families of SIC-POVM.
Consider $d=3$ and a fiducial vector
\begin{equation}
|\phi_{00}\rangle=\frac{1}{\sqrt{2}}[1,-1,0]^{\rm T}.
\end{equation}
Following~\cite{Za99} we define the Weyl-Heisenberg SIC-POVM
\begin{equation}
|\phi_{ab}\rangle=X^aZ^b|\phi_{00}\rangle,
\end{equation}
where, conventionally, $X|j\rangle=|j\oplus_3 1\rangle$, $Z|j\rangle=\exp\big\{\frac{2\pi i j}{3}\big\}|j\rangle$
and $\oplus_3$ denotes addition $\bmod\,3$.
After constructing the appropriate Gram matrix and solving associated system of linear equations, we arrive
at the restricted defect $\Delta=4$ and six one-dimensional families of SIC-POVM.
All solutions are equivalent --- see~\eqref{equivalence_relation} applied to the Gram matrix.
Moreover, they belong to the four-dimensional tangent plane described by Eq.~\eqref{lim_d_dt}
and cannot fit into a lower-dimensional space. Hence, the defect $\Delta$ cannot be decreased.
This solution is the most general SIC-POVM in $d=3$~\cite{Sz14} up to equivalence.
The family reads
\begin{equation}
\phi(\gamma) =\frac{1}{\sqrt{2}}\left[1,e^{i\gamma},0\right]^{\rm T} \quad : \quad\gamma\in[0,2\pi). 
\end{equation}
The value of the Hermitian defect $\Delta$ for a generic value of $\gamma\in[0,2\pi)$ is $2$, while
only for $|\phi_{00}\rangle$ we have $\Delta = 4$.
This is yet another example of a situation when the value of the defect is larger than a maximal possible family of solutions.

\section{Robustness of the Restricted Defect $\Delta$}

As we already mentioned, imperfect nature of physical systems
provides several obstacles in quantum computing.
On the other hand, calculations performed by classical machines are not free from their own issues either.
Vast majority of problems, even if they can be easily expressed in simple algebra, are usually
not computable analytically due to the complicated character of possible solutions.
Such numerical data can later serve as input for further analysis and,
independently of the initial precision set, the error propagation might
lead to unreliable results when they are subjected to many subsequent iterations and transformations.
The best example in the context of this Thesis was already presented in Chapter~\ref{chap:CHM}, where
we calculated several dozens of CHM (solving numbers of unitarity constraints numerically)
and claimed they all provide isolated examples of Hadamard matrices, as the unitary
defect vanishes for every matrix.
However, in that particular case such a statement is absolutely justified, due to the
properties of the numerical algorithm for the defect.

In this section we will study the robustness of the Hermitian unitary defect
under the presence of errors introduced to the POVM.
We only present the sketch of the reasoning %for the restricted defect $\Delta$
with all mathematical details left in Appendix of~$[\hyperlink{\paperslist}{\rm A3}]$.
Obviously, all arguments and calculations can be applied directly to the standard unitary defect too.

Consider a Gram matrix defined in~\eqref{Gram_matrix} by means of vectors $|\varphi_j\rangle\in\mathbb{C}^d$
and now we let these vectors be slightly perturbed,
$
|\varphi'_j\rangle = |\varphi_j+s\xi_j\rangle,
$
where $\xi_j$ is a random variable distributed uniformly in the interval $[-1,1]$ %^{\times d}$
and
\begin{equation}
s=\frac{1}{\sqrt{d}}\max_j\Big|\Big||\varphi'_j\rangle-|\varphi_j\rangle\Big|\Big|\label{inaccuracy_factor}
\end{equation}
is the factor quantifying\footnote{For example, if we set the working precision
to $k$ decimal digits, that is the components of $|\varphi'_j\rangle$
differ from the actual values $|\varphi_j\rangle$ on the $(k+1)^{\rm th}$ decimal place, 
the value of the inaccuracy factor is $s\approx 10^{-k}$. In definition of $s$ we use the standard Euclidean norm.}
the maximal allowed inaccuracy in components of $|\varphi_j\rangle$.

\begin{figure}[ht!]
\center
\includegraphics[width=5.4in]{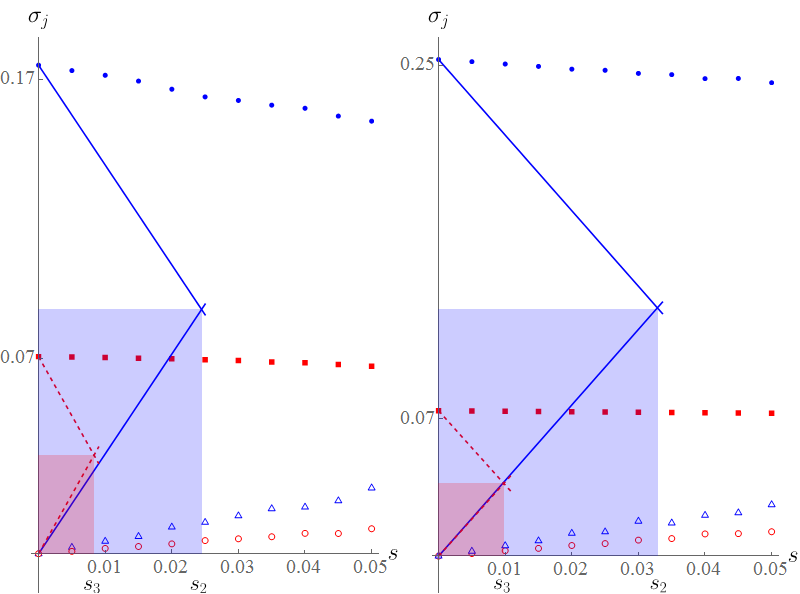}
\caption{Two smallest singular values $\sigma_0$ and $\sigma_1$
of the matrix $\mathcal{R}$ representing system~\eqref{2cUUURRL} associated with the SIC-POVM of order $d=4$ and $8$
are presented on the left panel.
Singular values $\sigma_j$ (for $j=0,1$) are functions of the inaccuracy factor $s$~\eqref{inaccuracy_factor}.
Open and solid symbols
represent $\sigma_0$ and $\sigma_1$, respectively, for two- ({\color{blue}$\bullet$~{\scriptsize $\triangle$}})
and three-qubit ({\color{red}{\scriptsize $\blacksquare$}~$\circ$}) systems. Each case is averaged over eight samples randomly
chosen from the highly accurate SIC-POVM~\cite{SG10} (we do not plot error bars for clarity).
The confidence regions depicted by light red and blue rectangles (color online) are described
by values of $0\leqslant s\leqslant s_{\rm max}$,
where $s_{\rm max}$ is determined by the intersection of the lower and upper bounds.
Values $s_2$ and $s_3$ represent two- and three-qubits systems, respectively.
Similar picture on the right panel holds for the full set of MUB of size $d=4$ and $8$
prepared from analytic formulas~\cite{WF89, Ch02, SK14}.}
\label{fig:confidence_regions}
\end{figure}

At the heart of calculations of the restricted defect $\Delta$ is the problem of determining
the rank of the system of equations~\eqref{2cUUURRL}. Hence, we are interested how
perturbation of the vectors $|\varphi'_j\rangle$ affects the values of the two smallest
singular values $\sigma_0=0$ and $\sigma_1>0$ of a matrix $\mathcal{R}$ associated
with system~\eqref{2cUUURRL}. As long as we can distinguish the two singular values,
we can correctly calculate the rank of the system and, in consequence, we know
the proper value of $\Delta$.
We must estimate an upper bound for the maximal perturbation of $\sigma_0$
and a lower bound for the maximal perturbation of $\sigma_1$ which leads 
to the inequality,
\begin{equation}
|\sigma'_{0,1}-\sigma_{0,1}|\leqslant \frac{64 s d^2}{N}\left(1-\frac{2d}{N}\right)^2\sqrt{\frac{N-d}{N(N-1)}}\label{sigma_bound},
\end{equation}
where $s$ is the inaccuracy factor~\eqref{inaccuracy_factor},
$N$ and $d$ such that $N>2d$ are appropriate dimensions introduced earlier,
the prime symbol indicates a perturbed singular value,
and the double subscript $\sigma_{0,1}$ denotes the fact that the above bound is valid for both $\sigma_0$ and $\sigma_1$.

The minimal value of $s$ such that the critical condition $\sigma'_0=\sigma'_1$ holds,
determines the upper bound for the confidence region of $\Delta$.
The value of $\Delta$ will not change if
\begin{equation}
0\leqslant s\leqslant s_{\max}=\frac{\sigma_1}{2f(d,N)}\leqslant\frac{\sigma'_1+f(d,N)s}{2f(d,N)},
\end{equation}
where $f(d,N)$ is the right-hand side of~\eqref{sigma_bound}
and the last inequality (since $\sigma_1$ depends on unavailable exact solution)
can be inferred from~\eqref{sigma_bound} too, under the assumption that $f(d,N)s\ll 1$.

In Figure~\ref{fig:confidence_regions} we depicted confidence regions for maximal sets of MUB and SIC-POVM
for two and three qubits. When leaving the confidence regions
we no longer can discriminate $\sigma_0$ and $\sigma_1$.
Numerical simulations show that setting an appropriate threshold we can
calculate the correct value of the (restricted) defect even for considerably disturbed
matrices. Nevertheless, in all
cases here and in previous chapter, we worked with highly accurate objects,
so that all presented results concerning ${\bf d}(U)$ and $\Delta$ are absolutely credible.

\section{Summary}

In this chapter we investigated the possibility to introduce free parameters
to a given generalized quantum measurement with a prescribed symmetry.
Two most relevant examples of generalized measurements
in low dimensions have been thoroughly examined, namely: mutually unbiased bases (MUB) and
symmetric informationally complete-positive operator-valued measures (SIC-POVM).

Based on the notion of unitary defect we introduced its symmetric analog $\Delta$
applicable to Hermitian unitary matrices with constant diagonal, which represent
Gram matrices corresponding to given POVM. This tool allows us to prove that
known maximal sets of MUB in squared dimensions $d\in\{4,8,9,16\}$ as well as
SIC-POVM in dimensions $d\in\{4,5,...,16\}$ are isolated. The result includes
a special class of the three-qubit system called Hoggar lines.
Also, we showed that three Kochen--Specker sets of vectors,
used in the theorem indicating contextuality of quantum mechanics, are isolated, so they
cannot be extended.

Moreover, calculating Hermitian unitary defect we can estimate the upper bound
for the maximal possible number of free parameters
that can exist in incomplete subsets of $2\leqslant m< d$ MUB
in dimension $d\in\{2,3,...,9\}$. Similar calculations
have been done for SIC-POVM and we recovered one-dimensional family
of this kind of POVM for $d=3$.

\medskip

Several open questions are in order:
\begin{enumerate}
\item Are maximal sets of $d+1$ MUB isolated in prime power dimensions?
\item Are SIC-POVM isolated in every dimension $d\neq 3$?
\item Can defect of a unitary matrix be extended over different classes of matrices?
\end{enumerate}

To answer the two first questions one needs a more efficient software to solve
very large systems of linear equations.
The last question should become clear after the next chapter where we will 
investigate a very special class of unitary matrices.
However, one cannot generalize the defect too far, because the idea of a tangent space,
that is the core of the defect, is well
established in differential and algebraic geometry and imposes natural constraints
on conditions that should be fulfilled by matrices under consideration.
One should focus on manifolds defined by matrix equations, where
the linear algebra machinery to manipulate an enormous number of equations can be used.
 %- it may happen that there are as many types of matrix equations as there are ordinary equations.

%\bigskip

\begin{figure}[ht!]
\center
\includegraphics[width=0.5in]{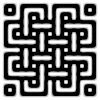}
\end{figure}

%\bigskip

We close the first part of this Thesis.
In the next chapter we recall the phenomenon of quantum entanglement and provide
an unexpected solution related to the particular state describing four-partite entanglement of six-level systems.

%%%%%%%%%%%%%%%%%%%%%%%%%%%%%%%%%%%%%%%%%%%%%%%%%%%%%%%%%%%%%%%%%%%%
\clearpage\null\thispagestyle{empty}
\chapter{Absolutely Maximally Entangled States (AME) of Four Quhexes}
\label{chap:AME46}
Chapter based on~$[\hyperlink{\paperslist}{\rm A4}]$.
\medskip
\medskip
\medskip

\noindent There were many twists and dead-ends in the history of searching for the AME$(4,6)$ state of four subsystems with six levels each,
which eventually have led us in~$[\hyperlink{\paperslist}{\rm A4}]$ to the conclusion in the form of

\begin{theorem}
There exists an {\rm AME}$(4,6)$ state.\label{AME_theorem}
\end{theorem}

Before we present a proof of this statement, we begin with preliminary results
that should be found interesting and might
shed some light on additional properties of such a state, which (at the time
of preparing this Thesis) are still obscured and not entirely understood.

\section{Quantum Entanglement}

Consider two linear spaces $\mathbb{C}^{d_1}$ and $\mathbb{C}^{d_2}$ of dimension $d_1$ and $d_2$, respectively.
It turns out, that a tensor product $\mathbb{C}^{d_1}\otimes \mathbb{C}^{d_2}$ is not equal to $\mathbb{C}^{d_1d_2}$
(individual tensors do not span the entire space)
and it can only be presented as a sum of two subsets,
\begin{align}
\mathbb{C}^{d_1d_2}=& \ \big\{x\otimes y : x\in\mathbb{C}^{d_1}, y\in\mathbb{C}^{d_2}\big\} \ \cup\nonumber\\
\cup & \ \Big\{\sum_{j=1}^r x_j\otimes y_j : x_j\in\mathbb{C}^{d_1}, y_j\in\mathbb{C}^{d_2}, r\geqslant 2\Big\}.
\end{align}
This means that not every vector $z\in\mathbb{C}^{d_1d_2}$ can be factorized onto the tensor product
$z=x\otimes y$ of two vectors $x\in\mathbb{C}^{d_1}$ and $y\in\mathbb{C}^{d_2}$.
To span entire space $\mathbb{C}^{d}$ (imposing a particular tensor structure allowed by the
factorization $d=d_1d_2$) we must include additional linear combinations of $x_j\otimes y_j$.
This observation can be generalized for any composite dimension $d=d_1d_2...d_n$. % $\displaystyle{d=\prod_j d_j}$.
The most celebrated example of this fact is usually illustrated by the bipartite Bell state
\begin{equation}
\big|\Psi^+\big\rangle=\frac{1}{\sqrt{2}}\big(|00\rangle+|11\rangle\big)=\frac{1}{\sqrt{2}}[1,0,0,1]^{\rm T}\neq
|x\rangle\otimes |y\rangle  \quad : \quad |x\rangle, |y\rangle\in\mathbb{C}^2. \label{Bell_state}
\end{equation}
This nontrivial property of tensor spaces
is a mathematical counterpart of the most intriguing non-classical phenomenon of the quantum world, called
{\sl quantum entanglement}, which describes strong correlations between composite
physical systems when subjected to a measurement process.

Every quantum system has an associated Hilbert space $\mathcal{H}_d\equiv\mathbb{C}^d$ (complex vector space) of size $d$.
Throughout this work we assume only finite-dimensional cases and composite systems, i.e
an appropriate Hilbert space takes the form of the tensor product of individual components (also finitely many)
%\begin{equation}
$\mathcal{H}_d=\mathcal{H}_{d_1}\otimes ... \otimes\mathcal{H}_{d_n}$ with $d=d_1d_2...d_n$ for some $n\geqslant 2$.
%\end{equation}
%with $\displaystyle{N=\prod_j N_j}$.
Each $\mathcal{H}_{d_j}$ is called a local subspace and corresponds to particular degrees of freedom of a given quantum object.
However, in a general picture, $\mathcal{H}_{d_j}$ might represent different microscopic objects or even distant laboratories.
Now, the information about the system carried by the quantum state, described by
the density matrix $\rho\in\mathbb{C}^{d\times d}$, is spread nonlocally over all
components (parties), which is a background for physical manifestation of entanglement.

Formally, we define a mixed state $\rho$ to be {\sl entangled}, if the following convex factorization does not exist,
\begin{equation}
\rho=\sum_k p_k\bigotimes_{j=1}^n\rho_{d_j,k},
\end{equation}
where operators $\rho_{d_j,k}$ belong to $\mathcal{L}(\mathcal{H}_{d_j})$ (space of linear maps in $\mathcal{H}_{d_j}$) and $p_k\geqslant 0$ are probabilities satisfying the normalization condition
$\sum_k p_k=1$.
In the case of a pure state this simplifies to the statement that
$|\psi\rangle\in\mathcal{H}_d$ is said to be {\sl entangled pure state}
if it cannot be expressed as the product of states from the individual subspaces
\begin{equation}
|\psi\rangle=\bigotimes_{j=1}^n|\psi_j\rangle \quad : \quad |\psi_j\rangle\in\mathcal{H}_{d_j}.
\end{equation}
Any state which is not entangled should be called {\sl separable}.

It must be emphasized that the entanglement makes sense only in the context
of the imposed tensor structure. For example,
consider a dimension $d=d_1d_2=d_3d_4$. States, which are entangled with respect to $\mathbb{C}^{d_1}\otimes\mathbb{C}^{d_2}$ might not
necessarily be entangled as elements of $\mathbb{C}^{d_3}\otimes\mathbb{C}^{d_4}$.

Certainly, we shall not delve into the problem of investigating and quantifying the problem of entanglement,
which is a good topic for a dozen of such theses~\cite{HHHH09, BFZ21}.
Instead, let us only recall the major applications of this quantum resource~\cite{Ho21, CG19}
shared by different parties of the entire system,
that can be used as a footing of other important processes like
dense coding schemes and teleportation~\cite{We01}, quantum cryptography~\cite{BB14, Ek91},
quantum thermodynamics~\cite{CG19} and many others, which would be impossible without existence of entangled states.

\section{Absolutely Maximally Entangled States}

Working with entanglement we distinguish two extrema.
Fully separable states for which entanglement vanishes, and the most entangled states
called maximally entangled, to which we devote the rest of this chapter.
What is in between goes beyond the scope of the story.
One should find many additional introductory information to the problem of measuring the amount
of entanglement in the survey~\cite{PV07}.

Suppose we have a system composed of $N$ parties,
and each party is represented by the Hilbert space $\mathcal{H}_d$ of size $d$, so in the QIT jargon
it is called a {\sl qudit} (being a natural extension of $2$-dimensional qubits).
We say that a pure state $|\psi\rangle$ of such multipartite system is {\sl $k$-uniform}~\cite{GALRZ15}, if
it maximizes the entanglement among all possible bipartitions. In other
words, we set $1\leqslant k\leqslant\lfloor N/2\rfloor$ and check
whether the state of $k$ subsystems becomes maximally mixed after tracing out $N-k$
other subsystems, that is, its reduced density matrix $\rho_{N-k}$ is proportional to identity,
\begin{equation}
{\rm Tr}_{N-k}\left(\rho_N\right)=\rho_{k}=\mathbb{I}_{d^k}/d^k,
\end{equation}
where $\rho_N=|\psi\rangle\langle\psi|$ is a density matrix of the entire system and ${\rm Tr}_{N-k}$ denotes
partial trace~\cite{Ma17} over $N-k$ subsystems.
We have already seen one such maximally entangled Bell state~\eqref{Bell_state} for which $N=2$ and $k=1$.
Finally, we define {\sl absolutely maximally entangled} (AME) states~\cite{GALRZ15, Fa09} as those for which $k=\lfloor N/2\rfloor$.

From now on, the symbol AME$(N,d)$ represents AME state of $N$ qudits.
In the work, we are interested in AME states of four quhexes, AME$(4,6)$, which describe\footnote{Actually, we should
say ``which should describe...'', because at this stage we do not know yet if such states exist.} a
physical system consisting of four parties with six degrees of freedom each.

As we did for CHM and POVM, let us briefly review the applications and existence aspects of AME states.
Most of all, highly entangled AME states are widely used in the field of quantum error correction~\cite{Sc04, LMPZ96}.
Each AME$(N,d)$ is related to a quantum pure error correcting code~\cite{MFGHS20}
\begin{equation}
{\rm AME}(N,d) \leftrightarrow (\!(N,1,N/2+1)\!)_d,
\end{equation}
where the triplet $(\!(N,K,D)\!)_d$ denotes $K$ codewords in $\mathcal{H}_d\otimes...\otimes\mathcal{H}_d$ ($N$ times)
characterized by distance $D$~\cite{Sc04}.
Such a code can be used in detection and recovering all errors acting on less than $D/2$ qudits.
Quantum error correcting codes (QECC) can be constructed from particular AME states (treated as codewords), and conversely
some AME states arise from QECC~\cite{RGRA18}.
Another applications of AME states include quantum secret sharing protocols~\cite{HCLRL12}, multipartite teleportation~\cite{HC13},
and even they can be found in such areas of modern physics like quantum holography~\cite{PYHP15}, modeling
the AdS/CFT correspondence~\cite{MFGHS20,Ma99,ADH15},
quantum gravity~\cite{Va19} or string theory~\cite{BDMR11}.

Existence of AME states is yet another open problem, which deserves a lot of effort~\cite{HESG18}.
They do not exist for four qubit system, $\mathcal{H}_2^{\otimes 4}$~\cite{HS00}, but they do exist for
four qudit systems $\mathcal{H}_d^{\otimes 4}$ with $d=3$, $4$, $5$, $7$, $8,...$~\cite{GZ14} (where number $6$ for quhex system is temporarily missing),
and for six qudit systems, $\mathcal{H}_d^{\otimes 6}$ with $d\geqslant 2$~\cite{Ra99}.
It has been also recently proved~\cite{HG20} that for all dimensions $N$ one 
can find arbitrarily large value of $d$ such that AME$(N,d)$ exists.
Comprehensive list including the most current state-of-the-art concerning AME states
is shown in the online {\sl Table of AME states and Perfect Tensors}~\cite{HW_table}.
        
\subsection{Multipartite Entanglement Algebra}

We slowly set the stage for the main result of this Thesis.
Without loss of generality, let us restrict to the scenario of four qudits (general 
discussion can be found in the supplementary material of~$[\hyperlink{\paperslist}{\rm A4}]$).
So, we assume the system composed of $N=4$ parties, each of dimension $d$, described by a pure state
$|\psi\rangle\in\mathcal{H}=\mathcal{H}_d\otimes\mathcal{H}_d\otimes\mathcal{H}_d\otimes\mathcal{H}_d$.
According to definition, $|\psi\rangle$ is AME$(4,d)$, if the partial trace of $|\psi\rangle\langle\psi|$
over any subsystem composed of $\lfloor N/2\rfloor=2$ parties is proportional to identity.
Let us expand $|\psi\rangle$ in the product basis of $\mathcal{H}$
\begin{equation}
|\psi\rangle=\sum_{j,k,l,m=1}^d\mathcal{T}_{jklm}|j\rangle|k\rangle|l\rangle|m\rangle,
\end{equation}
where a four-index tensor $\mathcal{T}_{jklm}$ can be transformed (reshaped) into a matrix $U\in\mathbb{C}^{d^2\times d^2}$.
addressed by $U_{\alpha,\beta}$ with $\alpha=k+d(j-1)$ and $\beta=m+d(l-1)$.
Note that it is actually possible to obtain six different matrices because there are six possible
ways of performing bipartition on a fourpartite system.
The tensor $\mathcal{T}$ describes AME$(4,d)$ iff every matrix obtained in this way is unitary~\cite{GALRZ15}.
Since unitarity is invariant under transposition, we can reduce considerations to only three matrices (dropping three
particular reorderings of the four-index $_{jklm}$), and it turns out, it is enough for being AME$(4,d)$
to require that $\mathcal{T}_{jklm}$, $\mathcal{T}_{jmlk}$ and $\mathcal{T}_{jlkm}$ give rise to three different
unitary matrices of size $d^2$.
Such tensors $\mathcal{T}$ are called {\sl perfect} while matrices $U\in\mathbb{U}(d^2)$ are called {\sl $2$-unitary}.
Analogously, in the general case of $N$ subsystems, where $N$ is even,
we can introduce the notion of $k$-unitarity\footnote{Index $k$ in $\mathcal{T}_{jklm}$ has nothing
in common with $\underline{k}$-unitarity.} of a matrix of order $d^k$ with $k=N/2$, see Appendix C in~\cite{GALRZ15}.

To describe entanglement qualitatively, 
we must recall two simple tools from the matrix algebra.
The first one is partial transpose operation taken on a given subsystem of
a bipartite system of size $d_1d_2$ described by a matrix $U=U_{d_1}\otimes U_{d_2}$.
It can be defined as
\begin{align}
U^{\rm T_1}&=({\rm T}\otimes \mathbb{I}_{d_2})(U_{d_1}\otimes U_{d_2}) = U_{d_1}^{\rm T}\otimes U_{d_2},\\
U^{\rm T_2}&=(\mathbb{I}_{d_1}\otimes {\rm T})(U_{d_1}\otimes U_{d_2}) = U_{d_1}\otimes U_{d_2}^{\rm T},
\end{align}
where ${\rm T}$ denotes the standard matrix transpose.
The second one is reshuffling (realignment), denoted by $U^{\rm R}$,
which transforms each block of $U$ into a consecutive row. For a schematic matrix
representing a two spin-$\frac{1}{2}$ system ($\mathbb{U}(2)\otimes\mathbb{U}(2)$) the reshuffling transformation looks like,
\begin{equation}
\left[\begin{array}{cc|cc}
a_1 & a_2 & b_1 & b_2\\
a_3 & a_4 & b_3 & b_4\\\hline
c_1 & c_2 & d_1 & d_2\\
c_3 & c_4 & d_3 & d_4
\end{array}\right]
\stackrel{\rm R}{\to}
\left[\begin{array}{cccc}
a_1 & a_2 & a_3 & a_4\\\hline
b_1 & b_2 & b_3 & b_4\\\hline
c_1 & c_2 & c_3 & c_4\\\hline
d_1 & d_2 & d_3 & d_4
\end{array}\right].
\end{equation}
Note, that while global transpose operation ${\rm T}$ is, by definition, dimension-agnostic,
its partial counterparts ${\rm T_1}$ and ${\rm T_2}$ (as well as ${\rm R}$) strongly
depend on the internal structure of the system.
However, in our case, the situation is quite simplified, because the size of two
local subsystems is equal and fixed $d=(d_1=d_2)=6$. Moreover, from
\begin{equation}
(U^{\rm T_2})^{\rm T_1}=U^{\rm T}\Longrightarrow U^{\rm T_2}=(U^{\rm T_1})^{\rm T}
\end{equation}
we infer that the spectra of $U^{\rm T_1}$ and $U^{\rm T_2}$ coincide. Therefore, we can concentrate only on one such operation
and, skipping irrelevant indices, we denote the partial transpose $\rm T_2$ by $\rm\Gamma$, as the symbol $\rm\Gamma$ resembles a part of $\rm T$. %we put ${\rm\Gamma} = {\rm T_2}$.
If $U$ is represented in a product basis, $U_{jklm}=\langle j,k|U|l,m\rangle$, 
both operations can be formally defined using multi-indices as
\begin{equation}
U_{jklm}^{\rm\Gamma}=U_{jmlk} \quad \text{and} \quad U_{jklm}^{\rm R}=U_{jlkm}\quad \text{for} \quad 1\leqslant j,l\leqslant d_1, \ 1\leqslant k,m\leqslant d_2.\label{RG_multiindex}
\end{equation}
%for $j,l\in\{1,...,d_1\}$ and $k,m\in\{1,...,d_2\}$.

Physical interpretation of ${\rm\Gamma}$ is adapted straightforwardly from the one of
${\rm T}$ and can be viewed as a time reversal process applied to
a single subsystem under consideration~\cite{BM05}.
It also plays the fundamental role in the ${\rm PPT}$
criterion of separability~\cite{Pe96, HHH96}.
In contrary, the transformation ${\rm R}$, which maps vectors of length $d^2$ into matrices of size $d$,
does not have any simple physical explanation.
Both operations ${\rm\Gamma}$ and ${\rm R}$ can be found in theory of superoperators~\cite{BZ17, BCSZ09}
or in dynamics in dual-unitary quantum circuits~\cite{PBCP20}.

Putting this all together, we can conclude that searching for AME$(4,d)$ state
is equivalent to searching a $2$-unitary matrix $U\in\mathbb{C}^{d^2\times d^2}$
such that it remains unitary after operations of reshuffling and partial transposition.
Three possible splittings of fourpartite system of qudits corresponding to the three configurations of indices of unitary matrix
are shown below
\begin{equation}       
        \begin{tabular}{cc}
        {\color{black}$\mathcal{A}$} & {\color{black}$\mathcal{B}$} \\
        \hline
        \cellcolor{lightblue}{\color{black}$\mathcal{C}$} & \cellcolor{lightblue}{\color{black}$\mathcal{D}$} \\
        \end{tabular} \ \leftrightarrow  \ U, 
        \qquad
        \begin{tabular}{c|c}
        {\color{black}$\mathcal{A}$} & \cellcolor{lightblue}{\color{black}$\mathcal{B}$} \\
        {\color{black}$\mathcal{C}$} & \cellcolor{lightblue}{\color{black}$\mathcal{D}$} \\
        \end{tabular} \ \leftrightarrow  \ U^{\rm\Gamma},
        \qquad
        \begin{tabular}{c|c}
        {\color{black}$\mathcal{A}$} & \cellcolor{lightblue}{\color{black}$\mathcal{B}$} \\
        \hline
        \cellcolor{lightblue}{\color{black}$\mathcal{C}$} & {\color{black}$\mathcal{D}$} \\
        \end{tabular} \ \leftrightarrow  \ U^{\rm R},
\end{equation}
where, for a better perception, all four subsystems have been assigned a name.

\subsection{Entangling Power and Gate Typicality}

To measure the entanglement, we are going to use the following linear entropy characterizing
distribution of singular values of an arbitrary matrix
$M\in\mathbb{C}^{d^2\times d^2}$,
\begin{equation}
E(M)=\frac{d^2}{d^2-1}\left(1-\frac{{\rm Tr}(MM^{\dagger}MM^{\dagger})}{{\rm Tr}^2(MM^{\dagger})}\right)
=\frac{d^2}{d^2-1}\left(1-\frac{1}{{\rm Tr}^2(MM^{\dagger})}\sum_{j=1}^{d^2}\lambda_j^2\right),\label{linear_entropy}
\end{equation}
where $\lambda_j$ is $j^{\rm th}$ eigenvalue of $MM^{\dagger}$ (squared singular value of $M$).
This entropy is normalized so that $E\in[0,1]$ and the maximum value
is attained for unitary matrices for which $\lambda_j=1$. 

{\sl Entangling power} of a bipartite unitary matrix $U$, can be defined by the entropy $E$,
\begin{equation}
e_p(U) = E\Big(U^{\rm R}\Big) + E\Big(US\Big)^{\rm R} - 1,
\end{equation}
where $S$ is the swap operator, $S(U_{d_1}\otimes U_{d_2})S=U_{d_2}\otimes U_{d_1}$.
This is very particular formulation, designed to facilitate operational applicability (especially
numerical investigation).
For a general definition of entangling power, refer to the original paper~\cite{Z01}.

Entangling power is interpreted as the average entanglement created when $U$
acts on a generic product state. The swap does not create any entanglement when acting on any bipartite
product state, which is a motivation for the definition of another indicator
called {\sl gate typicality}~\cite{JMZL17}, denoted by $g_t(U)$,
which distinguishes local operations from the swap
\begin{equation}
g_t(U) = E\Big(U^{\rm R}\Big) - E\Big(US\Big)^{\rm R} + 1.
\end{equation}
Both quantities are normalized,\footnote{Note that other authors can use different normalization for $e_p$, $g_t$ and $E$.} $e_p\in[0,1]$ and $g_t\in[0,2]$.
Notably, a hypothetical $2$-unitary matrix $T_{d^2}$ ($T$ for target\footnote{One should not be
confused when see transpose symbol ${\rm T}$ or $^{\rm T}$, matrix $T$, and
tensor $\mathcal{T}$ and also $R$, and ${\rm R}$ or $^{\rm R}$ for a rotation matrix and reshuffling operation, respectively.})
corresponding to AME$(N,d)$ state, if one exists, is characterized by $e_p(T_{d^2})=g_t(T_{d^2})=1$.

\begin{figure}[ht!]
\center
\includegraphics[width=4.2in]{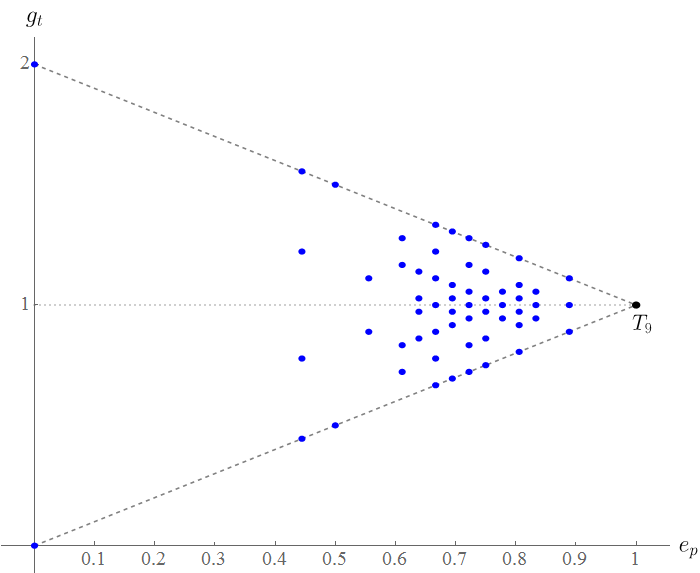}
\caption{Given $d>1$, all possible values of entangling power and gate typicality
form a characteristic triangular shape. Picture shows the case of $d=3$ with $60$ dots
representing possible classes of permutation matrices from $\mathbb{P}(9)$. One particular
point at $(e_p,g_t)=(1,1)$ denotes a $2$-unitary permutation matrix $T_9$ representing an AME$(4,3)$ state~\eqref{AME43}.}
\label{fig:epgt3}
\end{figure}

Given dimension $d>1$, all possible values of $e_p(U)$ and $g_t(U)$ for $U\in\mathbb{U}(d^2)$
form a characteristic triangle-shaped area. We call this informally as
an {\sl $(e_p,g_t)$-plane}. For $d=3$ it is sketched in Figure~\ref{fig:epgt3}.
In this particular case, all the dots within the triangular confinement depict
permutation matrices of size $9$. We see as, despite they do not cover the available
area uniformly, one distinguished point appears in the rightmost corner --- this is
$T_9$ representing an AME state of four qutrits~\cite{GALRZ15},
\begin{align}
|{\rm AME}(4,3)\rangle = \big(&|0000\rangle + |0112\rangle + |0221\rangle\,+\label{AME43} \\
&|1011\rangle +|1120\rangle +|1202\rangle\,+\nonumber\\
&|2022\rangle +|2101\rangle +|2210\rangle\big)/3,\nonumber
\end{align}
which in this case can also be expressed in the form of permutation
\begin{equation}
P_{\rm AME(4,3)}=[3,7,5,4,2,9,8,6,1],\label{PM_as_vector}
\end{equation}
where each number denotes the column with unity at $j^{\rm th}$ position.

Obviously, each point on the $(e_p,g_t)$-plane (not necessarily a permutation matrix)
does not describe a single matrix, but rather a whole continuum of matrices.
It is easy to see that entangling power is invariant with respect to local operations (rotations)
\begin{equation}
e_p(U)=e_p\left(\left(U_d^{(1)}\otimes U_d^{(2)}\right)U\left(U_d^{(3)}\otimes U_d^{(4)}\right)\right).\label{local_rotations}
\end{equation}
for $U_d^{(1)}$, $U_d^{(2)}$, $U_d^{(3)}$ and $U_d^{(4)}\in\mathbb{U}(d)$ and $U\in\mathbb{U}(d^2)$.
This fact will be extremely useful in Section~\ref{sec:AME_beautifier}.
Another property of $(e_p,g_t)$-plane is its symmetry along the bisector ($g_t\equiv 1$). Moreover,
the lower and upper edges %, containing dual unitary matrices\footnote{Matrix $U$ is called dual unitary,
are related by the mapping
\begin{equation}
\Big(X\in \text{lower edge} \Longrightarrow X\sqrt{S}\in \text{bisector}\Big) \Longrightarrow XS\in \text{upper edge},
\end{equation}
for $S$ be the swap. We will use $(e_p,g_t)$-plane to present pictorially
several interesting features of matrices from $\mathbb{O}(36)$ in Section~\ref{sec:matrix_V}.

\section{Quantization of the Problem of $36$ Officers of Euler}

Since searching for AME states can be performed
in the domain of unitary matrices and in particular, the state AME$(4,6)$ corresponds
to a $2$-unitary matrix from $\mathbb{U}(36)$, we should find such a matrix.
We start our search from the reminder of some classical problem
and its transformation to the quantum realm.

We have already recalled the definition of (orthogonal) Latin square in Section~\ref{sec:CHM_LS} of Chapter~\ref{chap:CHM}.
Problem of existence of OLS emerges at the very beginning for $d=2$.
In general, there are only two dimensions for which the answer concerning
orthogonality of Latin squares is negative; it is for $d=2$ and $d=6$~\cite{CD01, BSP60}, and
the latter case corresponds directly to the famous problem of thirty-six officers of Euler~\cite{Eu79}.
In the classical formulation, the question pertains to the arrangement of $36$ officers, six of different
rank, and six of different regiment (assignment), in a $6\times 6$ square such that no rank nor regiment
repeats in any row and column.
Equivalently, one asks about two Latin squares (one of ranks and another of regiments) related by orthogonality.
While the case of $d=2$ is trivial, it took more than a century to prove that no solution
of the $6$-dimensional problem exists~\cite{Ta01}.

Such problems can be conceivably solved in the quantum realm, which provides more room
for potential solutions as it was shown e.g. in the case of SudoQ~\cite{PWRBZ21}
or as we will see in Chapter~\ref{chap:EXCESS}. But, in the first
place, the issue must be properly quantized.

After~\cite{MV16} we recall the definition of {\sl quantum Latin square} (QLS), which is
a $d\times d$ array of elements of the Hilbert space $\mathcal{H}_d$ such that every row
and every column forms an orthonormal basis in $\mathcal{H}_d$. For example,
the following constellation of quantum states from $\mathcal{H}_4$ forms a QLS,
\begin{equation}
\left[\begin{array}{cccc}
|0\rangle & |1\rangle & |2\rangle & |3\rangle \\
|1\rangle & |0\rangle & |3\rangle & |2\rangle \\
|2\rangle & |3\rangle & |0\rangle & |1\rangle \\
|3\rangle & |2\rangle & |1\rangle & |0\rangle
\end{array}\right].\label{QLS_example}
\end{equation}
Two such quantum squares $\mathcal{Q}_1$ and $\mathcal{Q}_2$ are called {\sl orthogonal quantum Latin squares} (OQLS) iff
the pointwise inner product of any row from $\mathcal{Q}_1$ with any row from $\mathcal{Q}_2$ provides only a single $1$
among all zeros. Above QLS~\eqref{QLS_example} can be complemented\footnote{We apply standard orthogonality symbol $\perp$ to denote orthogonality of two QLS.} with the following one,
\begin{equation}
\left[\begin{array}{cccc}
|0\rangle & |1\rangle & |2\rangle & |3\rangle \\
|1\rangle & |0\rangle & |3\rangle & |2\rangle \\
|2\rangle & |3\rangle & |0\rangle & |1\rangle \\
|3\rangle & |2\rangle & |1\rangle & |0\rangle
\end{array}\right] \
\perp \
\left[\begin{array}{cccc}
|0\rangle & |2\rangle & |3\rangle & |1\rangle \\
|1\rangle & |3\rangle & |2\rangle & |0\rangle \\
|2\rangle & |0\rangle & |1\rangle & |3\rangle \\
|3\rangle & |1\rangle & |0\rangle & |2\rangle
\end{array}\right].
\end{equation}

One particular open problem concerns the question about existence of a pair of quantum
orthogonal LS in every dimension~\cite{MV19}.
If it was possible to find such OQLS of size six then, apart from filling one blank spot in quantum combinatorics discipline,
the quantum analog of the classical problem of Euler would be solved.

\section{Orthogonal Approximation of AME$(4,6)$}

Even though, there are no two orthogonal Latin squares of order six, one can examine
pairs of such objects, which are as close to be orthogonal as possible.
This problem has its roots in the studies of Horton~\cite{H74}.
Later, Lieven Clarisse et al. \cite{CGSS05}
investigated permutation matrices and Latin squares in the context of entangling power.
Here, we briefly recall one particular result, which from our perspective is the most interesting.
It states that there exists a permutation
matrix $P\in\mathbb{P}(36)$ such that\footnote{Note that our customized normalization differs from the one used in~\cite{CGSS05}.} $e_p(P) = 314/315\approx 0.9968$
and this value is
maximal possible over all permutation matrices of this size. 
This result approximates the solution of the Euler's problem in the best possible (classical) way.

\subsection{Permutation Matrix $P$ of Order $36$}

Matrix $P$ is given implicitly in the form
of two Latin squares, $L_1$ and $L_2$, which are ``almost'' orthogonal.
When superimposed, they form the following array $\widetilde{P}$ (we keep the original notation),
\begin{align}
\underbrace{\left[\begin{array}{cccccc}
1 & 2 & 3 & 4 & 5 & 6\\
2 & 1 & 4 & 3 & 6 & 5\\
3 & 4 & 6 & 5 & 1 & 2\\
4 & 3 & 5 & 6 & 2 & 1\\
5 & 6 & 2 & 1 & 4 & 3\\
6 & 5 & 1 & 2 & 3 & 4\\
\end{array}\right]}_{=\,L_1}\
&\cup \
\underbrace{\left[\begin{array}{cccccc}
1 & 2 & 3 & 4 & 5 & 6\\
3 & 4 & 5 & 6 & 1 & 2\\
2 & 1 & 4 & 3 & 6 & 5\\
6 & 5 & 1 & 2 & 4 & 3\\
4 & 3 & 6 & 5 & 2 & 1\\
5 & 6 & 2 & 1 & 3 & 4\\
\end{array}\right]}_{=\,L_2}=\nonumber
\\
&=
\left[\begin{array}{cccccc}
11 & 22 & {\bf 33} & {\bf 44} & 55 & 66\\
24 & 14 & 45 & 36 & 61 & 52\\
32 & 41 & 64 & 53 & 16 & 25\\
46 & 35 & 51 & 62 & 24 & 13\\
54 & 63 & 26 & 15 & 42 & 31\\
65 & 56 & 12 & 21 & {\bf 33} & {\bf 44}
\end{array}\right]=\widetilde{P}.\label{P_tilde}
\end{align}

Array in~\eqref{P_tilde} encodes aforementioned permutation matrix $P\in\mathbb{P}(36)$
in the following way. Every ordered pair $jk\in\widetilde{P}$ can be viewed as a $6\times 6$ matrix
$|j\rangle\langle k|$ and forms an appropriate block of $P$.
In other words, if $P$ is seen as a block matrix with $36$ blocks
of size $6\times 6$, then
\begin{equation}
P=\sum_{a,b=1}^6|aj\rangle\langle bk|,
\end{equation}
where $j$ and $k$ correspond to a proper element of $\widetilde{P}$, i.e. $jk=\widetilde{P}_{ab}$
(note that $jk$ is not a product but a string of symbols).
Apparent conflicts in two pairs of blocks, ${\bf 33}$ and ${\bf 44}$
prevent $\widetilde{P}$ from being a six-dimensional OLS. Note also that
the symbols $34$ and $43$ do not appear in the square.

\begin{figure}[ht]
\center
\includegraphics[width=2.6in]{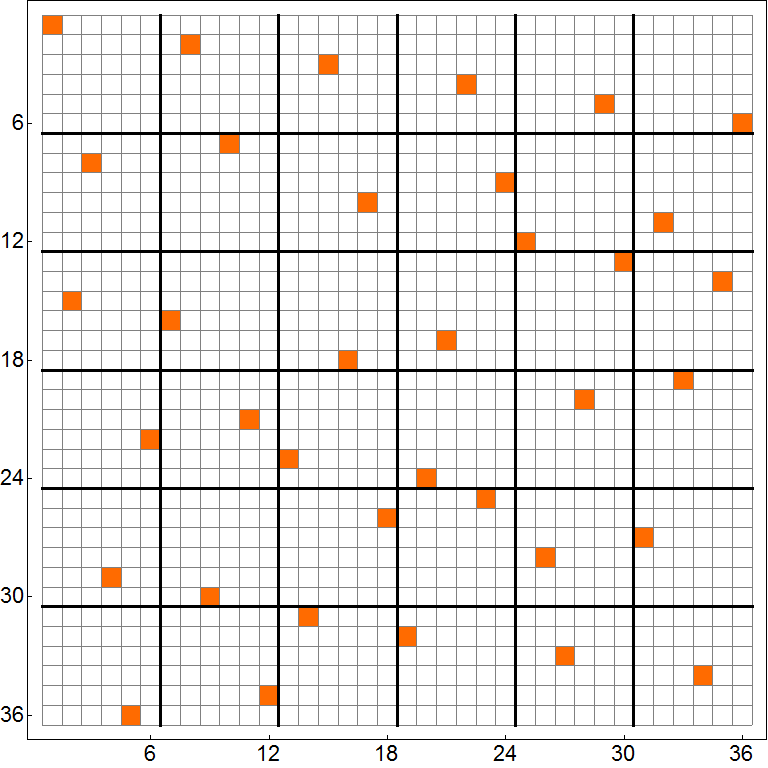}
\caption{Permutation matrix $P$ of order $36$ associated with two ``almost'' orthogonal Latin squares~\eqref{P_tilde} that maximizes
entangling power in the set $\mathbb{P}(36)$. Blank squares represent zeros, while filled squares represent ones.}
\label{fig:P_tilde}
\end{figure}

Matrix $P$ in the decoded form is shown in Figure~\ref{fig:P_tilde}.
It has additional property as it lies on the lower edge of the
$(e_p,g_t)$-plane, so both $P$ and $P^{\rm\Gamma}$ are permutations too.
Explicit vectorized, alas unhandy, form of permutation matrix $P$ reads
\begin{equation}
\begin{split}
P = [1, 15, 8, 29, 36, 22, 16, 2, 30, 7, 21, 35,23, 31, 3, 18, 10, 26,\\
 32, 24, 17, 4, 25, 9, 12, 28, 33, 20, 5, 13, 27, 11, 19, 34, 14, 6],
\end{split}
\end{equation}
which should be understood like~\eqref{PM_as_vector}.

Authors conclude the paper~\cite{CGSS05} with a sequence of open questions. One of them
concerns the existence of unitary matrix for which entangling power would exceed the value $e_p(P)$.
Hence, the next natural step is to ask about extension of the domain of searching over
orthogonal or unitary matrices.
We will follow the gradual path: $\mathbb{P}(36) \to\mathbb{O}(36) \to\mathbb{C}^{36\times 36} \to\mathbb{U}(36)$
with one unnatural deviation in the penultimate step.
Meanwhile, let us start with the real case, and take permutation $P$ as the starting point for our research.

\subsection{Orthogonal Matrix $Q$ of Order $36$}

Two pairs of blocks, $|3\rangle\langle 3|$ and $|4\rangle\langle 4|$, in $\widetilde{P}$ are in conflict
of (Latin) orthogonality.
This observation might tentatively suggest a sort hidden functional dependence that
has been temporarily suppressed when projected onto the meager set of permutations.
As we already noticed, sometimes such obstacles turn to be resolvable
when the problem is embedded into a more general environment.
One obvious thing that comes for consideration is a notion
of entanglement between particular elements of the matrix $P$, or even better, of $\widetilde{P}$,
where it would be easier to visualize inseparability in the form of combinations of particular
elements from $\widetilde{P}$.
However, at this stage it is nothing but pure intuition.

Let us then introduce additional structures in $P$ to leave
permutations and immerse into the set of orthogonal matrices, to check whether
this can somehow improve the value of entangling power.
To this end, take two problematic blocks, ${\bf 33}$ and ${\bf 44}$ in $P$ and
extend them in the following way (allowing entangled states):
\begin{align}
{\bf 33}&\leftrightarrow |3\rangle\langle 3|\to|3\rangle\langle 3|\cos \omega_1-|4\rangle\langle 3|\sin \omega_1,\\
{\bf 44}&\leftrightarrow |3\rangle\langle 4|\to|3\rangle\langle 4|\sin \omega_1+|4\rangle\langle 4|\cos \omega_1,
\end{align}
so that each unity is overridden by $\cos \omega_1$.
Then, we complete appropriate spaces with $\mp\sin\omega_1$ to preserve global orthogonality.
Now, the fragment of the first block-row (first six rows) of $P$ looks like ($\bullet=0$)
\begin{equation}
\cdots \begin{array}{|rr>{\columncolor{gray!10}}rrrr|rr>{\columncolor{gray!10}}rrrr|}
\hline
\bullet  & \bullet  & \bullet & \bullet  & \bullet  & \bullet  & \bullet  & \bullet  & \bullet  & \bullet  & \bullet  & \bullet \\
\bullet  & \bullet  & \bullet & \bullet  & \bullet  & \bullet  & \bullet  & \bullet  & \bullet  & \bullet  & \bullet  & \bullet \\
\rowcolor{gray!10}
\bullet & \bullet & \cos \omega_1 & \bullet & \bullet & \bullet & \bullet & \bullet & -\sin \omega_1 & \bullet & \bullet & \bullet\\
\rowcolor{gray!10}
\bullet & \bullet & \sin \omega_1 & \bullet & \bullet & \bullet & \bullet & \bullet & \cos \omega_1 & \bullet & \bullet & \bullet\\
\bullet  & \bullet  & \bullet & \bullet  & \bullet  & \bullet  & \bullet  & \bullet  & \bullet  & \bullet  & \bullet  & \bullet \\
\bullet  & \bullet  & \bullet & \bullet  & \bullet  & \bullet  & \bullet  & \bullet  & \bullet  & \bullet  & \bullet  & \bullet \\
\hline
\end{array}
\cdots
\end{equation}
If we repeat this replacement for two lower ${\bf 33}$ and ${\bf 44}$ blocks,
matrix $P$ becomes\footnote{$Q$ is the next unused character in the Latin alphabet. Next in
the line are $V$ and $W$, as $R$, $S$, $T$ and $U$ are reserved for
rotation, swap, target and unitary matrices, respectively.} $Q$
\begin{equation}
P\to Q=Q(\omega_1,\omega_2)\in\mathbb{O}(36),
\end{equation}
which is orthogonal and depends on two arbitrary phases $\omega_j\in[0,2\pi)$.

This procedure can be rephrased formally. We start with a {\sl template} matrix $Q_0$ of the form
\begin{equation}
Q_0(\omega_1,\omega_2)=\mathbb{I}_2\oplus R(\omega_1)\oplus\mathbb{I}_{28}\oplus R(\omega_2)\oplus\mathbb{I}_2,\label{template_Q0}
\end{equation}
where
$$
R(\omega_j)=\left[\begin{array}{rr}
\cos \omega_j & -\sin \omega_j\\
\sin \omega_j &  \cos \omega_j 
\end{array}
\right]
$$
is an ordinary two-dimensional rotation matrix,
and we apply the permutation matrix $P$ from the previous section
to $Q_0(\omega_1,\omega_2)$ obtaining
\begin{equation}
Q(\omega_1,\omega_2) = Q_0(\omega_1,\omega_2)P.
\end{equation}
One can readily validate the correctness of this construction.
The structure of the block matrix $Q_0$ corresponds directly to the form we wish to
impose on $P$. Here, two orthogonal blocks $R(\omega_1)$ and $R(\omega_2)$
are going to overlap colliding unities in two pairs of blocks in $P$.

\begin{figure}[ht]
\center
\includegraphics[width=5in]{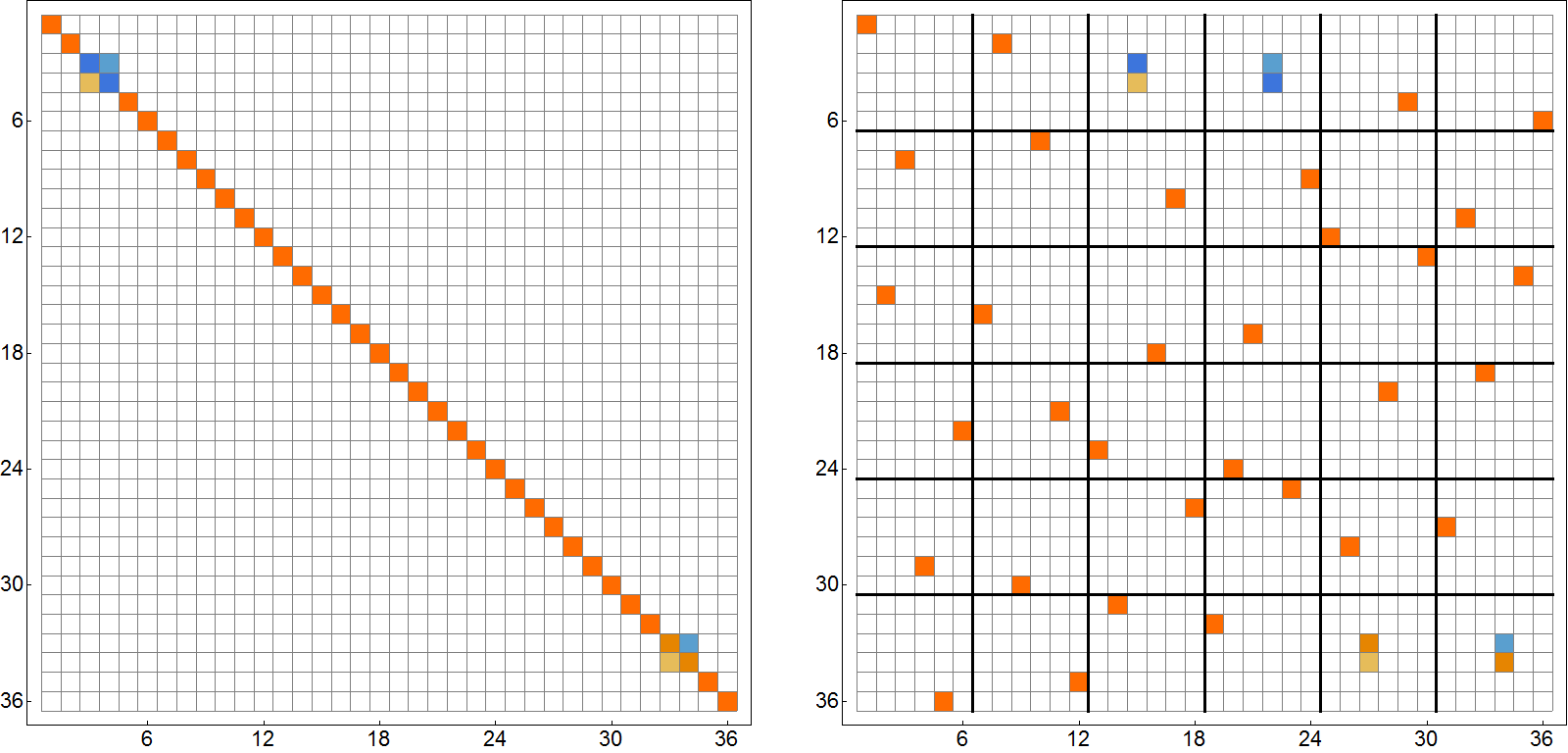}
\caption{Template $Q^{\star}_0$~\eqref{template_Q0} (left panel) and its final form $Q^{\star}$ (right panel) after optimization procedure.
Note the difference with respect to Figure~\ref{fig:P_tilde}, as the conflict in entries ${\bf 33}$ and ${\bf 44}$ is partially
cured by allowing entangled states.}
\label{fig:matrix_Q0_Q}
\end{figure}

Matrix $Q$ can be immediately optimized over all possible phases to maximize the value
of the entangling power $e_p(Q(\omega_1,\omega_2))$. It turns out that for
\begin{equation}
\omega^{\star}=(\omega^{\star}_1,\omega^{\star}_2)=\pi\left({\bf \frac{5}{6}},\frac{1}{6}\right)
\end{equation}
we observe slightly increasing trend
\begin{equation}
0.9968=e_p(P)<e_p\left(Q^{\star}\right)=\frac{419}{420}\approx 0.9976.
\end{equation}
We will denote the best representation of $Q$ by $Q^{\star}$ for some $\omega^{\star}$.\footnote{Symbol of star $^\star$ is not a complex conjugate $^*$.}
The reason the first phase is in bold will be explained below.
Matrices $Q^{\star}_0$ and $Q^{\star}$ of order $36$ are shown in Figure~\ref{fig:matrix_Q0_Q}.

This is the first minor achievement on the way to AME$(4,6)$
that, at least, answers a one 15-years-old
question from~\cite{CGSS05} in the affirmative way by means of
quite primitive machinery. But, it is possible to get even more.

Instead of probing two particular blocks, one can do this for a different pair,
or even, for any sequence of different blocks.
The problem is, which setup to choose to start with, because clearly
the scale of the problem prevents any sequential work, to go through
every possible initial configuration, in any reasonable time.
To omit this problem, we shall try to attack this issue globally.

\subsection{Orthogonal Matrix $V$ of Order $36$}
\label{sec:matrix_V}

Let us prepare an orthogonal block-diagonal matrix
\begin{equation}
V_0\left(\omega\right)=\bigoplus_{j=1}^{18}R(\omega_j)\label{template_V0},
\end{equation}
where $\omega=(\omega_1,\dots,\omega_{18})$ denotes the vector of phases.
And then, as in the case of $Q$, we put
\begin{equation}
V= V(\omega)=V_0(\omega) P.
\end{equation}

\begin{figure}[ht]
\center
\includegraphics[width=5in]{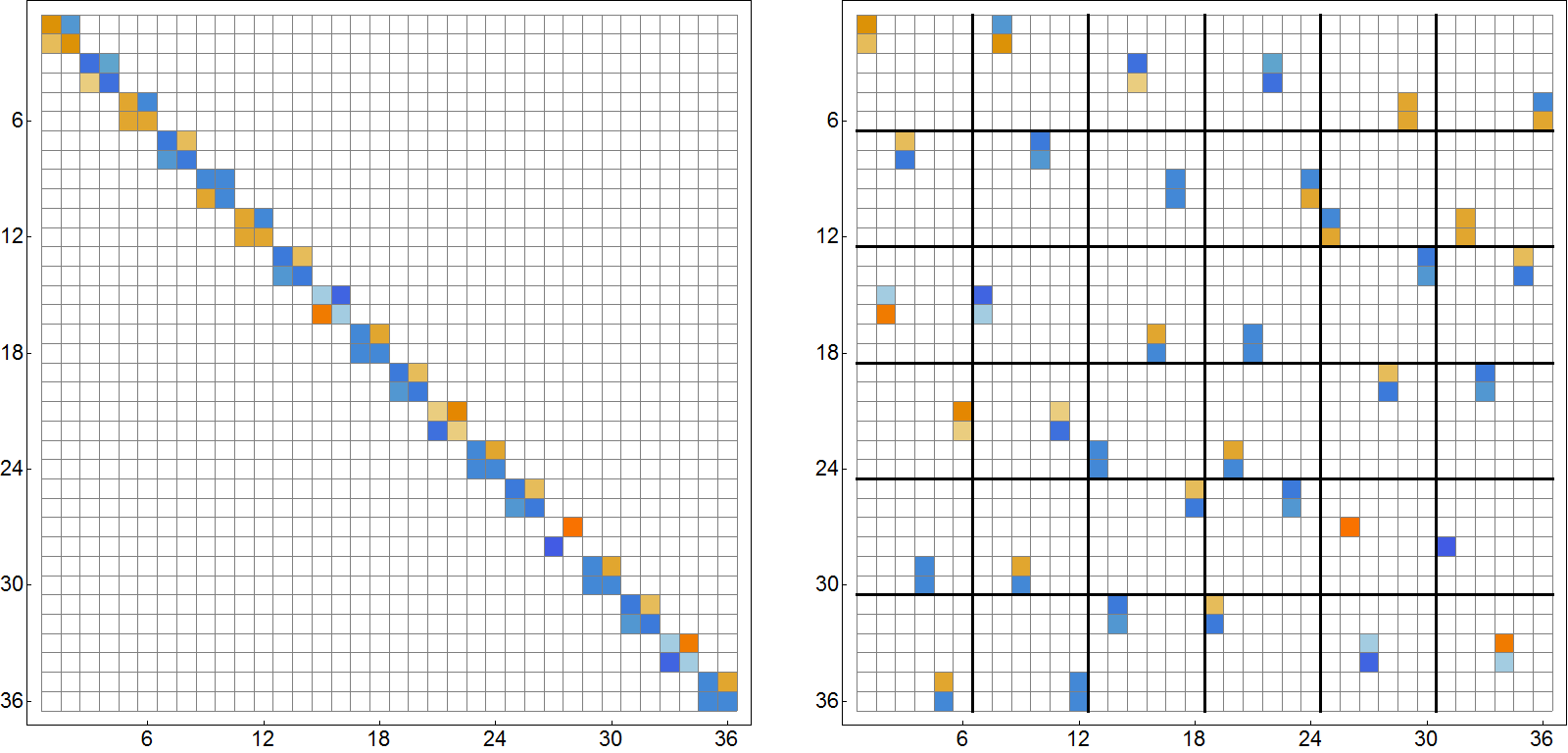}
\caption{Block diagonal template $V^{\star}_0$~\eqref{template_V0} (left panel) and its final form $V^{\star}$ (right panel).}
\label{fig:matrix_V0_V}
\end{figure}

Such construction obviously overlaps all direct regions containing unities in $P$. By introducing $18$ orthogonal
variables we hope to overstep the current value of $e_p(Q^{\star})$.
Indeed, it is possible to improve previous results and to find unitary matrix $V^{\star}$ such that
\begin{equation}
0.9976=e_p(Q^{\star})<e_p(V^{\star})=\frac{208+\sqrt{3}}{210}\approx 0.9987
\end{equation}
for some phases $\omega^{\star}$, which will be provided below.
Additionally we have $g_t(V^{\star})=1$ so this matrix lies exactly on the bisector of $(e_p,g_t)$-plane.
The structure of matrices $V^{\star}_0$ and $V^{\star}$ is presented in Figure~\ref{fig:matrix_V0_V}.

\begin{figure}[ht!]
\center
\includegraphics[width=5in]{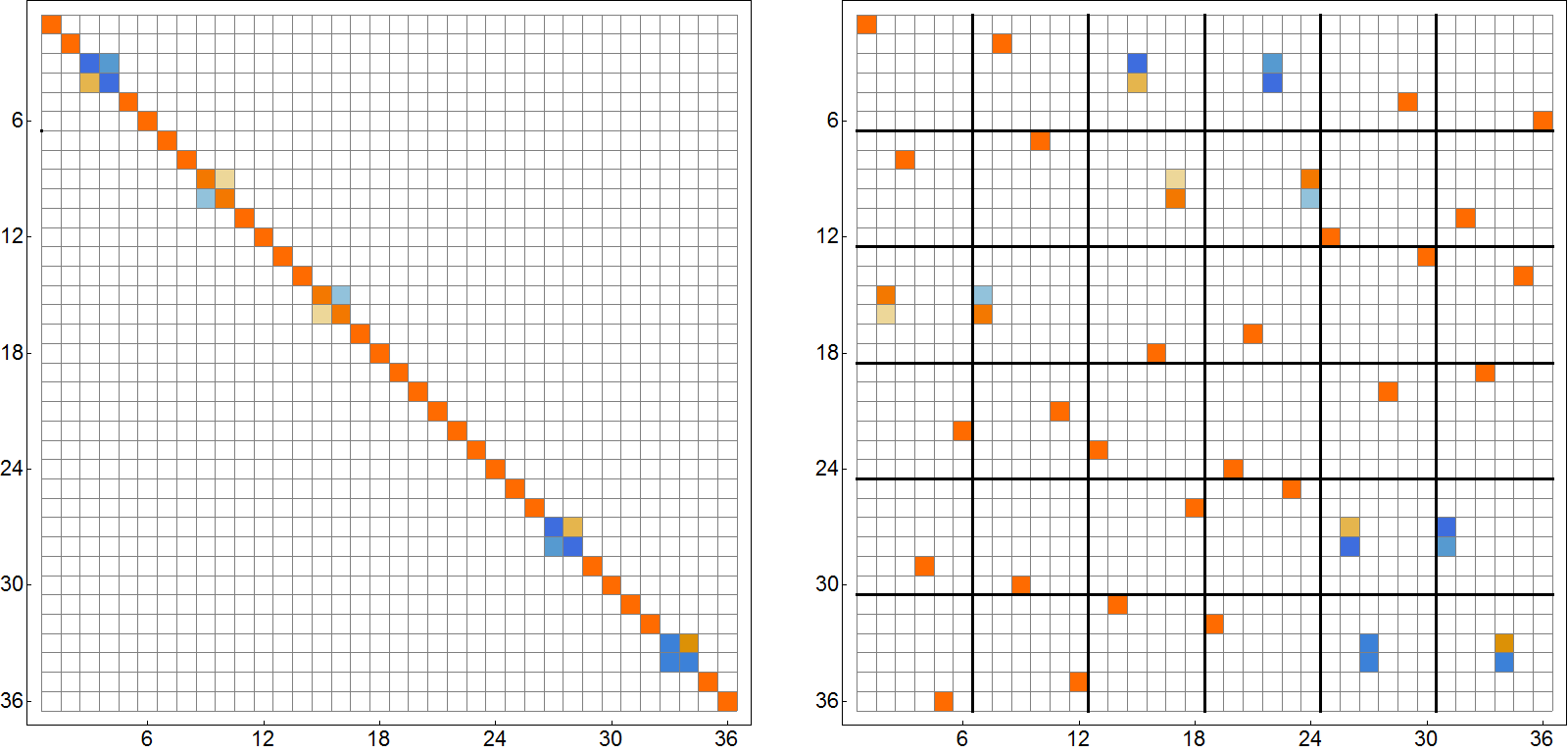}
\caption{Template $W_0$~\eqref{W_reduced} (left panel) and its final form $W$ (right panel), which represents $V^{\star}$ reduced to only $5$ nontrivial phases.}
\label{fig:matrix_W0_W}
\end{figure}

%\medskip

Let us elaborate on this result. After all, actually almost identical random walk routine
over multidimensional real space has already been described in Chapter~\ref{chap:CHM}.
The distinction is that here we optimize (maximize) slightly different target
function $\mathcal{Z}(M)\propto e_p(M)$.
Regardless of the initial choice of phases $\omega\in[0,2\pi)^{\times 18}$,
such optimization procedure is almost always convergent
to a configuration, which is usually much better than previously
considered matrices.
It very rarely gets stuck on $V(\omega_{\rm stuck})$
such that $e_p\left(V(\omega_{\rm stuck})\right)\leqslant e_p(Q^{\star})$
for some set of parameters~$\omega_{\rm stuck}$.

The space of phases is quite large and this allows us to
fix some particular values of $\omega$ to recover all other
components analytically. One particular example of $\omega^{\star}$, out of continuum possibilities,
that maximizes entangling power, is
\begin{equation}
\omega^{\star}=\pi\left(\frac{1}{5}, {\bf \frac{5}{6}}, \frac{1}{4}, \frac{6}{5}, \frac{3}{4}, \frac{1}{4}, \frac{6}{5}, \frac{7}{12}, \frac{5}{4}, \frac{6}{5}, \frac{5}{3}, \frac{5}{4}, \frac{6}{5}, \frac{3}{2}, \frac{5}{4}, \frac{6}{5}, \frac{17}{12}, \frac{5}{4}\right).\label{p_star}
\end{equation}
Note, that for simplicity, in $Q^{\star}$, here, and in the next example, the second $2\times 2$ block in the template matrix has been
fixed to $R(5\pi/6)$.

In particular, one can reduce the set of non-zero elements in $\omega$ not decreasing the
value of $e_p$. This leads to the following simplified form of $V$ called $W$,
\begin{equation}
W=W_0\left(\pi\left({\bf \frac{5}{6}}, \frac{23}{12}, \frac{1}{12}, 0, \frac{7}{6}, \frac{5}{4}\right)\right)P,
\end{equation}
where
\begin{equation}
W_0(\omega_1,\dots,\omega_6)=\bigoplus_{j=1}^6\big(\mathbb{I}_2\oplus R(\omega_j)\oplus\mathbb{I}_{2}\big)\label{W_reduced}
\end{equation}
so that $e_p\left(W\right)=e_p\left(V^{\star}\right)$ and effectively, we have only $5$ degrees of freedom.
It is not known, whether it is possible to further shrink the number of phases down,
or to reduce the number of nonvanishing entries of $V$ in any other way,
keeping the value of entangling power unchanged.
Matrices $W_0$ and $W$ are drawn in Figure~\ref{fig:matrix_W0_W} and from now on we will refer to $W$ (instead of $V^{\star}$) as
the best orthogonal approximation of $T_{36}$.
%we will be using $V^{\star}$ and $W$ interchangeably throughout this chapter.

\begin{figure}[hbp!]
\center
\includegraphics[width=3.9in]{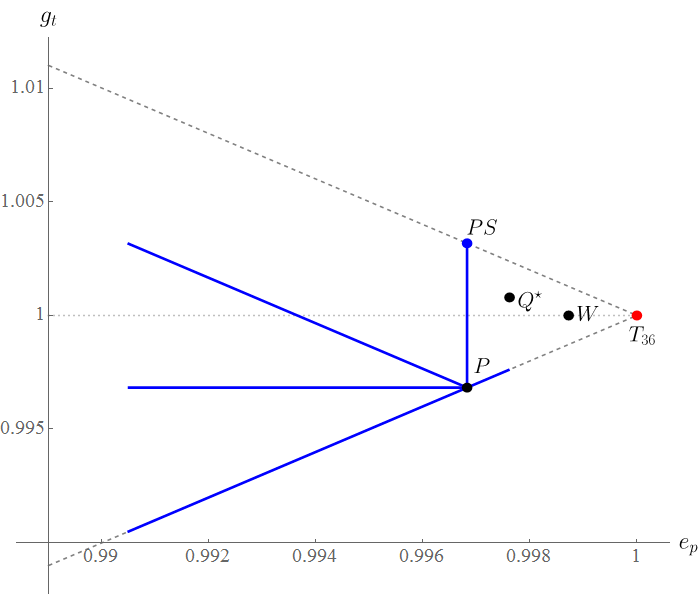}
\caption{Unitary matrices of order $36$ projected onto the $(e_p,g_t)$-plane.
Four simple trajectories starting at $P$, obtained by means of $V(\omega)$ for different
parameterizations of the vector of phases $\omega$ in Eq.~\eqref{W_reduced}, are drawn.
In particular, we can thoroughly describe the families
lying on the edges of the triangle, however, no trajectory can
get close enough to the corner of the triangle representing the desired $2$-unitary matrix $T_{36}$.}
\label{fig:trajectories_general}
\end{figure}

\begin{figure}[ht!]
\center
\includegraphics[width=3.9in]{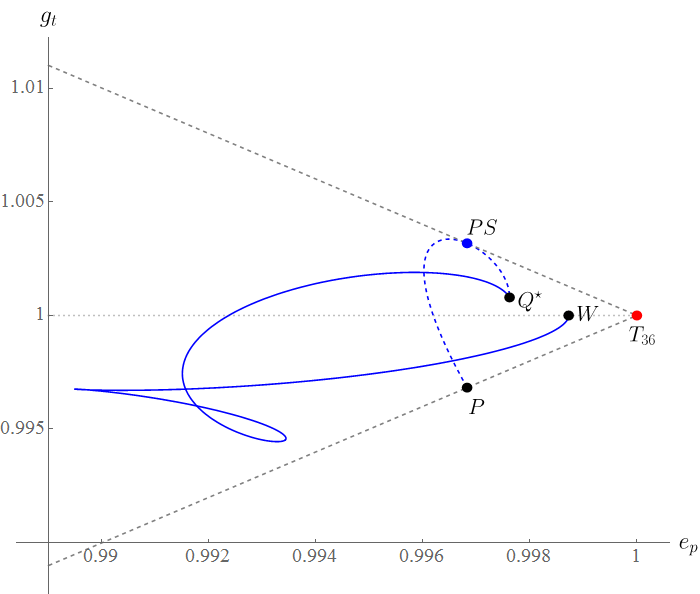}
\caption{Example of two trajectories~\eqref{trajectory_PQ} and~\eqref{trajectory_QW} connecting $P$ with $Q^{\star}$ (dashed line) and $Q^{\star}$ with $W$ (solid line). 
Note that the first trajectory crosses $PS$ on the upper edge before it gets to $Q^{\star}$. Both trajectories representing orthogonal matrices
are relatively far from the target $2$-unitary matrix $T_{36}$.}
\label{fig:trajectories}
\end{figure}

Such a wealth of phases allows us to penetrate the $(e_p,g_t)$-plane in many previously unknown directions.
In some sense, the family of $V$ includes $Q$, $P$ and other interesting objects, for instance
those laying on the edges of the triangle plotted in Figure~\ref{fig:trajectories_general}.
Knowing exact phases, we can easily interpolate between $P$, $Q$ and $W$.
For example, using the pa\-ra\-meterization
\begin{equation}
\omega(t)=\frac{\pi}{6} t\left(5, 1\right) \quad : \quad t\in[0,1]\label{trajectory_PQ}
\end{equation}
applied to $Q_0$ we can smoothly travel from $P$ to $Q^{\star}$,
while
\begin{equation}
\omega(t)=\frac{\pi}{12} \left( 10, 23 t, t, 0, 14 t, 13 t +2\right) \quad : \quad t\in[0,1]\label{trajectory_QW}
\end{equation}
applied to $W_0$ connects $Q^{\star}$ and $W$ (or $V^{\star}$), see Figure~\ref{fig:trajectories}.

\begin{figure}[ht!]
\center
\includegraphics[width=4.5in]{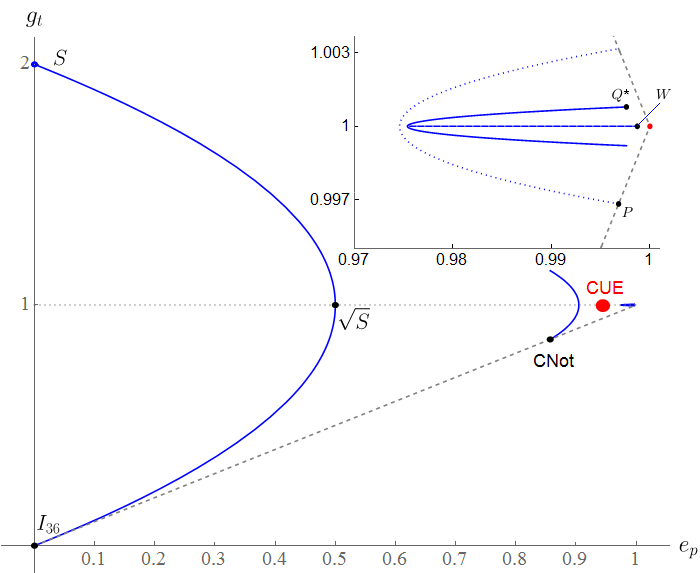}
\caption{
The $(e_p,g_t)$-plane for unitary matrices of order $36$.
Orbits of identity matrix and representation of the CNot gate
under the powers of swap~\eqref{swap_map} exhibit attractive character towards the CUE area.
This is universal observation also for other randomly taken matrices (not plotted here for clarity).
In the magnified vicinity of the corner around $(1,1)$
one recognizes similar behavior for $P$ (dotted), $Q^{\star}$ (solid) and $W$ (dashed line) matrices. Upper
edge of $(e_p,g_t)$-plane and $T_{36}$ are not drawn.}
\label{fig:CUE_attractor}
\end{figure}

Another curious feature of all matrices from the family of $V$
is an attractive character towards {\rm CUE}\footnote{By CUE ({\sl circulant unitary ensemble}) we mean a class of random
unitary matrices
drawn according to the Haar measure on the unitary group $\mathbb{U}(36)$, for which
the average values of $e_p$ and $g_t$ read $\displaystyle{\big\langle(e_p,g_t)\big\rangle=\left(\frac{d^2-1}{d^2+1},1\right)\approx (0.9459,1)}$ in case of $d=6$.}
under the action of swap, which is represented by the following map,
\begin{equation}
s:\mathbb{U}(36)\ni U \mapsto US^t\in\mathbb{U}(36) \quad \text{for} \quad t\in[0,1].\label{swap_map}
\end{equation}
This seems to be a general attribute of any unitary matrix
since no matter what we take as a starting point, we always observe either parabolas
bending to the center of the $(e_p,g_t)$-plane, or straight lines, when we start from
points of $g_t\equiv 1$, see~Figure~\ref{fig:CUE_attractor}.
Such behavior, which makes CUE to be a kind of an attractor, is very suggestive
and might lead one to a conclusion that a hypothetical solution $T_{36}$, if exists, must
be extremely isolated and it differs from a generic form of an orthogonal matrix significantly.

Next generalization of the permutation matrix $P$ consist of gradually increasing the size of the blocks used to overlap
unities in $P$.
We have already checked the case of eighteen $2\times 2$ orthogonal blocks spread
over an $36\times 36$ array according to the permutation matrix~$P$. Factorization the number $36$ provides
$9$ such possibilities. We start with $36$ blocks of size $1\times 1$, through
$18$ blocks of size $2\times 2$, then $12$ blocks of size $3\times 3$, ... up to a single
block of size $36\times 36$.
In each case we can apply optimization procedure to maximize the entangling power.
The results are collected in Table~\ref{tab:blocks_ep}.
\begin{table}[ht!]
\center
\begin{tabular}{c|c|l}
$\#$\text{blocks} & \text{block size} & $\max\{e_p\}$\\
\hline
%36 & 1\times 1 & = e_p(P_{36})\\
$18$ & $2\times 2$ & = $e_p(W)$\\
$12$ & $3\times 3$ & = $e_p(Q^{\star})$\\
 $9$ & $4\times 4$ & = $e_p(W)$\\
 $6$ & $6\times 6$ & = $e_p(W)$\\
 $4$ & $9\times 9$ & = $e_p(W)$\\
 $3$ &$12\times 12$& = $e_p(W)$\\
 $2$ &$18\times 18$& $\ll e_p(Q^{\star})$\\
 $1$ &$36\times 36$& $\ll e_p(Q^{\star})$
\end{tabular}
\caption{Numerical simulations using block structures of different size introduced in the permutation matrix $P$.
In each case the highest value of entangling power cannot exceed $e_p(W)$, which might suggest that $W$ is the global maximum.}
\label{tab:blocks_ep}
\end{table}

In the first case $36\times(1\times 1)$ (not included in Table~\ref{tab:blocks_ep}), we cannot get
a matrix with a greater entangling power, just by definition of $P$.
Only in the configuration of $18$ blocks of size two we used parameterization of blocks in the form of
rotation matrix. In other cases, we mainly used generic orthogonal matrices obtained
from polar decomposition procedure. Some attempts with analytical parameterization
of orthogonal matrices of order $3$ and $4$ did not bring any better result.
In the last two cases, it was even impossible to recover the best values because the space accessible for a random searching was too large.
In general, despite many different attempts and putting much computational effort, we were
unable to find anything better than $W$.
Interestingly, in the case $9\times(4\times 4)$ we observe as optimal $4\times 4$ blocks
almost always simplify to the form 
$
R(\alpha_j)\oplus R(\beta_k)
$
for some $\alpha_j, \beta_k\in[0,2\pi)$,
which reduces this case to the ``best'' one of $18\times(2\times 2)$. This gently suggests,
that $W$ might be the global optimum.

Permutation matrix $P$, as a base for further calculations,
has been chosen unquestionably because of its distinguished role in the set of all
permutations matrices. Nevertheless, one can naturally ask
about picking something different than $P$.
We checked ``a few'' other permutations out of
\begin{equation}
36! = 371\,993\,326\,789\,901\,217\,467\,999\,448\,150\,835\,200\,000\,000
\end{equation}
--- the total number of possibilities, and
in no case any matrix with better $e_p$ signature
was found. Different starting points did not lead to any better findings than those
represented by $Q^{\star}$ or $W$.
Neither, no better results are obtained
when we temporarily switch to complex domain,
replacing orthogonal rotations $R(\omega_j)$ by unitary counterparts
or similar complex structures. % in higher dimensions.
This last remark is exceptionally puzzling, particularly
in the context of the results described in Section~\ref{sec:AME_map}!

Exhaustive numerical and theoretical research did not lead to anything beyond $W$ in
the class of orthogonal matrices.
Instead, matrices $P$, $Q^{\star}$, $W$ and others have been thoroughly examined
in the matter of being local extrema of $e_p$,
by Grzegorz Rajchel-Mieldzio\'{c} --- the coauthor of~$[\hyperlink{\paperslist}{\rm A4}]$.
Grzegorz also performed detailed analysis of the reduced family $W$, which was roughly presented
in Figure~\ref{fig:trajectories_general}.
Please refer to his thesis\footnote{Under preparation.} for more interesting insights and details concerning
analytical description of the $(e_p,g_t)$-plane.

\newpage

We briefly conclude this section with the following chain of inequalities
\begin{equation}
0<\underbrace{e_p\left(P\right)}_{\approx 0.9968} < \underbrace{e_p\left(Q^{\star}\right)}_{\approx 0.9976} < \underbrace{e_p\left(W\right)}_{\approx 0.9987} < 1
\end{equation}
that significantly improves the state of knowledge about the set of orthogonal matrices $\mathbb{O}(36)$
in the context of entangling power and approximation of AME$(4,6)$ state.
In view of the numerical evidence, we tend to believe that a real
AME$(4,6)$ state does not exist. And the matrix $W$ is, so far, the best approximation of such an object.
In the next section we present another prospective idea,
which might help to support the fact of nonexistence of a real AME$(4,6)$ state.

\section{Isoentropic Triplets of Matrices}
\label{sec:iso_triplets}
So far, we have proved that the set of orthogonal matrices can be investigated
in new directions, which provided much better metrics for entangling power, getting us closer
to the hypothetical target $T_{36}$ at the cusp in the $(e_p,g_t)$-plane.

Let us look at the vectors of singular values corresponding to matrices $P$, $Q^{\star}$ and $W$
and how they progressively flatten.
We have, for $P$
\begin{align}
{\rm svd}(P)={\rm svd}\big(P^{\rm\Gamma}\big)&=\left[\{1\}^{36}\right]^{\rm T},\label{flat_svd}\\
{\rm svd}\big(P^{\rm R}\big)&=\left[\{\sqrt{2}\}^2,\{1\}^{34}\right]^{\rm T},
\end{align}
for $Q^{\star}$
\begin{align}
{\rm svd}(Q^{\star})&=\left[\{1\}^{36}\right]^{\rm T},\\
{\rm svd}\big((Q^{\star})^{\rm R}\big)&=\left[\Big\{\sqrt{\frac{3}{2}}\Big\}^{2},\{1\}^{32},\Big\{\frac{1}{\sqrt{2}}\Big\}^{2}\right]^{\rm T},\\
{\rm svd}\big(Q^{\star})^{\rm\Gamma}\big)&=\left[\Big\{\sqrt{\frac{3}{2}}\Big\}^{4},\{1\}^{28},\Big\{\frac{1}{\sqrt{2}}\Big\}^{4}\right]^{\rm T},
\end{align}
and for $W$ % V^*
\begin{align}
{\rm svd}(W)&=\left[\{1\}^{36}\right]^{\rm T},\\
{\rm svd}\big(W^{\rm R}\big)&={\rm svd}\big(W^{\rm\Gamma}\big)=\\
%&=\left[\Bigg\{\sqrt{\frac{2+\sqrt{2-\sqrt{3}}}{2}}\Bigg\}^{6},\{1\}^{20},\Bigg\{\sqrt{\frac{2-\sqrt{2-\sqrt{3}}}{2}}\Bigg\}^{6}\right]^{\rm T},
&=\left[\Bigg\{\frac{1}{2}\sqrt{4-\sqrt{2}+\sqrt{6}}\Bigg\}^{6},\{1\}^{20},\Bigg\{\frac{1}{2}\sqrt{4+\sqrt{2}-\sqrt{6}}\Bigg\}^{6}\right]^{\rm T},
\end{align}
where symbol $\{x\}^m=\overbrace{x,...,x}^{\times m}$ denotes $m$ repetitions of the number $x$.
We observe as the number of non-unity components gradually grows
but their values get equalized and
obviously on the way to $T_{36}$ these vectors should be all ones also for $^{\rm R}$ and $^{\rm\Gamma}$ realignments.
To see this better, let us look at the suitably presented numerical values:
\begin{align*}
&\Big[\{1.4142\}^2, \{1\}^{34}\Big]^{\rm T}\\
&\Big[\{1.2247\}^2, \{1\}^{32}, \{0.7071\}^2\Big]^{\rm T}\\
&\Big[\{1.2247\}^4, \{1\}^{28}, \{0.7071\}^4\Big]^{\rm T}\\
&\Big[\{1.1219\}^6, \{1\}^{20}, \{0.8609\}^6\Big]^{\rm T}\\
&\ \ \ \downarrow\\
&\Big[\{1\mp\varepsilon\}^{36}\Big]^{\rm T} \ \text{for} \ \varepsilon\text{-neighborhood of} \ T_{36},
\end{align*}
where the last entry makes sense (for $\varepsilon\ll 1$) provided that $T_{36}$ is not a topologically isolated point.

All previous calculations assumed restriction to the set of orthogonal matrices $\mathbb{O}(36)$.
We can still stay in the real domain but relaxing the constraint of orthogonality.
This should allow us to examine a broader class of generic matrices of size $36$ and possibly find another path to $T_{36}$.
However, even in the set of matrices ``without any structure'', we should focus our attention
on a special subclass. We will consider the set of matrices
\begin{equation}
\mathcal{J}_{36} = \Big\{ M\in\mathbb{R}^{36\times 36} : {\rm svd}(M)={\rm svd}\left(M^{\rm\Gamma}\right)={\rm svd}\left(M^{\rm R}\right) \Big\}\label{isoentropic}
\end{equation}
for which each realignment yields the same singular value decomposition, or equivalently, the same entropy $E$.
A matrix $M\in\mathcal{J}_{36}$ will be called {\sl isoentropic}.

\begin{figure}[ht!]
\center
\includegraphics[width=4.0in]{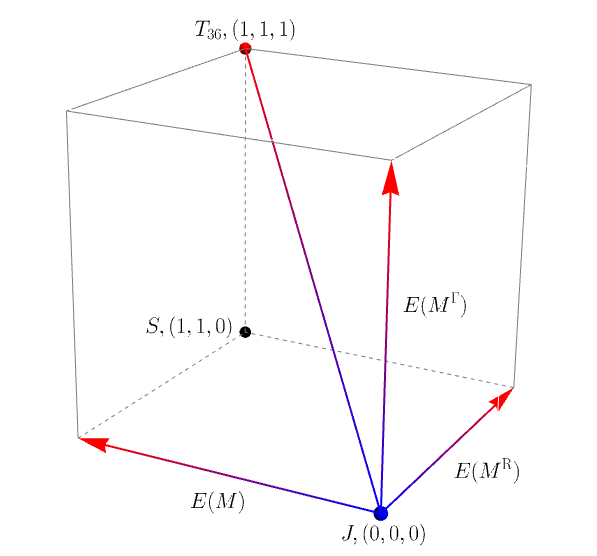}
\caption{Visualization of the $E$-cube with its main diagonal. Two antipodal points of $E$-diagonal are occupied by
$T_{36}$ and the flat matrix $J$. Swap matrix $S$ for which $E(S)=E(S^{\rm R})=1$ and $E(S^{\rm\Gamma})=0$ is located in the $(1,1,0)$ vertex.}
\label{fig:sl_diagonal}
\end{figure}

As we will no longer sample the set $\mathbb{O}(36)$, we cannot use
$e_p$ and $g_t$. Thus we should switch to a general measure in form of
the linear entropy~\eqref{linear_entropy} adapted for $d=6$ to real domain
\begin{equation}
E(M)=\frac{36}{35}\left(1-\frac{{\rm Tr}\big(MM^{\rm T}MM^{\rm T}\big)}{{\rm Tr}^2\big(MM^{\rm T}\big)}\right).\label{SL_restricted}
\end{equation}

One must be careful and notice that if all vectors of singular value of $M$, its partial transpose $M^{\rm \Gamma}$,
and reshuffling $M^{\rm R}$ are
equal then $M$ has equal entropies for each realignment too. But the converse statement does not hold true.

The set $\mathcal{J}_{36}$ is not empty, as for example zero-entropy flat matrix $J_{jk}={\rm const}\in\mathbb{R}\,(\forall\,j,k)$ belongs there.
Soon it will be clear that $\mathcal{J}_{36}$ has a much richer internal structure
with continuum of other nontrivial elements.
We can envision $\mathcal{J}_{36}$ as the main diagonal of the unit cube for which each principal edge
corresponds to $E(M)$, $E\big(M^{\rm R}\big)$ and $E\big(M^{\rm\Gamma}\big)$, see Figure~\ref{fig:sl_diagonal}.
We will call it {\sl $E$-cube} and {\sl $E$-diagonal}, respectively.
So $J$ lies at the beginning of the $E$-diagonal, while the real AME$(4,6)$
representation in the form of $T_{36}$, if exists, would obviously be found
on the other end.

Suppose for a moment a general case and put $d>1$.
In the remaining part of this section we will investigate a simple map in $\mathbb{R}^{d^2\times d^2}$
that transforms any real matrix $M$ into an isoentropic one.
We will check the properties of this map and draw a conjecture concerning the existence of real a AME$(4,6)$ state.

Take any real matrix $X=X_0$ of size $d^2\times d^2$ and
define a recursive sequence
\begin{equation}
X_{k+1}=\frac{1}{2}X_k^{\rm R}+\frac{1}{2}X_k^{\rm\Gamma} \quad \text{for} \quad k\in\mathbb{N}_0.\label{Pi_sequence}
\end{equation}
Suppose the limit of $X_k$ for $k\to\infty$ exists, put
$\displaystyle{Y=\lim_{k\to\infty} X_k}$ and denote the limit operation by $\Xi$,
hence $Y=\Xi(X)$.
\begin{proposition}
$\forall \ d\in\mathbb{N}$ and $\forall \ X\in\mathbb{R}^{d^2\times d^2}$ the limit of the sequence~\eqref{Pi_sequence}
exists and the image of $\,\Xi(X)$ is isoentropic.\label{iso_proposition}
\end{proposition}
We present the proof of this Proposition at the end of this section.
The reasoning is technical and does not bring any additional information to the subject.
Meanwhile, let us examine the map $\Xi$ in details.
Explicit expression for $X_k$ takes the form of
\begin{align}
X_1&=\frac{1}{2}\Big(X_0^{\rm R}+X_0^{\rm\Gamma}\Big),\\
X_2&=\frac{1}{4}\Big(X_0^{\rm RR}+X_0^{\rm R\Gamma}+X_0^{\rm \Gamma R}+X_0^{\rm\Gamma\Gamma}\Big),\\
X_3&=\frac{1}{8}\Big(X_0^{\rm RRR}+X_0^{\rm R\Gamma R}+X_0^{\rm \Gamma RR}+X_0^{\rm \Gamma \Gamma R}+X_0^{\rm R R \Gamma }+X_0^{\rm R\Gamma\Gamma}+X_0^{\rm \Gamma R\Gamma }+X_0^{\rm \Gamma \Gamma \Gamma}\Big),\\
\vdots&\nonumber
\end{align}
In general,
\begin{equation}
X_k=\frac{1}{2^k}\sum_{\alpha_j\in\{\rm R,\Gamma\}^{\times k}}X_0^{\alpha_1\alpha_2\cdots\alpha_k}.\label{Xk_general}
\end{equation}

For simplicity, we informally rewrite operations $X^{\rm R}$ and $X^{\rm\Gamma}$ in the form of operators %(no hats),
${\rm R}X = X^{\rm R}$ and $\Gamma X = X^{\Gamma}$.
Then,~\eqref{Xk_general} can be expressed as the sum of
compositions of ${\rm R}$ and $\Gamma$
\begin{equation}
X_k=\frac{1}{2^k}\left(\sum_{m=1}^{2^k}\prod_{\alpha_{j_m}\in\{{\rm R, \Gamma}\}^{\times k}} \alpha_{j_m}\right)X_0=A_k X_0.\label{RG_composition}
\end{equation}

Operations ${\rm R}$ and $\rm\Gamma$ do not commute in general, however when applied to a given matrix, they can only
produce six different combinations\footnote{Recall six different ways of producing unitary matrix out of the splitting of a fourpartite system.}
\begin{equation}
\Big\{\mathbb{I}_{d^2}, {\rm R}, {\rm\Gamma}, {\rm R\Gamma}, {\rm\Gamma R}, {\rm R\Gamma R}={\rm\Gamma R \Gamma}\Big\}.\label{base_RGamma}
\end{equation}
This is an immediate consequence of definition ${\rm R}$ and ${\rm\Gamma}$ by means of multi-indices, see~\eqref{RG_multiindex}.
All other sequences of ${\rm R}$ and ${\rm\Gamma}$ reproduce one particular element from~\eqref{base_RGamma}.
Thus, for $k\to\infty$, one readily sees that $A_k$ in~\eqref{RG_composition} reduces to
\begin{equation}
A_{k}\to\frac{1}{3}\mathbb{I}_{d^2}+\frac{1}{3}{\rm R\Gamma}+\frac{1}{3}{\rm\Gamma R}\label{Pi_simplified_1}
\end{equation}
or, equivalently,
\begin{equation}
A_k\to\frac{1}{3}{\rm R}+\frac{1}{3}{\rm\Gamma}+\frac{1}{6}{\rm R\Gamma R}+\frac{1}{6}{\rm \Gamma R\Gamma}
=\frac{1}{3}{\rm R}+\frac{1}{3}{\rm\Gamma}+\frac{1}{3}{\rm R\Gamma R}.\label{Pi_simplified_2}
\end{equation}
This provides the final, compact and elegant form for~\eqref{Pi_sequence}
\begin{align}
\Xi(X)=Y&=\frac{1}{3}\left(X+X^{\rm R\Gamma}+X^{\rm\Gamma R}\right) \quad \text{or}\\
Y&=\frac{1}{3}\left(X^{\rm R}+X^{\rm\Gamma}+X^{\rm R\Gamma R}\right),
\end{align}
where $X\in\mathbb{R}^{d^2\times d^2}$. Hence, infinite (albeit quickly convergent) sequence can be
replaced by a simple algebraic formula.
Using relations~\eqref{base_RGamma} and the fact that ${\rm R}$ and ${\rm\Gamma}$ are
involutions, ${\rm R^2=\Gamma^2} = \mathbb{I}_{d^2}$, one can easily infer that $\Xi=\Xi^2$ is
a projection onto $E$-diagonal, which is in full accordance with expectations.

\begin{figure}[ht!]
\center
\includegraphics[width=4.6in]{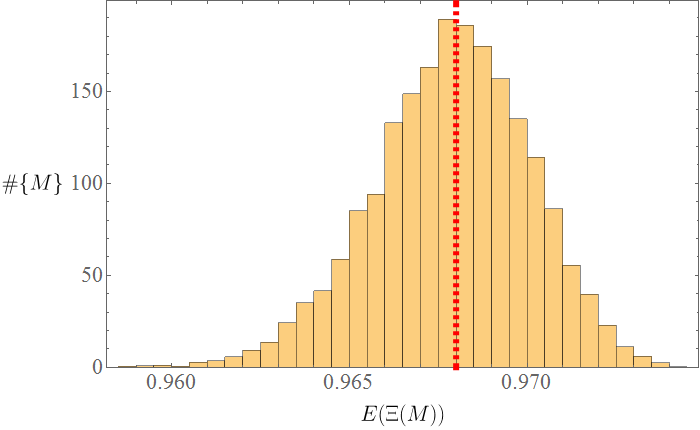}
\caption{Normalized histogram of entropy distribution for $2^{14}$ samples of matrices of order $36$,
drawn at random according
to real normal distribution $M\sim\mathcal{N}(0,1)^{\times 36^2}$ and transformed by $\Xi$ onto $E$-diagonal.
Mean value $\big\langle E(\Xi(M))\big\rangle=\big\langle E\big(\Xi^{\rm R}(M)\big)\big\rangle=\big\langle E\big(\Xi^{\rm\Gamma}(M)\big)\big\rangle$ (red dotted line) concentrates
around $0.9678$ with standard deviation of $0.0022$.}\label{Pi_random_probing}     % std. dev = sqrt{variance}
\end{figure}

Values of $\Xi$ for random arguments $M$ reflect similar behavior
like probing entangling power with generic unitary matrices (CUE) --- they concentrate
quite close to the corner where $T_{36}$ should be, but without any occasion to reach it.
The average value of entropy %(no matter which one we take as they all are equal when $M$ is projected on $E$-diagonal)
oscillates around $0.9678$, see Figure~\ref{Pi_random_probing}.
Slightly better results are obtained when one systematically tries to increase the entropy
performing a random walk over $\mathbb{R}^{36\times 36}$ and maximizing the target function $\mathcal{Z}(M)\propto E(M)$.
Unfortunately, in no case can it reach a direct vicinity of $T_{36}$.
The best output from $\Xi$ subjected to such procedure
is given by a matrix $Z$ for which $E(\Xi(Z))=0.9982$.
Matrix $Z$ has no apparent structure and cannot be presented in any straightforward analytic form.

Additional evidence supporting impossibility of reaching the second verge of $E$-diagonal
can be carried out be means of semi-analytical calculations.
We would like to have $X\in\mathbb{R}^{36\times 36}$ such that $\Xi(X)=T_{36}$.
As it was mentioned, the maximal value of $E$~\eqref{SL_restricted} is attained by orthogonal matrices.
Hence, for $Y=\Xi(X)$ one should check the condition $YY^{\rm T}=\mathbb{I}_{36}$, which translates into
the system of $d^2(d^2+1)/2$ nonlinear equations involving entries of $X$ (or $Y$).
Attempting to solve such system for $d=2$ yields empty set, which confirms the lack of four-qubit AME$(4,2)$ states~\cite{HS00}.
In turn, numerical calculations for $d=3$ provide a solution in compliance with
the situation concerning four-qutrit systems~\cite{GALRZ15}.
For $d=4$ and higher everything complicates rapidly and really smart numerical methods are needed to get a credible solution.
We were only able to roughly estimate the anticipated fact that (perhaps)
\begin{equation}
\big\{YY^{\rm T}=\mathbb{I}_{36}\big\}=\emptyset.
\end{equation}
But this approach still requires more computational power. Needless to say, that
any analytical hack is beyond reach for such complicated expressions.

To exemplify the problem of analyzing the map $\Xi$,
let us write down explicit form of image of $\Xi(X)$ for some $X\in\mathbb{R}^{d^2\times d^2}$.
For $d=2$ it can be presented as
\begin{equation}
\Xi(X)=\frac{1}{3}\left[\begin{array}{rrrr}
      3x_1 & x_2 & x_2 & x_3\\
      x_2 & x_3 & x_3 & 3x_4\\
      3x_5 & x_6 & x_6 & x_7\\
      x_6 & x_7 & x_7 & 3x_8\\
\end{array}\right]
\end{equation}
for $x_j\in\mathbb{R}$.
This implies the rank $r(\Xi(X))\leqslant 3$. Indeed, when assuming that all non-zero singular values
are all equal, ${\rm svd}(\Xi(X))=\left[\{\sigma\}^3,0\right]^{\rm T}$, for some $\sigma\neq 0$, then one has $E(\Xi(X))=8/9$,
which is immediately confirmed numerically.
Here we use alternative (equivalent) formula for entropy~\eqref{linear_entropy}, expressed by singular values
\begin{equation}
E(\sigma_1,...,\sigma_{d^2})=\frac{d^2}{d^2-1}\left(1-\frac{\sum_{j=1}^{d^2}\sigma_j^4}{\left(\sum_{j=1}^{d^2}\sigma_j^2\right)^2}\right),
\end{equation}
which is directly related to the rank of the matrix.

For $d>2$, a generic matrix $\Xi(X)$ is no longer singular, i.e. the rank takes its maximal value, $r(\Xi(X))=d^2$.
When trying to recover any pattern for $d=6$, one finds a great difficulty.
Consider $X\in\mathbb{R}^{36\times 36}$ and its realignments, where only six first, additionally stripped (to save space) rows are shown
\newpage
%{\small
\begin{align}
X&=\left[\begin{array}{cccccc|ccc|cccc}
a_{11} & a_{12} & a_{13} & a_{14} & a_{15} & a_{16}                   & b_{11} & b_{12} & ... & f_{11} & {\bf\underline{f_{12}}} & f_{13} & ... \\
a_{21} & a_{22} & a_{23} & a_{24} & a_{25} & {\bf\underline{a_{26}}}  & b_{21} & b_{22} & ... & f_{21} & f_{22} & f_{23}                  & ... \\
a_{31} & a_{32} & a_{33} & a_{34} & a_{35} & a_{36}                   & b_{31} & b_{32} & ... & f_{31} & f_{32} & f_{33}                  & ... \\
a_{41} & a_{42} & a_{43} & a_{44} & a_{45} & a_{46}                   & b_{41} & b_{42} & ... & f_{41} & f_{42} & f_{43}                  & ... \\
a_{51} & a_{52} & a_{53} & a_{54} & a_{55} & a_{56}                   & b_{51} & b_{52} & ... & f_{51} & f_{52} & f_{53}                  & ... \\
a_{61} & a_{62} & a_{63} & a_{64} & a_{65} & a_{66} & {\bf\underline{b_{61}}}& b_{62} & ... & f_{61} & f_{62} & f_{63}                  & ... \\
\end{array}\right],\label{general_X6_1}\\
%\end{align}
%\begin{align}
X^{\rm R\Gamma}&=\left[\begin{array}{cccccc|ccc|cccc}
a_{11} & b_{11} & c_{11} & d_{11} & e_{11} & f_{11} & a_{21}                  & b_{21} & ... & a_{61} & {\bf\underline{b_{61}}} & c_{61} & ...\\
a_{12} & b_{12} & c_{12} & d_{12} & e_{12} & {\bf\underline{f_{12}}} & a_{22} & b_{22} & ... & a_{62} & b_{62} & c_{62}                  & ...\\
a_{13} & b_{13} & c_{13} & d_{13} & e_{13} & f_{13} & a_{23}                  & b_{23} & ... & a_{63} & b_{63} & c_{63}                  & ...\\
a_{14} & b_{14} & c_{14} & d_{14} & e_{14} & f_{14} & a_{24}                  & b_{24} & ... & a_{64} & b_{64} & c_{64}                  & ...\\
a_{15} & b_{15} & c_{15} & d_{15} & e_{15} & f_{15} & a_{25}                  & b_{25} & ... & a_{65} & b_{65} & c_{65}                  & ...\\
a_{16} & b_{16} & c_{16} & d_{16} & e_{16} & f_{16} & {\bf\underline{a_{26}}} & b_{26} & ... & a_{66} & b_{66} & c_{66}                  & ...\\
\end{array}\right],\label{general_X6_2}\\
%\end{align}
%\begin{align}
X^{\rm\Gamma R}&=\left[\begin{array}{cccccc|ccc|cccc}
a_{11} & a_{21} & a_{31} & a_{41} & a_{51} & a_{61} & a_{12} & a_{22} & ... & a_{16} & {\bf\underline{a_{26}}} & a_{36} & ...\\
b_{11} & b_{21} & b_{31} & b_{41} & b_{51} & {\bf\underline{b_{61}}} & b_{12} & b_{22} & ... & b_{16} & b_{26} & b_{36} & ...\\
c_{11} & c_{21} & c_{31} & c_{41} & c_{51} & c_{61} & c_{12} & c_{22} & ... & c_{16} & c_{26} & c_{36}                  & ...\\
d_{11} & d_{21} & d_{31} & d_{41} & d_{51} & d_{61} & d_{12} & d_{22} & ... & d_{16} & d_{26} & d_{36}                  & ...\\
e_{11} & e_{21} & e_{31} & e_{41} & e_{51} & e_{61} & e_{12} & e_{22} & ... & e_{16} & e_{26} & e_{36}                  & ...\\
f_{11} & f_{21} & f_{31} & f_{41} & f_{51} & f_{61} & {\bf\underline{f_{12}}} & f_{22} & ... & f_{16} & f_{26} & f_{36} & ...\\
\end{array}\right].\label{general_X6_3}
\end{align}
%}

\noindent After summing them up, the explicit form of $Y=\Xi(X)$ for $d=6$ reads (we intentionally omit the factor $1/3$ and change indexing of $y$, hence
one should not be surprised that indices with value $7$ do appear)
%{\scriptsize
\begin{align}
Y&=X+X^{\rm R\Gamma}+X^{\rm\Gamma R}=\nonumber\\
&=\left[\begin{array}{cccccc|ccc|cccccc}
{\bf a_{11}} & y_{02} & y_{03} & y_{04} & y_{05} &                 y_{06}  &                 y_{02} &      y_{07} & ...  & y_{06} & {\bf\underline{y_{27}}} & y_{28} & y_{29} & y_{30} &      y_{31}\\
     y_{02}  & y_{07} & y_{12} & y_{17} & y_{22} & {\bf\underline{y_{27}}} &                 y_{07} & {\bf b_{22}} & ... & y_{11} &                 y_{36}  & y_{49} & y_{50} & y_{51} &      y_{52}\\
     y_{03}  & y_{08} & y_{13} & y_{18} & y_{23} &                 y_{28}  &                 y_{12} &      y_{33} & ...  & y_{16} &                 y_{40}  & y_{56} & y_{63} & y_{64} &      y_{65}\\
     y_{04}  & y_{09} & y_{14} & y_{19} & y_{24} &                 y_{29}  &                 y_{17} &      y_{34} & ...  & y_{21} &                 y_{44}  & y_{59} & y_{68} & y_{71} &      y_{72}\\
     y_{05}  & y_{10} & y_{15} & y_{20} & y_{25} &                 y_{30}  &                 y_{22} &      y_{35} & ...  & y_{26} &                 y_{48}  & y_{62} & y_{70} & y_{74} &      y_{75}\\
     y_{06}  & y_{11} & y_{16} & y_{21} & y_{26} &                 y_{31}  & {\bf\underline{y_{27}}} &     y_{36} & ...  & y_{31} &                 y_{52}  & y_{65} & y_{72} & y_{75} & {\bf f_{66}}\\
\vdots
\end{array}\right].\label{general_Y6}
\end{align}
%}
One should recognize the pattern:
$6$ independent invariants depicted by bold font in~\eqref{general_Y6},
and $70$ independent sums of three entries, e.g. $y_{27}=a_{26}+f_{12}+b_{61}$ in
underlined-bold font in~\eqref{general_X6_1},~\eqref{general_X6_2} and ~\eqref{general_X6_3}.
After including remaining rows (not drawn above) it gives
$36$ invariants and $70\times6=420$ others independent entries in total.

As stated before, a generic matrix $Y$ of such form has full rank.
However, it is not so hard to decrease the rank of $Y$ by setting appropriate $y_{jk}$ values.
One must be careful to choose columns wisely, since some pairs of columns might have repeated values.
Assuming $Y$ has rank $35$, $34$ or $33$, (no need to take lower values, because we can get such matrices using other methods), numerical analysis cannot
provide the best theoretical matrix $T_{36}$ in any case. We do not know whether there exists any ``magic''
combination of columns that make them linearly dependent, so that it recovers
the best hypothetical value for a given rank, and in consequence the entropy exceeds, say $0.999$.
For example, suppose that the rank of $Y$ is $34$.
Then, the expected maximal value of $E(Y) = \frac{36}{35}\left(1-\frac{1}{34}\right) = 0.9983$ cannot
be reached (so far) by any matrix $X$, being transformed by $\Xi$. Currently the best value that can be achieved is $0.9982$.
It must be stressed, that all these considerations make sense assuming that all singular values are equal or ``almost'' equal
(recall the comment concerning ``flatness'' of singular value vectors in~\eqref{flat_svd} and subsequent formulas).

Here comes one final remark. All above attempts when applied to dimensions $d=3$, $4$, $5$, and $7$ always lead
to a solution for AME$(4,d)$. For $d=3$, the target~is~being found almost immediately,
while for $d>3$ the process of searching gradually slows down, but one can always reach the point that
resembles $T_{d^2}$ up to an arbitrary precision. This is not the case for the analyzed problem $d=6$(!).

We stop here the investigation of potential real AME states.
There are many other ideas, which are beyond the scope of this Thesis.
None of them brought us any closer to a real AME state of four quhexes.
But as usual, several open questions are in order:
\begin{enumerate}
\item Is there a real matrix $Y^{\star}\in\mathbb{R}^{36\times 36}$ such that either $e_p(Y^{\star})>e_p(W)$ or $E(\Xi(Y^{\star}))>0.9982$?
\item Is it possible to calculate the preimage of $\Xi$ to prepare a full analysis of this map?
\item Are there other isoentropic maps, which lead to a higher value of $e_p$?
\end{enumerate}

%\medskip

Based on the observations from this and previous sections,
let us conclude this part with the following
\begin{conjecture}
There is no real state {\rm AME}$(4,6)$.
\end{conjecture}

\newpage

\subsection*{Proof of the Proposition~\ref{iso_proposition}}
\label{sec:proposition_proof}

Below we present the sketch of proof of the Proposition~\ref{iso_proposition}. %presented at the beginning of Section~\ref{sec:iso_triplets}.
%\begin{proof}
First consider the case of $d=2$ and a $4\times 4$ real matrix
$$X_0=\left[\begin{array}{cccc}
      x_{11} & x_{12} & x_{13} & x_{14}\\
      x_{21} & x_{22} & x_{23} & x_{24}\\
      \ast & \ast & \ast & \ast \\
      \ast & \ast & \ast & \ast \\
\end{array}\right].
$$  
Without loss of generality, consider only two first rows of $X_0$, since $X_0^{\rm R}$ and $X_0^{\rm\Gamma}$
affect only entire $d$ contiguous blocks of size $d\times d^2$ in $X_0$.
Write down a few initial elements of $X_{k>0}$ to recognize the general pattern (for brevity,
only two first rows are shown), $X_0=X$,
%{\scriptsize
\begin{align}
%X_0 &= X,\\
X_1 &= \frac{1}{2}\left[\begin{array}{cccc}
	    2x_{11}      &     x_{12}+x_{13}      &    x_{13}+x_{21}    &      x_{14}+x_{23}\\
	    x_{12}+x_{21}     &     x_{14}+x_{22}      &    x_{22}+x_{23}    &      2x_{24}\\
\end{array}\right],\\
X_2 &= \frac{1}{4}\left[\begin{array}{cccc}
	    4x_{11}      &     2x_{12}+x_{13}+x_{21}    &   x_{12}+2x_{13}+x_{21}   &    2x_{14}+x_{22}+x_{23}\\
	    x_{12}+x_{13}+2x_{21}  &     x_{14}+2x_{22}+x_{23}    &   x_{14}+2x_{23}+x_{22}   &    4x_{24}\\
\end{array}\right],\\\nonumber
%X_3 &= \frac{1}{8}\left[\begin{array}{cccc}
%	    8x_{11}       &    3x_{12}+2x_{13}+3x_{21}  &   2x_{12}+3x_{13}+3x_{21}  &   2x_{14}+3x_{23}+3x_{22}\\
%	    3x_{12}+3x_{13}+2x_{21} &    3x_{14}+3x_{22}+2x_{23}  &   3x_{14}+2x_{22}+3x_{23}  &   8x_{24}\\
%\end{array}\right]\\
%X_4 &= \frac{1}{16}\left[\begin{array}{cccc}
%	    16x_{11}      &    6x_{12}+5x_{13}+5x_{21}  &   5x_{12}+6x_{13}+5x_{21}  &   6x_{14}+5x_{22}+5x_{23}\\
%	    5x_{12}+5x_{13}+6x_{21} &    5x_{14}+6x_{22}+5x_{23}  &   5x_{14}+6x_{23}+5x_{22}  &   16x_{24}\\
%\end{array}\right]\\
%\vdots&\\
%X_5 &= \frac{1}{32}\left[\begin{array}{cccc}
%	    32x_{11}         & 11x_{12}+10x_{13}+11x_{21} & 10x_{12}+11x_{13}+11x_{21} & 10x_{14}+11x_{22}+11x_{23}\\
%	    11x_{12}+11x_{13}+10x_{21} & 11x_{14}+11x_{22}+10x_{23} & 11x_{14}+10x_{22}+11x_{23} & 32x_{24}\\
%\end{array}\right]\\
\vdots&\nonumber
\end{align}
%}
Observe that matrix elements $\{x_{11}, x_{12}, ..., x_{24}\}$ tend to group into three classes:
$\{x_{11}, x_{24}\}$, $\{x_{12}, x_{13}, x_{21}\}$ and $\{x_{14}, x_{22}, x_{23}\}$.
Group $\{x_{11}, x_{24}\}$ is invariant with respect to $\Xi$.
Group $\{x_{12}, x_{13}, x_{21}\}$ is of the form of $(k_1x_{12} + k_2x_{13} + k_3x_{21})/2^{k}$ for $k^{\rm th}$ step of $\Xi$
where the (unordered) triplet $\{k_1, k_2, k_3\}$ takes the values $\{\kappa, \kappa, \kappa \pm 1\}$ for
$\kappa$ being the limit of the recursive sequence
\begin{equation}
\begin{cases}
\kappa(1) = 1,\\
\kappa(j) = 2 \kappa(j-1) - \delta \quad \text{with} \quad \delta = (j \bmod 2) \quad \text{and} \quad j \to k.
\end{cases}
\end{equation}
Note that $\kappa(k)$ is not greater than $2^k$. A similarly property holds for $\{x_{14}, x_{22}, x_{23}\}$.
Hence, as $k \to\infty$, one has
\begin{equation}
    X_k \to \left[\begin{array}{cccc}
        x_{11}      &    \xi(x_{12},x_{13},x_{21}) &  \xi(x_{12},x_{13},x_{21}) &  \zeta(x_{14},x_{22},x_{23}) \\
        \xi(x_{12},x_{13},x_{21}) &  \zeta(x_{14},x_{22},x_{23}) &  \zeta(x_{14},x_{22},x_{23}) &  x_{24} \\
        \ast      &    \ast    &     \ast   &       \ast \\
        \ast      &    \ast    &     \ast   &       \ast
    \end{array}\right],
\end{equation}
where $\xi$ and $\zeta$ are limits of combinations of $\{x_{12}, x_{13}, x_{21}\}$ and $\{x_{14}, x_{22}, x_{23}\}$,
respectively.
Both limits exist because appropriate sequences are bounded and monotonic.
The limit of $\kappa/2^k$ is $1/3$ or $1/6$, depending on the associated group of elements, which
is useful in derivation of the simplified version of the map $\Xi$, see~\eqref{Pi_simplified_1} or~\eqref{Pi_simplified_2}.

The matrix $\displaystyle{Y=\lim_{k\to\infty} X_k}$ is invariant with respect
to reshuffling and partial transpose, $Y=Y^{\rm R}=Y^{\rm\Gamma}$.
However, last equality (as a result of~$\,\Xi$) holds only for $d=2$.
Hence, we see that
    ${\rm svd}(Y) = {\rm svd}(Y^{\rm R}) = {\rm svd}(Y^{\rm\Gamma})$
and additionally, 
    ${\rm eig}(Y) = {\rm eig}(Y^{\rm R}) = {\rm eig}(Y^{\rm\Gamma})$.
In the case $d>2$, one has only $Y \ne Y^{\rm R} = Y^{\rm\Gamma}$ but the structure of the matrix $Y$
reflects similar symmetries like for $d=2$.
To end the proof, one exploits the fact that $YY^{\dagger} = Y^{\rm R}(Y^{\rm R})^{\dagger} = Y^{\rm\Gamma}(Y^{\rm\Gamma})^{\dagger}$.
Thus ${\rm svd}(Y) = \sqrt{{\rm eig}(YY^{\dagger})} = {\rm svd}(Y^{\rm R}) = {\rm svd}(Y^{\rm\Gamma})$,
and $Y\in\mathcal{J}_{d^2}$~\eqref{isoentropic}.$\hfill\square$
%\end{proof}

%%%%%%%%%%%%%%%%%%%%%%%%%%%%%%%%%%%%%%%%%%%%%%%%%%%%%%%%%%%%%%%%%%%%%%%%%%%%%%%%
%%%%%%%%%%%%%%%%%%%%%%%%%%%%%%%%%%%%%%%%%%%%%%%%%%%%%%%%%%%%%%%%%%%%%%%%%%%%%%%%
%%%%%%%%%%%%%%%%%%%%%%%%%%%%%%%%%%%%%%%%%%%%%%%%%%%%%%%%%%%%%%%%%%%%%%%%%%%%%%%%

\section{Unitary Approximation of AME$(4,6)$}

Let us finally switch to the complex domain.
Being the richest structure in our research, it should provide the best results
in search of the desired AME state.

\subsection{Dynamical Map}
\label{sec:AME_map}

Recall the Sinkhorn alternating algorithm~\cite{Si64, SK67}
used extensively in Section~\ref{sec:Sinkhorn_AA} of the second chapter
to generate complex Hadamard matrices.
The use of such algorithm, in the case when we are dealing with a very similar
pattern of ``projections'' in the form of two consecutive operations, ${\rm R}\to{\rm R\Gamma}\to{\rm R\Gamma R}\to...$,
is even imposed as a natural method of numerical search for a solution.

Formally, we define a nonlinear dynamical map
\begin{equation}
\mathcal{M}:\mathbb{C}^{d^2\times d^2}\ni M\mapsto \mathcal{M}(M)\in\mathbb{U}(d^2)\label{SS_map},
\end{equation}
where $\mathcal{M}(M)$ is the limit of the following recursive sequence (provided that it is convergent),
\begin{equation}
\begin{cases}
M_0 \in\mathbb{C}^{d^2\times d^2}, \\
M_{j+1} = \pi\big(M_j^{\rm R\Gamma}\big) \ : \ j\in\mathbb{N}_0,\label{M_map}
\end{cases}
\end{equation}
where $M_0$ is any complex starting point (a seed) and $\pi$ denotes polar decomposition procedure
to restrict the mapping to the set of unitary matrices. That is,
$\pi(X)=X/\sqrt{X^{\dagger}X}$
(cf. Section~\ref{sec:Sinkhorn_AA} in Chapter~\ref{chap:CHM}).
Actually, $\pi$ projects any matrix $M_{j>1}$ onto a nearest unitary 
one.
%If $\mathcal{M}_j$ is convergent %, and appropriate limit exists,
%we call it
%$\displaystyle{\mathcal{M}=\lim_{j\to\infty}\mathcal{M}_j(M)}$.
Our goal is to find such a seed $M_0$ that $\mathcal{M}(M_0)=T_{36}$ is a $2$-unitary matrix.
%$\displaystyle{T_{d^2}=\lim_{j\to\infty}\mathcal{M}(M_j)}$.

Encouraged by the fact that for $d=3$, $4$, and $5$ the algorithm
converges without much impediments, we hope to recover similar solution for $d=6$.
But the main problem that kept us from solving it for a number of years was the right form of the
starting point. Again, a natural choice for a seed seems to be the matrix $P$, as the one
with help of whom, we already obtained most, previously undiscovered, results.
But this is not the case anymore.
A~detailed search of the available space with a special focus on $P$, $Q^{\star}$, $W$
and other seemingly distinguished points remained unsuccessful for a bit long,
when, eventually, Suhail Ahmad Rather (another crucial contributor to~$[\hyperlink{\paperslist}{\rm A4}]$),
worked out that a very particular starting point should be taken.
Surprisingly, it is a permutation $P^{\star}$, slightly disturbed
in the complex direction
\begin{equation}
M^{\star}_0 = P^{\star} \exp\Big\{ i \frac{\varepsilon}{2}\Big(G+G^{\rm T}\Big)\Big\} \quad : \quad \varepsilon\ll 1,
\end{equation}
where $G\in\mathbb{R}^{36\times 36}$ is an appropriately chosen real random matrix.
Counterintuitively, $P^{\star}$ is characterized by a bit worse value of entangling power than $P$,
which makes it closer to ${\rm CUE}$, which as we saw in Figure~\ref{fig:CUE_attractor}, is the very strong attractor for any evolution
on the $(e_p,g_t)$-plane.

\begin{figure}[hbp!]
\center
\includegraphics[width=5in]{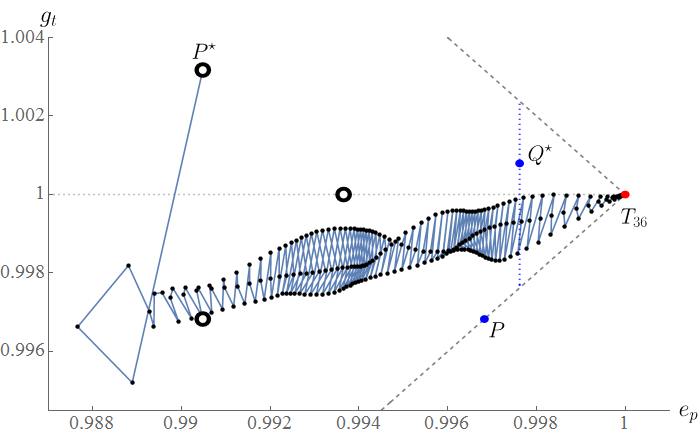}
\caption{Numerical procedure to find a $2$-unitary matrix $T_{36}\in\mathbb{U}(36)$, which leads to AME$(4,6)$.
Example of a successful trajectory of $\mathcal{M}$ for a given starting point $P^{\star}$ (seed) in the $(e_p,g_t)$-plane.
For this particular seed, convergence is very fast, and approximately after $200$ iterations
the desired solution $T_{36}$ (with arbitrary precision) is reached. 
Other two possible seeds (black rings), which are closest to $P$ are depicted too. 
Every third iteration follows individual sub-path (small black dots), which might correspond to
three operations: $\mathbb{I}$, ${\rm R}$ and ${\rm\Gamma}$.
Vertical line crossing $Q^{\star}$ denotes the area of strong attraction for the map $\mathcal{M}$.}
\label{fig:SEED_trajectory}
\end{figure}

Hence, an AME state of four subsystems, each of local dimension six, does exist! At least numerically.
This is a rather surprising observation in the light of many facts that over the years
could prove that such a state is unlikely to be constructed. 
Having this done finally opened a more systematic way of numerical generation of AME states for $d=6$.
In Figure~\ref{fig:SEED_trajectory} we present three locations 
of seeds that, together with a proper choice of perturbation $G$, provide a solution when supplied to $\mathcal{M}$.
One particular trajectory of consecutive steps of $\mathcal{M}$
showing quite irregular behavior is drawn in the same figure.

\newpage

It is worth to mention that really few matrices have a property of being a good initial point $M_0$, which generates
a trajectory converging to $2$-unitary matrix $T_{36}$.
This was an annoying obstacle for many researchers who tried to tackle this problem
and it led to the common belief that such a construction is not possible.
Now, at the time of preparing this Thesis, we know about $20$ different seeds
from the entire set $\mathbb{P}(36)$ that, mixed with appropriate perturbations $G$,
can serve as the seeds for $\mathcal{M}$. Vast majority
of permutation matrices, even those close\footnote{We say that a permutation matrix $P'$ is close to $P$, if
an arbitrarily small number (say $1$ or $2$) of transpositions of rows (or columns) is required to obtain
$P$ from $P'$.} to $P$, provided as arguments for $\mathcal{M}$,
give $e_p(\mathcal{M}(P_{\rm generic}))=e_p(Q^{\star})$, which makes matrices on the vertical
line $e_p=e_p(Q^{\star})$ of the $(e_p,g_t)$-plane extremely strong attractors for $\mathcal{M}$. No other matrix, beside carefully
selected and perturbed permutation, was found so far as an appropriate seed for $\mathcal{M}$.
We recommend the Reader to follow~$[\hyperlink{\paperslist}{\rm A4}]$
for more interesting features of the map $\mathcal{M}$.

\subsection{Golden AME$(4,6)$ State}
\label{sec:AME_beautifier}

\begin{figure}[ht]
\center
\includegraphics[width=5in]{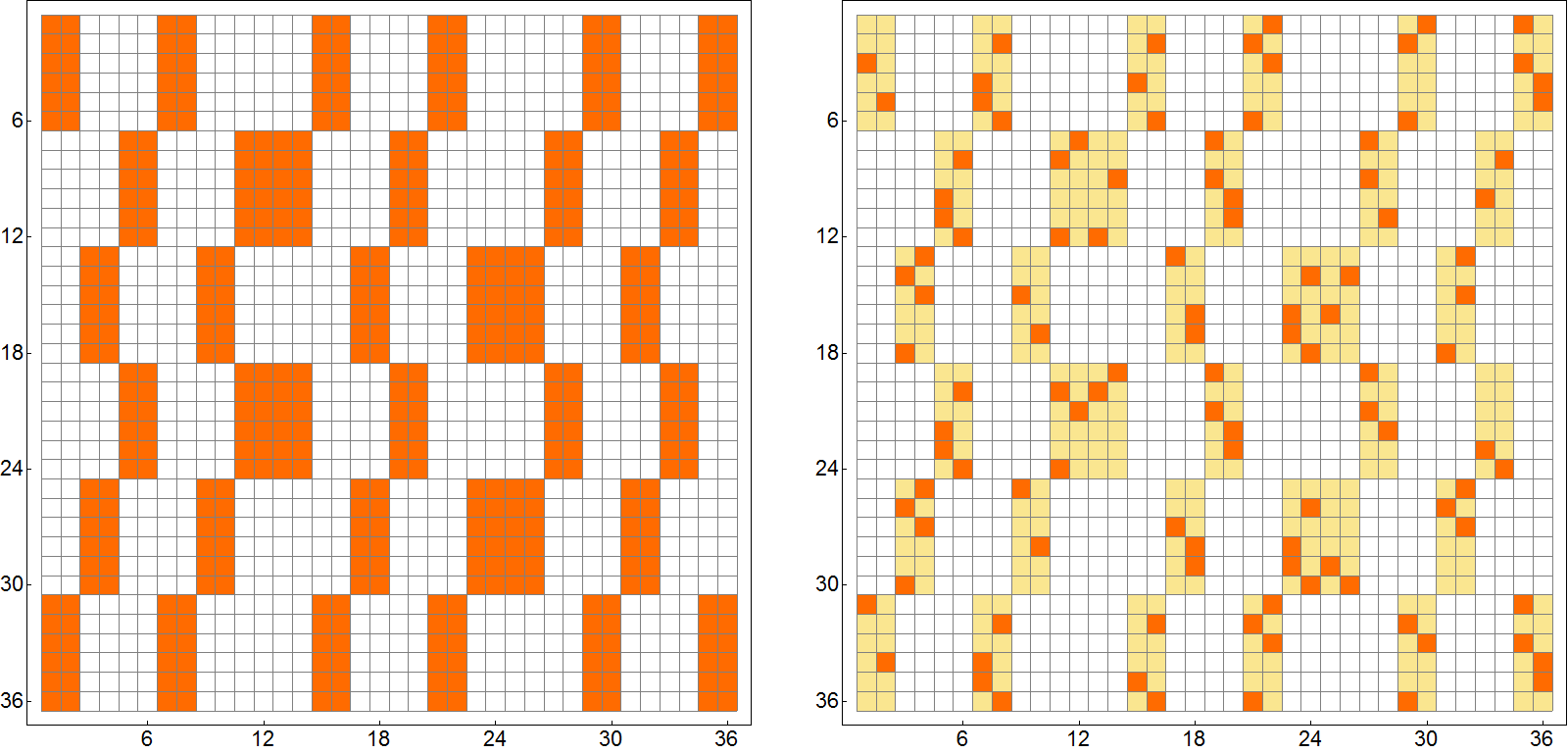}
\caption{Left panel represents $432$ nonvanishing elements of a numerical representation of $T_{36}$. On
the right panel we observe as local unitary rotations~\eqref{T_local_rotation} transform $T_{36}$ into analytical form
where only $112$ entries (dark squares) are nontrivial, while the rest eventually vanish (light squares).
Such a sparse and scattered form partially explains the fact that we were unable to find a complex representation of $T_{36}$
in Section~\ref{sec:matrix_V}.}
\label{fig:numerical_AME_superimposed}
\end{figure}

Representations of AME states, $T_{36}$, being the output from the map $\mathcal{M}$~\eqref{M_map} are usually given
in the form of $1296$ complex numbers arranged without any order into an array of size $36$ by $36$.
Sometimes, algorithm converges to a simpler matrix with only $432$ non-zero elements that can be
rearranged\footnote{Be cautious at this stage! The state AME$(4,6)$ is very sensitive to the tiniest value
changes and element shuffling, including highly devastating process of permutations of rows or columns.
Such rearrangements %destroy the property of being AME immediately and
should be considered only as temporary visual representation. The only operations, presented in this Thesis,
that preserve the AME property are: ${\rm R}$, ${\rm\Gamma}$, global
transpose ${\rm T}$ and local unitary rotations from $\mathbb{U}(6)\otimes\mathbb{U}(6)$.}
into three blocks of size $12\times 12$; see left panel in Figure~\ref{fig:numerical_AME_superimposed}.
Despite the fact that we can set the precision of the output arbitrarily,
i.e. $e_p(T_{36})=1$ within the limits of machine precision, this cannot in any way be
considered a formal proof of Theorem~\ref{AME_theorem}. Thus we have to follow a sort of a similar path
as in Chapter~\ref{chap:CHM}
when we had to work out the analytical formulas for entries of CHM out of their numerical forms.

We already mentioned the invariant operations~\eqref{local_rotations}
that act locally and do not change the value of $e_p$,
while at the same time, they can significantly divert the numerical representation.
We modified the optimization procedure, and redesigned it to seek
four local rotations $U^{(1)}_6$, $U^{(2)}_6$, $U^{(3)}_6$, and $U^{(4)}_6$ from $\mathbb{U}(6)$,
which transform $T_{36}$ in the following way 
\begin{equation}
T_{36}\mapsto\left(U^{(1)}_6\otimes U^{(2)}_6\right)T_{36}\left(U^{(3)}_6\otimes U^{(4)}_6\right)\label{T_local_rotation}
\end{equation}
and eventually provided the analytic formulas for elements of $T_{36}$.
This time, the function $\mathcal{Z}$ to be optimized took the form
of complicated dependencies between certain groups of elements of $T_{36}$.
The general idea was to make as many different $6\times 2$ blocks of $T_{36}$ (see Figure~\ref{fig:numerical_AME_superimposed})
mutually orthogonal as possible, and eventually turn some of them into
a standard (canonical) basis $|j\rangle\in\mathbb{C}^6$ to simplify the entire structure.
Several dozens of different forms of the target functions $\mathcal{Z}$'s, combining different areas of $T_{36}$,
applied one after another, have led in the end to such a representation
that allowed us to guess some analytical values and the rest was easily calculated from unitarity constraints.
As partially expected, during the optimization procedure, it turned out that only $112$ entries 
remained with non-zero values, which makes the final analytic form of $T_{36}$ quite sparse, see right panel in Figure~\ref{fig:numerical_AME_superimposed}.

Let us present the AME$(4,6)$ state in its full grandness.
There are many ways of presenting such state: as a matrix, a vector, a (perfect) tensor, or as a pair of OQLS.
Each of these methods has its advantages and disadvantages, however, the main problem
that emerges in all cases is always the same --- its size.
Drawing a $36\times 36$ matrix on any surface is always a challenge.
Here, using algebraic tricks, we have prepared a special version, which should consume a decent amount of space
and simultaneously to reveal some peculiar properties.

Put $\omega=\exp\{i\pi/10\}$ and\footnote{Asterisk $^*$ in $b$ and $c$ denotes complex conjugate.}
\begin{equation}
a=\frac{1}{\sqrt{2}}, \ \ \ \ \
b=\frac{a}{\omega+\omega^*}=\frac{1}{\sqrt{5+\sqrt{5}}},  \ \ \ \ \
c=\frac{a}{\omega^3+(\omega^*)^{3}}=\frac{1}{2\sqrt{5}}\sqrt{5+\sqrt{5}},
\end{equation}
and note that $a^2=1/2=b^2+c^2$. This implies unitarity of the following blocks ($\bullet=0$): %\footnote{We draw the
%Reader's attention that the dots $\bullet$ in $B_1$, $B_2$, and $B_3$
%represent zeros. See~\cite{TZ06, CHM_catalogue}.} %three square blocks $B_1$, $B_2$ and $B_3$ of size $12$,
%depending on constants $a$, $b$, and $c$ and appropriate phases,

\begin{align}
B_1&=\left[\begin{array}{llllllllllll}
\bullet & \bullet & c \omega^{19} & \bullet & \bullet & \bullet & b \omega^{18} & \bullet & \bullet & \bullet & \bullet & a \omega^6\\
\bullet & c \omega^9 & \bullet & \bullet & \bullet & \bullet & \bullet & a \omega^3 & b & \bullet & \bullet & \bullet\\
\bullet & \bullet & \bullet & c \omega^7 & c \omega^{15} & \bullet & \bullet & \bullet & \bullet & b \omega^{13} & b \omega^{14} & \bullet\\
\bullet & \bullet & \bullet & b \omega^2 & b \omega^{14} & \bullet & \bullet & \bullet & \bullet & c \omega^2 & c \omega^{19} & \bullet\\
\bullet & b \omega^5 & \bullet & \bullet & \bullet & a \omega^{19} & \bullet & \bullet & c \omega^6 & \bullet & \bullet & \bullet\\
 a \omega^{19} & \bullet & b \omega^{17} & \bullet & \bullet & \bullet & c \omega^6 & \bullet & \bullet & \bullet & \bullet & \bullet\\
 a & \bullet & b \omega^8 & \bullet & \bullet & \bullet & c \omega^{17} & \bullet & \bullet & \bullet & \bullet & \bullet\\
\bullet & b \omega^6 & \bullet & \bullet & \bullet & a \omega^{10} & \bullet & \bullet & c \omega^7 & \bullet & \bullet & \bullet\\
\bullet & \bullet & \bullet & b \omega^{16} & b \omega^{12} & \bullet & \bullet & \bullet & \bullet & c \omega^{12} & c \omega & \bullet\\
\bullet & \bullet & \bullet & c \omega & c \omega & \bullet & \bullet & \bullet & \bullet & b \omega^{11} & b \omega^{16} & \bullet\\
\bullet & c \omega^{12} & \bullet & \bullet & \bullet & \bullet & \bullet & a \omega^{16} & b \omega^3 & \bullet & \bullet & \bullet\\
\bullet & \bullet & c \omega^{16} & \bullet & \bullet & \bullet & b \omega^{15} & \bullet & \bullet & \bullet & \bullet & a \omega^{13}
\end{array}\right],\\
%\end{align}
%\begin{align}
B_2&=\left[\begin{array}{llllllllllll}
\bullet &  \bullet &  b &  \bullet &  \bullet &  \bullet &  \bullet &  a \omega^{13} &  c \omega^{16} &  \bullet &  \bullet &  \bullet\\
 b \omega^{18} &  \bullet &  \bullet &  \bullet &     c \omega &  a \omega^{14} &  \bullet &  \bullet &  \bullet &  \bullet &  \bullet &  \bullet\\
\bullet &  \bullet &  \bullet &  a \omega^{15} &  \bullet &  \bullet &  \bullet &  \bullet &  \bullet &     b \omega &  \bullet &  c \omega^3\\
\bullet &  \bullet &  \bullet &  \bullet &  \bullet &  \bullet &  a \omega^{18} &  \bullet &  \bullet &  c \omega^2 &  \bullet &     b \omega^{14}\\
 c \omega^8 &  a \omega^{14} &  \bullet &  \bullet &  b \omega &  \bullet &  \bullet &  \bullet &  \bullet &  \bullet &  \bullet &  \bullet\\
\bullet &  \bullet &  c &     \bullet &  \bullet &  \bullet &  \bullet &  \bullet &  b \omega^6 &  \bullet &  a &  \bullet\\
\bullet &  \bullet &  c \omega^{11} &  \bullet &  \bullet &  \bullet &  \bullet &  \bullet &     b \omega^{17} &  \bullet &  a \omega &  \bullet\\
 c \omega &  a \omega^{17} &  \bullet &  \bullet &  b \omega^{14} &  \bullet &  \bullet &  \bullet &  \bullet &  \bullet &  \bullet &     \bullet\\
\bullet &  \bullet &  \bullet &  \bullet &  \bullet &  \bullet &  a \omega^6 &  \bullet &  \bullet &  c &  \bullet &  b \omega^{12}\\
\bullet &  \bullet &  \bullet &     a \omega^{15} &  \bullet &  \bullet &  \bullet &  \bullet &  \bullet &  b \omega^{11} &  \bullet &  c \omega^{13}\\
 b \omega &  \bullet &  \bullet &  \bullet &  c \omega^4 &     a \omega^7 &  \bullet &  \bullet &  \bullet &  \bullet &  \bullet &  \bullet\\
\bullet &  \bullet &  b \omega &  \bullet &  \bullet &  \bullet &  \bullet &  a \omega^4 &  c \omega^{17} &  \bullet &     \bullet &  \bullet
\end{array}\right],\\
%\end{align}
%\begin{align}
B_3&=\left[\begin{array}{llllllllllll}
 b \omega^4 &  \bullet &  \bullet &  \bullet &  \bullet &  \bullet &  a \omega^8 &  \bullet &  \bullet &  c \omega^{14} &  \bullet &  \bullet\\
\bullet &  \bullet &  a \omega^{14} &  \bullet &     \bullet &  \bullet &  \bullet &  b \omega^6 &  \bullet &  \bullet &  \bullet &  c \omega^{17}\\
\bullet &  b \omega^6 &  \bullet &  \bullet &  c \omega^{16} &  \bullet &  \bullet &  \bullet &     a &  \bullet &  \bullet &  \bullet\\
\bullet &  c \omega^{19} &  \bullet &  \bullet &  b \omega^{19} &  \bullet &  \bullet &  \bullet &  \bullet &  \bullet &  a \omega^{15} &     \bullet\\
\bullet &  \bullet &  \bullet &  \bullet &  \bullet &  a \omega &  \bullet &  c \omega^8 &  \bullet &  \bullet &  \bullet &  b \omega^9\\
 c &  \bullet &  \bullet &     a \omega^{14} &  \bullet &  \bullet &  \bullet &  \bullet &  \bullet &  b &  \bullet &  \bullet\\
 c \omega^2 &  \bullet &  \bullet &  a \omega^6 &  \bullet &  \bullet &  \bullet &  \bullet &     \bullet &  b \omega^2 &  \bullet &  \bullet\\
\bullet &  \bullet &  \bullet &  \bullet &  \bullet &  a \omega^{19} &  \bullet &  c \omega^{16} &  \bullet &  \bullet &  \bullet &     b \omega^{17}\\
\bullet &  c \omega^{12} &  \bullet &  \bullet &  b \omega^{12} &  \bullet &  \bullet &  \bullet &  \bullet &  \bullet &  a \omega^{18} &  \bullet\\
\bullet &     b \omega^{15} &  \bullet &  \bullet &  c \omega^5 &  \bullet &  \bullet &  \bullet &  a \omega^{19} &  \bullet &  \bullet &  \bullet\\
\bullet &  \bullet &  a \omega^6 &  \bullet &  \bullet &     \bullet &  \bullet &  b \omega^8 &  \bullet &  \bullet &  \bullet &  c \omega^{19}\\
 b \omega^8 &  \bullet &  \bullet &  \bullet &  \bullet &  \bullet &  a \omega^2 &  \bullet &  \bullet &     c \omega^{18} &  \bullet &  \bullet
\end{array}\right].
\end{align}

Now, the absolutely maximally entangled state of four quhexes can be concisely represented 
by a unitary matrix $A$ of order $36$,
\begin{equation}
A = \left(B_1\oplus B_2\oplus B_3\right)P^{\rm T}\label{A_AME},
\end{equation}
with help of the inverse of $P$~\eqref{P_tilde} to recover desired structure.
With some effort one can confirm that
\begin{equation}
E(A)=E\big(A^{\rm R}\big)=E\big(A^{\rm\Gamma}\big)=e_p(A)=g_t(A)=1
\end{equation}
which implicitly ends the proof of Theorem~\ref{AME_theorem}.\hfill$\square$

\noindent The state AME$(4,6)$ can be written as
\begin{equation}
|{\rm AME}(4,6)\rangle=\frac{1}{6}\sum_{j=1}^6\sum_{k=1}^6|j\rangle\otimes |k\rangle \otimes \big(A|j\rangle\otimes |k\rangle\big)\in\mathcal{H}_{6}^{\otimes 4}.
\end{equation}

\begin{figure}[ht!]
\center
\includegraphics[width=5in]{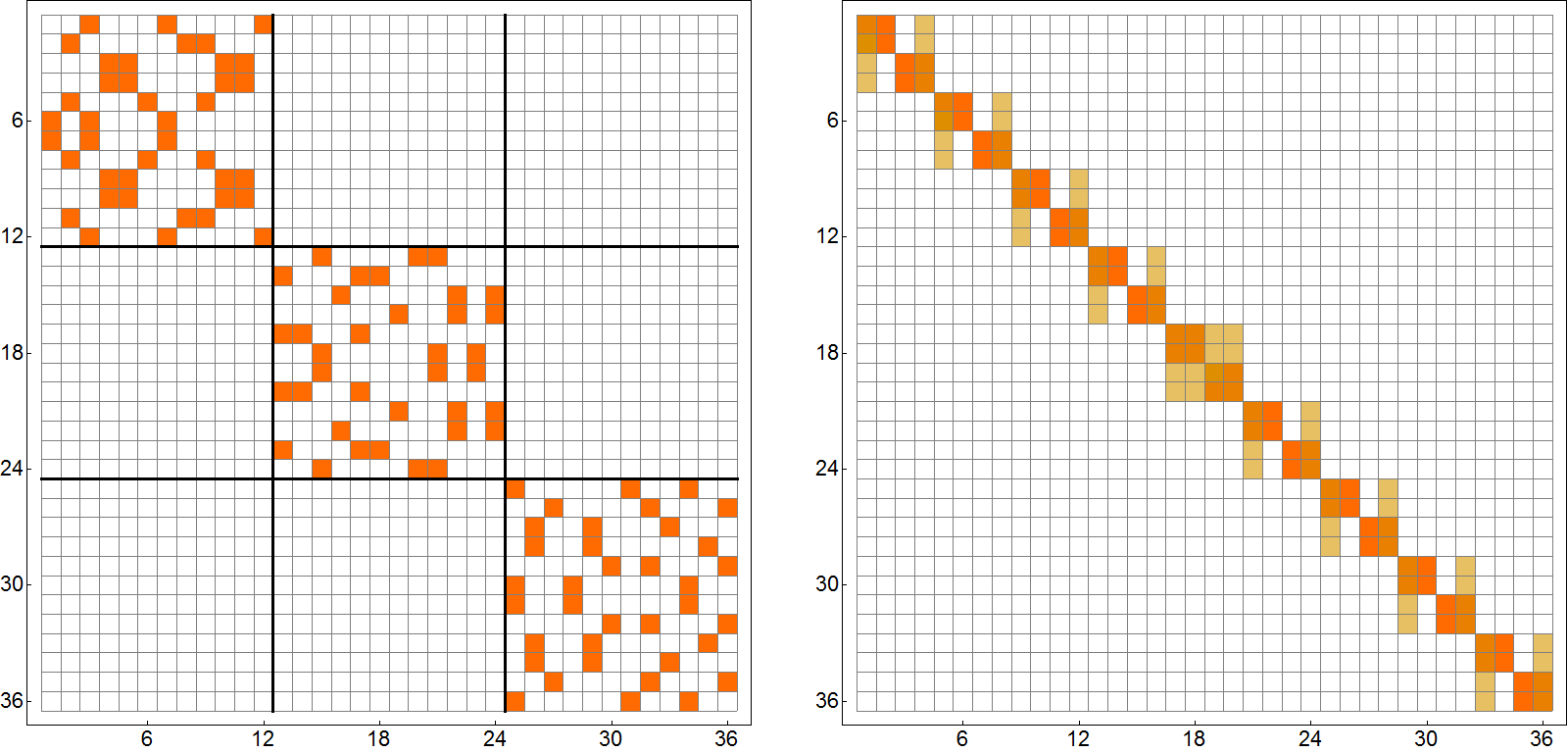}
\caption{Left panel shows non-zero elements of $AP$ in the form of a block-diagonal structure $3\times(12\times 12)$ with
additional internal vertical symmetries. On the right panel
we see absolute values of entries of $\widetilde{A}=P_{\rm r}AP_{\rm c}$, forming another extra-ordinary
symmetric block-diagonal $9\times(4\times 4)$ matrix. This pattern implies that $36$ officers of Euler can be divided into $9$ groups
of $4$ entangled officers.
Matrices $P_{\rm r}$ and $P_{\rm c}$ are shown in Figure~\ref{fig:final_AME46_PRPC}.}
\label{fig:final_AME46_3_9}
\end{figure}

\begin{figure}[ht!]
\center
\includegraphics[width=5in]{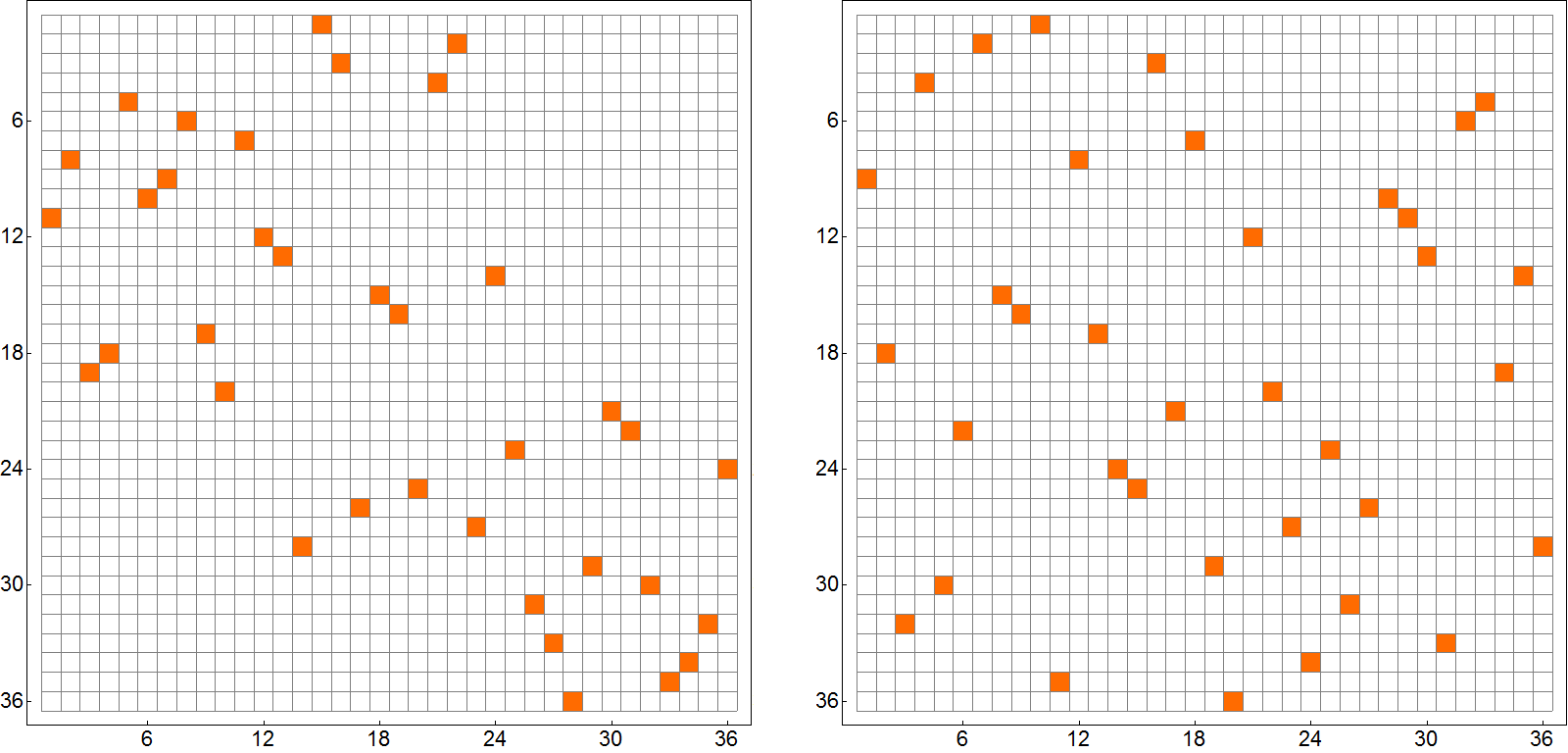}
\caption{Permutation matrices $P_{\rm r}$ (left panel) and $P_{\rm c}$ (right panel) that bring $A$ to a block-diagonal $9\times(4\times 4)$ form
$\widetilde{A}=P_{\rm r}AP_{\rm c}$.}
\label{fig:final_AME46_PRPC}
\end{figure}

Blocks $B_1$, $B_2$ and $B_3$ are presented on the left panel of Figure~\ref{fig:final_AME46_3_9}.
Such representation can be further visually simplified with help of two permutations,
$\widetilde{A}=P_{\rm r}AP_{\rm c}$,
which is depicted on the right panel of Figure~\ref{fig:final_AME46_3_9}, while
the two permutations $P_{\rm r}$ and $P_{\rm c}$ are shown in Figure~\ref{fig:final_AME46_PRPC}.

Intriguing relation between the building blocks of $A$ in
the form of the golden ratio, $c/b=\varphi$,
should explain the proposition to call $A$ the {\sl golden AME state}.
We are not going to describe the internal structure of $A$,
which has been studied in details by Adam Burchardt in his thesis\footnote{In preparation.} --- the third
major contributor to~$[\hyperlink{\paperslist}{\rm A4}]$ --- who, armed with the
optimizing software, written %and customized to this very problem
by the Author, was patiently recovering the analytical form of the matrix $T_{36}$ turning
it finally to a representation equivalent to the matrix $A$.

\section{Summary}

In this chapter we have proven the unexpected existence of the absolutely maximally entangled state of four quhexes.
Many groups tried to solve this problem over the years and almost everybody (including us!)
shared the belief that such states do not exist.
Unclear difficulties related to the relatively small number $6$ are common to other objects of this size: MUB in $\mathbb{C}^6$ and CHM (incomplete classification
of $\mathbb{H}(6)$).
But we believe that our result will bring us closer to a full understanding of this puzzling property.

With analytic construction of AME$(4,6)$ we solved several equivalent problems. They are:
existence of a $2$-unitary~matrix of size $36$ with maximal entangling power,
existence of a perfect tensor with four indices (each running from $1$ to $6$),
existence of pure quantum error correcting code $(\!(4,1,3)\!)_6$, and
existence of a pair of orthogonal quantum Latin squares of order $6$.
Note that this last result disproves Conjecture 1 from ~\cite{GRMZ18}.
However, despite putting many efforts we failed to resolve the problem of finding an orthogonal counterpart of the complex state AME$(4,6)$.

\newpage
Further open questions include:
\begin{enumerate}
\item Is it possible to redefine the notion of the defect (Chapters~\ref{chap:CHM} and~\ref{chap:DELTA}) so that it is applicable to AME?
If so, would it be possible to stem a family of inequivalent states from the discovered state AME$(4,6)$
or to prove that this is an isolated point, up to local unitary rotations?
\item What about different seeds that can be supplied to the numerical procedure~\eqref{SS_map}?
Is there any systematic method of constructing AME$(4,6)$ state(s)? Or, how to systematize and simplify
the sequence of optimizations that led us after many laborious months to the analytic form of $A$ given in Eq.~\eqref{A_AME}?
How to make it applicable for other allowed dimensions $N$ and $d$?
\item Is the existence of the state AME$(4,6)$ somehow related to the MUB problem in $\mathbb{C}^6$?
\item What is the most optimal way to represent a quantum circuit corresponding to the AME state four quhexes?
\end{enumerate}

%\medskip

\begin{figure}[ht!]
\center
\includegraphics[width=0.5in]{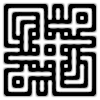}
\end{figure}

%\bigskip

In the next, and the last chapter of this Thesis, we are going to continue research concerning
the concept of quantum entanglement from the perspective of the foundations of quantum mechanics and Bell's theorem.

%%%%%%%%%%%%%%%%%%%%%%%%%%%%%%%%%%%%%%%%%%%%%%%%%%%%%%%%%%%%%%%%%%%%
\clearpage\null\thispagestyle{empty}
%\clearpage\null%\thispagestyle{empty}
\chapter{Excess of a Matrix and Bell Inequalities}
\label{chap:EXCESS}
Chapter based on~$[\hyperlink{\paperslist}{\rm A5}]$.
\medskip
\medskip
\medskip

\noindent The preprint~$[\hyperlink{\paperslist}{\rm A5}]$ on which this chapter is based,
is rooted on ideas of collaborators thus, in contrary to all
previous chapters, this short one has mostly descriptive character.
As promised in the introduction, we restrict
the work merely to original contributions of the Author, which in this case
were mainly focused on numerical research to confirm analytical predictions
and scrutinous studying the literature in search for a similar
approach (to exclude the possibility of accidental reuse of the same concept).
One particular example that we found interesting will be used in Section~\ref{sec:QA_example}
to illustrate Bell inequalities with quantum advantage induced by a special class of matrices.
The presence of this chapter is justified from the perspective
of the integrity and general appearance of the Thesis.
Its content refers to many raised topics and neatly concludes
some problems mentioned previously. For example, we will
recall MUB, (real) Hadamard, unitary, and circulant matrices for the last time
and examine them for completely different purposes.

\medskip

We already signalized the ``problems'' with quantum mechanics (QM) in Chapter~\ref{chap:DELTA} during the investigation
a possibility of introduction of free parameters in the set of Kochen--Specker vectors (which, anyway, turned out to be isolated structures).
QM undoubtedly reached a status of a successful theory as it provides many explanations
of physical processes at the microscopic level,
and opens a great variety of possible applications unavailable in the classical approach.
However, many questions concerning the foundations and self-consistency of QM that arose almost immediately with its advent,
still remain unanswered.
Needless to say, we refer to the famous EPR Gedankenexperiment~\cite{EPR35} as the origin of the long-standing disputes~\cite{Br17}
regarding the interpretation of quantum mechanics.

In short, it all began with the problem of allegedly immediate transfer of information between spatially separated objects.
Consider two entangled\footnote{Now we know that entanglement is crucial to understand
this paradox.} photons traveling in two opposite directions.
It turns out that when we measure the polarization of one photon, no matter how much they are apart,
the polarization of the other one is always strongly correlated to the first one.
Famous authors of~\cite{EPR35} questioned this fact,
as it openly contradicts the rules of the special theory of relativity (STR),
and they postulated the notion of {\sl local realism}, according to which objects can interact only within their neighborhood.
So, if there is a correlation between distant objects (events), invoked by some interaction (measurement),
then local realism implies existence of some {\sl local hidden variables} (LHV),
which carry additional information.
In consequence, something is wrong with the Copenhagen interpretation of quantum mechanics, because
the associated wave function apparently does not provide full description of the system.
In general, QM seems incomplete.

In 1964 John Bell~\cite{Be64}
suggested a solution to the paradox of LHV in a form of a simple arithmetic relation, which can be verified experimentally.
If we assume the local realism and hidden variables model then one can conclude a set of inequalities, {\sl Bell inequalities} (BI),
which must hold in the classical case.
In the simplest formulation, it involves linear combinations of correlations of measurement outputs that, when combined into a single number,
it must not exceed a particular bound.
However, in the quantum world one observes violation of these inequalities, that is the quantum value of BI
is strictly bigger than its classical counterpart.
Hence, local realism is not a proper assumption to explain physics and separated objects can ``communicate'' over long distances.
Formally this means that measurement affects entire physical system, no matter how huge, not only its parts.
It is worth to note that the concept of Bell does not refute the existence of hidden variables,
it only demonstrates that correlations at the quantum level cannot be derived from deterministic and LHV models.
Moreover, there is no conflict with STR (no ``spooky action at a distance'') because,
despite the fact we observe strong correlations, they appear in a purely random fashion and as such they cannot
transfer any information.
Simply speaking, quantum mechanics describes {\sl nonlocal} Universe.

First experiments confirming violation of Bell inequalities were performed in the early eighties of the last century~\cite{AGR81}.
Since then several practical modifications at the very high level were proposed in order to avoid possible loopholes,
for example~\cite{HBD15},
and also many different classes (families of) Bell inequalities were investigated %and axioms
for testing QM and its possible generalizations~\cite{GBPM21}.
There are several technological applications of quantum nonlocality, which significantly outperforms similar classical tasks.
This includes: device-independent protocols~\cite{Li13,SVXATW13}, communication~\cite{CB97, CLMW11},
quantum key distribution and general quantum cryptography~\cite{AGM06,MPA11,VV14,KW16}, 
testing quantumness~\cite{AL16,GS18}, and random numbers generation~\cite{CK11,PAMLMMOHLMM10,CR12}.

The main open question in quantum nonlocality concerns the existence of
quantum advantage in comparison with the classical (LHV) scenario for a given Bell inequality.
Seeking a solution to this problem, one  must find the biggest possible value of the Bell inequality in LHV\footnote{Following
the convention from the preprint, in place of the ``classical scenario'' we should use the abbreviation LHV.} theories and in the quantum case.
But there is at least one crucial impediment:
finding the LHV value of a given BI is NP-hard problem~\cite{ALMSS98} --- the complexity grows with the number
of measurement settings $m$ and the number of outcomes $q$. Even the case of two-party scenarios
for relatively small values of $m$ and $q$ remains unsolved~\cite{BCPSW14, RBG14}.
Thus, it would be of great importance to introduce a method of estimating the LHV value in a simple way.
In order to address this task, we propose a connection between
the LHV value of bipartite BI and mathematical notion of the excess of a given matrix. 
We replace the problem of finding LHV value associated with a bipartite BI with
equivalent problem of finding matrices with the maximal excess (a number associated with matrix, to be defined in Section~\ref{sec:SigmaDef}).
Having this, we are able to characterize infinite number of families of bipartite BI describing a variety of scenarios.
Since the properties of the excess have been studied in mathematics for decades,
we can have many results ``out of the box'' and apply them in the field of quantum nonlocality.

\section{Initial Observation}

Consider a special case of BI --- the celebrated CHSH inequality~\cite{CHSH69}. The associated Bell operator reads
\begin{equation}
\mathbb{B}=A_0\otimes B_0+A_0\otimes B_1+A_1\otimes B_0-A_1\otimes B_1\label{CHSH},
\end{equation}
where $A_j$ and $B_k$ are observables represented by Hermitian operators having $\mp 1$ eigenvalues.
To calculate the classical value $\mathcal{C}$ of $\mathbb{B}$, we replace the operators with variables %\footnote{Without loss of generality we assume that variables take values from the set}
$a_j$, $b_k\in\{-1,1\}$, so
\begin{equation}
\mathcal{C} = \max_{a_j,b_k=\mp 1}\big\{{\bf +}a_0b_0{\bf +}a_0b_1{\bf +}a_1b_0{\bf -}a_1b_1\big\}=2\label{C_CHSH}.
\end{equation}
Note that~\eqref{C_CHSH} can be equivalently expressed in the form of
\begin{equation}
\mathcal{C} = \max_{a_j,b_k=\mp1}\Bigg\{
\sum
\left[\begin{array}{cc}
{\bf +}a_0b_0 & {\bf +}a_0b_1\\
{\bf +}a_1b_0 & {\bf -}a_1b_1
\end{array}\right]\Bigg\} =
\sum
\left[\begin{array}{cc}
{\bf +} & {\bf +}\\
{\bf +} & {\bf -}
\end{array}\right]=2,
\end{equation}
where $\sum$ denotes the sum of all matrix elements, and the last equality assumes such relabeling of values of $a_j$ and $b_k$ that their products are equal to unity.
One readily recognizes that the matrix under the operator $\sum$ is just a real Hadamard matrix $H_2$ of order~$2$.
This observation suggests that instead of looking for a maximum in~\eqref{C_CHSH},
we can find a matrix with certain properties to obtain (or estimate) the classical value of the Bell inequality immediately.
We will formalize this procedure in the next sections.

\section{Bell Inequalities}
\label{sec:BI}

We assume a scenario of two parties ($A$ and $B$), $m$ measurement settings (for each party)
and $q$ outcomes for each measurement.
Let $P(a,b|x,y)$ denote a joint probability distribution of outcomes $a,b\in\{0,\dots,q-1\}$ for $A$ and $B$,
given measurement settings $x,y\in\{0,\dots,m-1\}$, respectively. The value of $P(a,b|x,y)$ can be obtained from a sequence
of identically prepared quantum states subjected to a measurement process.
As shown in~\cite{F82}, single correlation $P(a,b|x,y)$ does not provide a contradiction with LHV models, but
a linear combination of $P(a,b|x,y)$, which appears in a Bell inequality, might attain values
that are out of reach in any LHV models~\cite{Be64}.
Any Bell inequality can be defined in the following way~\cite{BCPSW14}
\begin{equation}\label{Bell}
\sum_{x,y=0}^{m-1}\sum_{a,b=0}^{q-1}S^{ab}_{xy}\,P(a,b|x,y)\leq\mathcal{C},
\end{equation}
where $S^{ab}_{xy}$ is a real-valued function and $\mathcal{C}$ is the maximal possible value that can be achieved
in a local deterministic theory\footnote{This means that there is
statistical independence between the results of $A$ and $B$ expressed as $P(a,b|x,y)=P(a|x)P(b|y)$,
and the deterministic character of the outcomes, say $P(a|x),P(b|y)\in\{0,1\}$, for every pair of measurement
settings $x,y\in\{0,...,m-1\}$ and outcomes $a,b\in\{0,...,q-1\}$, respectively for $A$ and $B$.}.
We call $\mathcal{C}$ a {\sl classical} value or a {\sl local hidden variable} (LHV) value.

Furthermore, the {\sl quantum value} $\mathcal{Q}$ denotes the maximal possible value of the left-hand side of~\eqref{Bell},
when optimization takes place over all possible joint probability distributions
allowed in quantum realm, provided that two parties $A$ and $B$ do not communicate (to exclude
the possibility of ``instantaneous'' transfer of information).
As described in Chapter~\ref{chap:DELTA},
quantum probabilities take the form $P(a,b,|x,y)=\mathrm{Tr}\big((\Pi_a^x\otimes\Pi_b^y) \rho_{AB}\big)$,
where $\{\Pi_a^x\}$ and $\{\Pi_b^y\}$ represent POVM\footnote{Positive-operator valued measure.}, and $\rho_{AB}$ is
a bipartite quantum state shared by parties $A$ and $B$.
Using discrete Fourier transform, we can write probabilities in~\eqref{Bell} as expectation
values of correlators~\cite{SATWAP17}
\begin{equation}
P(a,b|x,y)=\frac{1}{\sqrt{q}}\sum_{s,t=0}^{q-1}\omega^{as+bt} \langle A^s_x\otimes B^t_y\rangle,\label{FourierP}
\end{equation}
where $\omega=\exp\big\{\frac{2\pi i}{q}\big\}$.
Here, $A^s_x$ (similarly for $B^t_y$) denotes the $s^{\rm th}$ power of observable $A_x$ related to $A$.
Both observables are chosen to have $q$ eigenvalues, associated with $q$ different roots of unity. 
Now Eq.~\eqref{Bell} can be expressed as
\begin{equation}\label{BellM}
\sum_{x,y=0}^{m-1}\sum_{s,t=0}^{q-1}M_{ms+x,mt+y}\,\langle \\
A^{s}_{x}\otimes B^{t}_{y}\rangle\leq\mathcal{C},
\end{equation} 
where $M$ is a square matrix of size $N=qm$ with elements defined as
\begin{equation}
M_{ms+x,mt+y}=\frac{1}{\sqrt{q}}\sum_{a,b=0}^{q-1}\omega^{sa+tb}\,S^{ab}_{xy},\label{def_M_q}
\end{equation}
for every $0\leqslant s,t<q$ and $0\leqslant x, y < m$.

We must assert that the left-hand side of~\eqref{BellM} is real for any $q\geqslant 2$ and $m\geqslant 2$.
Provided that both $A^s_x$ and $B^t_y$ are unitary and imposing the following symmetry constraint
\begin{equation}
M_{m[q-s]_q+x,m[q-t]_q+y}=M^*_{ms+x,mt+y},\label{symm}
\end{equation}
for $0\leqslant s,t<q$ and $0\leqslant x,y< m$,
with $^*$ representing the complex conjugate and $[x]_q$ denoting $x$ modulo $q$,
the left-hand side of~\eqref{BellM} indeed remains real for any value of $q\geqslant 2$.
In particular, if $q=2$ (two outcome scenario) we have $[2-s]_2=s$, for any $s\in\{0, 1\}$.
This means, that condition~\eqref{symm} guarantees that $M$ is a real matrix for any value of $m$.
%and the left-hand side of~\eqref{BellM} is real for any $M$.
The constant prefactor $\frac{1}{\sqrt{q}}$ in~\eqref{def_M_q} is sometimes absorbed into $S_{xy}^{ab}$, but to remain consistent
with some mathematical results presented in~$[\hyperlink{\paperslist}{\rm A5}]$, one must explicitly adjust its value.

We distinguish a special class of matrices $M$ corresponding to Bell inequalities~\eqref{BellM}
without marginal terms, i.e. those for which $s=0$ or $t=0$. In other words, we 
take into account only such scenarios, where there is a correlation between parties $A$ and $B$ ($s>0$ and $t>0$),
so that we discard zeroth powers of observables $A_x$ and $B_y$.
We call such a matrix $M$ {\sl correlation matrix}.
Condition~\eqref{symm} implies that the marginal terms are always located at the first $m$ rows and first $m$ columns of matrix $M$.
What remains is the nontrivial {\sl core} of order $n=(q-1)m$.
Note the similarity to the core of the complex Hadamard matrix, defined in the context
of its dephased (normalized) form --- see Chapter~\ref{chap:CHM}.

For further convenience, in the real case ($q=2$), $M\in\mathbb{R}^{2m\times 2m}$,
as matrix $M$ fulfills condition~\eqref{symm} trivially, we can
remove all rows and columns from $M$ corresponding to the marginal terms, and consider only its core.
Again, let us take for example the case of CHSH. Corresponding matrix $M$ of order $N=qm=2\cdot 2=4$ has the following form
\begin{equation}
M=\frac{1}{\sqrt{2}}\left[
\begin{array}{c|c|c|c}
{\rm t}_{00}^{00}  & {\rm t}_{01}^{00}  & {\rm t}_{00}^{01} &      {\rm t}_{01}^{01}\\
\hline
{\rm t}_{10}^{00}  & {\rm t}_{11}^{00}  & {\rm t}_{10}^{01} &      {\rm t}_{11}^{01} \\
\hline
{\rm t}_{00}^{10} &  {\rm t}_{01}^{10}  & {\bf t_{00}^{11}} & {\bf t_{01}^{11}}\\
\hline
{\rm t}_{10}^{10} &  {\rm t}_{11}^{10}  & {\bf t_{10}^{11}} & {\bf t_{11}^{11}} 
\end{array}\right],
\end{equation}
where every matrix element, denoted by ${\rm t}_{xy}^{st}$ is an appropriate sum of products of powers of $\omega$ and $S_{xy}^{st}$
--- here all are real.
Therefore, only the core (of order $n=2$) 
entries marked in bold contribute to~\eqref{CHSH}, whereas all other elements
represent irrelevant numbers contributing to marginal terms.
In two examples that will be presented later, we will consider
only the core of real matrices $M$, hence the actual dimension $N=qm$ will be reduced
to $n=(q-1)m$, which for $q=2$ simplifies to $n=m$.
%the actual size $N$ of a real matrix $M$ used
%to induce particular Bell inequalities,
%reads $N=(q-1)m$, which reduces to $N=m$ for $q=2$. So, again, we will only consider the core of $M$.

In a general case, the matrix $M\in\mathbb{C}^{qm\times qm}$ obeying~\eqref{symm}
has a form of a tensor structure with $q^2$ blocks each of size $m\times m$.
For even values of $q=2$, $4$, $6$, ... it contains four $\mathbb{R}$-valued blocks of size $m\times m$ located
at $M_{0,0}$, $M_{0,qm/2}$, $M_{qm/2,0}$, and $M_{qm/2,qm/2}$ (matrix is indexed from $0$ to $N-1=qm-1$).
After removing marginal terms, the remaining $(q-1)^2$ blocks reveal a central symmetric structure with respect
to the $\mathbb{R}$-valued block at $(qm/2,qm/2)$ which now occupies the central place in $M$.
For example, for $(q,m)=(6,2)$ we have
% fancy braces around array: https://www.latex4technics.com/?note=35RA
\newcommand\coolover[2]{\mathrlap{\smash{\overbrace{\phantom{%
    \begin{matrix} #2 \end{matrix}}}^{\mbox{$#1$}}}}#2}
\newcommand\coolunder[2]{\mathrlap{\smash{\underbrace{\phantom{%
    \begin{matrix} #2 \end{matrix}}}_{\mbox{$#1$}}}}#2}
\newcommand\coolleftbrace[2]{%
#1\left\{\vphantom{\begin{matrix} #2 \end{matrix}}\right.}
\newcommand\coolrightbrace[2]{%
\left.\vphantom{\begin{matrix} #1 \end{matrix}}\right\}#2}
\begin{equation}
\vphantom{% phantom stuff for correct box dimensions
    \begin{matrix}
    \overbrace{XYZ}^{\mbox{$R$}}\\ \\ \\ \\ \\ \\
    \underbrace{pqr}_{\mbox{$S$}}
    \end{matrix}}%
M=\left[\begin{array}{cc|cc|cc|cc|cc|cc}
\rowcolor{gray!20}\mathbb{R} & \mathbb{R} & {\rm t}_{00}^{01} & {\rm t}_{01}^{01} & {\rm t}_{00}^{02} & {\rm t}_{01}^{02} & \mathbb{R} & \mathbb{R} & {\rm t}_{00}^{04} & {\rm t}_{01}^{04} & {\rm t}_{00}^{05} & {\rm t}_{01}^{05}\\
\rowcolor{gray!20}\mathbb{R} & \mathbb{R} & {\rm t}_{10}^{01} & {\rm t}_{11}^{01} & {\rm t}_{10}^{02} & {\rm t}_{11}^{02} & \mathbb{R} & \mathbb{R} & {\rm t}_{10}^{04} & {\rm t}_{11}^{04} & {\rm t}_{10}^{05} & {\rm t}_{11}^{05}\\
\hline
\cellcolor{gray!20}{\rm t}_{00}^{10} & \cellcolor{gray!20}{\rm t}_{01}^{10} & {\rm t}_{00}^{11} & {\rm t}_{01}^{11} & {\rm t}_{00}^{12} & {\rm t}_{01}^{12} & {\rm t}_{00}^{13} & {\rm t}_{01}^{13} & {\rm t}_{00}^{14} & {\rm t}_{01}^{14} & {\rm t}_{00}^{15} & {\rm t}_{01}^{15}\\
\cellcolor{gray!20}{\rm t}_{10}^{10} & \cellcolor{gray!20}{\rm t}_{11}^{10} & {\rm t}_{10}^{11} & {\rm t}_{11}^{11} & {\rm t}_{10}^{12} & {\rm t}_{11}^{12} & {\rm t}_{10}^{13} & {\rm t}_{11}^{13} & {\rm t}_{10}^{14} & {\rm t}_{11}^{14} & {\rm t}_{10}^{15} & {\rm t}_{11}^{15}\\
\hline
\cellcolor{gray!20}{\rm t}_{00}^{20} & \cellcolor{gray!20}{\rm t}_{01}^{20} & {\rm t}_{00}^{21} & {\rm t}_{01}^{21} & {\rm t}_{00}^{22} & {\rm t}_{01}^{22} & {\rm t}_{00}^{23} & {\rm t}_{01}^{23} & {\rm t}_{00}^{24} & {\rm t}_{01}^{24} & {\rm t}_{00}^{25} & {\rm t}_{01}^{25}\\
\cellcolor{gray!20}{\rm t}_{10}^{20} & \cellcolor{gray!20}{\rm t}_{11}^{20} & {\rm t}_{10}^{21} & {\rm t}_{11}^{21} & {\rm t}_{10}^{22} & {\rm t}_{11}^{22} & {\rm t}_{10}^{23} & {\rm t}_{11}^{23} & {\rm t}_{10}^{24} & {\rm t}_{11}^{24} & {\rm t}_{10}^{25} & {\rm t}_{11}^{25}\\
\hline
\cellcolor{gray!20}\mathbb{R} & \cellcolor{gray!20}\mathbb{R} & {\rm t}_{00}^{31} & {\rm t}_{01}^{31} & {\rm t}_{00}^{32} & {\rm t}_{01}^{32} & \mathbb{R} & \mathbb{R} & {\rm t}_{00}^{34} & {\rm t}_{01}^{34} & {\rm t}_{00}^{35} & {\rm t}_{01}^{35}\\
\cellcolor{gray!20}\mathbb{R} & \cellcolor{gray!20}\mathbb{R} & {\rm t}_{10}^{31} & {\rm t}_{11}^{31} & {\rm t}_{10}^{32} & {\rm t}_{11}^{32} & \mathbb{R} & \mathbb{R} & {\rm t}_{10}^{34} & {\rm t}_{11}^{34} & {\rm t}_{10}^{35} & {\rm t}_{11}^{35}\\
\hline
\cellcolor{gray!20}{\rm t}_{00}^{40} & \cellcolor{gray!20}{\rm t}_{01}^{40} & {\rm t}_{00}^{41} & {\rm t}_{01}^{41} & {\rm t}_{00}^{42} & {\rm t}_{01}^{42} & {\rm t}_{00}^{43} & {\rm t}_{01}^{43} & {\rm t}_{00}^{44} & {\rm t}_{01}^{44} & {\rm t}_{00}^{45} & {\rm t}_{01}^{45}\\
\cellcolor{gray!20}{\rm t}_{10}^{40} & \cellcolor{gray!20}{\rm t}_{11}^{40} & {\rm t}_{10}^{41} & {\rm t}_{11}^{41} & {\rm t}_{10}^{42} & {\rm t}_{11}^{42} & {\rm t}_{10}^{43} & {\rm t}_{11}^{43} & {\rm t}_{10}^{44} & {\rm t}_{11}^{44} & {\rm t}_{10}^{45} & {\rm t}_{11}^{45}\\
\hline
                              \cellcolor{gray!20}{\rm t}_{00}^{50} & \cellcolor{gray!20}{\rm t}_{01}^{50} & {\rm t}_{00}^{51} & {\rm t}_{01}^{51} & {\rm t}_{00}^{52} & {\rm t}_{01}^{52} & {\rm t}_{00}^{53} & {\rm t}_{01}^{53} & {\rm t}_{00}^{54} & {\rm t}_{01}^{54} & {\rm t}_{00}^{55} & {\rm t}_{01}^{55}\\
\coolunder{\text{marginals}}{\cellcolor{gray!20}{\rm t}_{10}^{50} & \cellcolor{gray!20}{\rm t}_{11}^{50}} & {\rm t}_{10}^{51} & {\rm t}_{11}^{51} & {\rm t}_{10}^{52} & {\rm t}_{11}^{52} & {\rm t}_{10}^{53} & {\rm t}_{11}^{53} & {\rm t}_{10}^{54} & {\rm t}_{11}^{54} & {\rm t}_{10}^{55} & {\rm t}_{11}^{55}\\
\end{array}\right].
\begin{matrix}% matrix for right braces
    \coolrightbrace{d \\ x}{\text{marginals}}\\
    {\color{white}\coolrightbrace{y \\ y \\ y \\ x \\ x \\ x \\ x \\ x \\ x \\ x}{\text{marginals}}}
\end{matrix}\label{Mqm62}
\end{equation}

\bigskip
\bigskip

\noindent After removing the marginal terms, 
and if we temporarily change the reference point and assume that the central block is located
at the relative position $(0,0)$, then all other blocks denoted by $b_{j,k}$ indexed by $j$ and $k$ are in relation $b_{j,k}=b^*_{-j,-k}$.
Therefore, the core of the matrix in~\eqref{Mqm62} of order $n=(q-1)m$ can be symbolically represented as
\begin{align}
{\rm core}\big(M\big)&=\left[\begin{array}{l|l|l|l|l}
b_{-2, 2} & b_{-1, 2} & b_{0, 2} & b_{1, 2} & b_{2, 2}\\
\hline
b_{-2, 1} & b_{-1, 1} & b_{0, 1} & b_{1, 1} & b_{2, 1}\\
\hline
b_{-2, 0} & b_{-1, 0} & b_{0, 0} & b_{1, 0} & b_{2, 0}\\
\hline
b_{-2,-1} & b_{-1,-1} & b_{0,-1} & b_{1,-1} & b_{2,-1}\\
\hline
b_{-2,-2} & b_{-1,-2} & b_{0,-2} & b_{1,-2} & b_{2,-2}\\
\end{array}\right]\\
&=\left[\begin{array}{l|l|l|l|l}
b_{-2, 2} & b_{-1, 2} & b_{0, 2} & b_{1, 2} & b_{2, 2}\\
\hline
b_{-2, 1} & b_{-1, 1} & b_{0, 1} & b_{1, 1} & b_{2, 1}\\
\hline
b_{-2, 0} & b_{-1, 0} & b_{0, 0} & b^*_{-1, 0} & b^*_{-2, 0}\\
\hline
b^*_{2,1} & b^*_{1,1} & b^*_{0,1} & b^*_{-1,1} & b^*_{-2,1}\\
\hline
b^*_{2,2} & b^*_{1,2} & b^*_{0,2} & b^*_{-1,2} & b^*_{-2,2}\\
\end{array}\right],
\end{align}
where each $b_{j,k}$ for $(j,k)\in [-2,2]\times[-2,2]$ is a $m\times m=2\times 2$ block
with one particular representative $b_{0,0}$ with all four real entries, being a center of symmetry
for relation~\eqref{symm}.

For odd values of $q\in\{3,5,7,...\}$ matrix $M$ has only one $\mathbb{R}$-valued block of size $m\times m$
located at the upper-leftmost corner of $M$ (at $(0,0)$), which lies exactly at the marginal region outside the core of $M$.
Remaining blocks exhibit the same central symmetry according to condition~\eqref{symm}.

\section{Excess of a Matrix $\Sigma$}
\label{sec:SigmaDef}

Recall the notion of (real) Hadamard matrix $H$, which is
a square $N\times N$ matrix
$H$ with entries $\mp1$ and pairwise orthogonal columns. We will use the symbol
${\rm B}\mathbb{H}(N,2)$ (introduced in Chapter~\ref{chap:CHM}) to indicate the set of real Hadamard matrices of size $N$.

Given a real Hadamard matrix $H$, the {\sl excess} of $H$, denoted by $\Sigma(H)$,
is defined as the difference between the number of the positive and negative entries of $H$~\cite{S73}.
For any dimension $N$, for which real Hadamard matrix exists, we consider the
{\sl maximal excess}, maximized over all Hadamard matrices of this order.
It is known~\cite{B77} that the maximal excess obeys the following bounds
\begin{equation}
\frac{N^2}{2^{N}}\binom{N}{N/2}\leq \max_{H \in {\rm B}\mathbb{H}(N,2)}\Sigma(H)\leq N\sqrt{N}.\label{best_bound}
\end{equation}
It can be shown~\cite{B77} that for a Hadamard matrix $H$, the upper bound~\eqref{best_bound} is attained
iff the sum of all elements in each row of $H$ yields the same value. We call such matrix
a {\sl regular} or {\sl constant row sum} Hadamard matrix. It is easy to prove
that such Hadamard matrices with a constant row sum exist only in squared dimensions $N=k^2$ for some $k>1$.
One can equivalently define the excess of a $H\in{\rm B}\mathbb{H}(N,2)$ as
the sum of all entries of $H$,
\begin{equation}
\Sigma(H)=\sum_{j,k=0}^{N-1}H_{jk} \quad : \quad H\in {\rm B}\mathbb{H}(N,2).%\label{Sigma}
\end{equation}

In analogy to~\eqref{equivalence_relation}, we say that two matrices $M_1$ and $M_2$ of order $N$ obeying~\eqref{symm}
are {\sl $q$-equivalent} if there exist %two permutation matrices $P_1$, $P_2\in\mathbb{P}(N)$ and
two diagonal unitary matrices
$D_1$ and $D_2$ whose $(ms+x)^{\rm th}$ diagonal entries are $s^{\rm th}$ power of a $q^{\rm th}$ root of the unity $\omega=\exp\{\frac{2\pi i}{q}\}$,
for every $s\in\{0,...,q-1\}$ and $x\in\{0,...,m-1\}$,
such that $M_1=D_1M_2D_2$. We write $M_1\simeq_q M_2$.
For example, the explicit form of the diagonal matrix for $q=3$ and $m=2$ reads
\begin{equation}
D_j  = {\rm diag}\left[1,1,\omega,\omega,\omega^2,\omega^2\right] = {\rm diag}\big[1,\omega,\omega^2\big]\otimes{\rm diag}\big[1,1\big]\ \ : \ \ j\in\{1,2\},
\end{equation}
where the emphasized tensor structure corresponds to the internal form of matrix $M_j$, $j\in\{1, 2\}$.
Observe that two $q$-equivalent matrices $M_1$ and $M_2$ induce the same Bell inequality~\eqref{BellM}
up to a relabeling of the outcomes for $A$ and $B$.
We do not include permutations (permutation matrices) in the definition of $q$-equivalence, because the
excess does not depend on the order of summands --- entries of a matrix. Such permutation matrices would correspond
to a relabeling of measurement settings, which are irrelevant in this scenario.

Finally, we extend the definition of the standard maximal excess over any matrices beyond Hadamard ones.
Given $q\geqslant 2$, we say that matrix $M$ of order $N=qm$ has {\sl maximal excess} with respect to $q$-equivalence, if
\begin{equation}
\Sigma(M)=\max\Big\{\Sigma\big(\widetilde{M}\big) : \widetilde{M}\simeq_q M\Big\}.
\end{equation}
Note that for any matrix $M$ related to a Bell inequality~\eqref{BellM},
there exists a relabeling of outputs that produces another inequality, related to a matrix $\widetilde{M}$, with maximal excess.
If $M$ is a real Hadamard matrix and $q=2$, above definition coincides with the standard maximal excess for Hadamard matrices.

\section{Excess of a Matrix and Classical Value of Bell Inequality}

There is a one-to-one link between the notion of excess of a matrix applied to a certain class of matrices and the classical (LHV)
value of Bell inequality (BI).

\begin{proposition}
If $M$ is a matrix of dimension $N=qm$ having maximal excess with respect to $q$-equivalence then,
the LHV value of Bell inequality~\eqref{BellM} induced by $M$, having $q$ outcomes for each setting,
equals to the excess of $M$, i.e. $\mathcal{C}(M)=\Sigma(M)$.
Conversely, for any Bell inequality~\eqref{BellM} induced
by $M$, there exists a $q$-equivalent matrix $\widetilde{M}$ satisfying~\eqref{symm} such that $\mathcal{C}(\widetilde{M})=\Sigma(\widetilde{M})$.\label{excessLHV}
\end{proposition}
This can be explained by noting that given a BI~\eqref{BellM} induced by $M$, there exist two diagonal matrices $D_1$, $D_2$ composed of
$q^{\rm th}$ roots of the unity such that a $q$-equivalent matrix $\widetilde{M}=D_1MD_2$ provides an
optimal LHV strategy if all outputs for $A$ and $B$ are equal to $+1$.
Changing $M\rightarrow \widetilde{M}$ in~\eqref{BellM} imposes a relabeling of outputs of the measurement devices for $A$ and $B$.
Hence, the LHV value of BI can be calculated as the excess of $\widetilde{M}$.
Clearly, for matrices $\widetilde{M}$ with this property we have $\Sigma\big(\widetilde{M}\big)=\Sigma(M)=\mathcal{C}$.

There are infinitely many known cases for which the maximum excess of Hadamard matrix can be calculated
exactly~\cite{FK87,KF88,KS91,K91,XXS03}. Also another classes of matrices having maximal
excess have been deeply examined: complex Hadamard matrices~\cite{KS93}, orthogonal designs~\cite{FG05},
and tensors~\cite{HLS78}.
This last class of objects provides a way of extension the area of research over multipartite systems.
This implies the importance of Proposition~\ref{excessLHV}, because we can simply
extend the set of BI for which the LHV value is known.
Supported by many mathematical results related to maximal excess,
we can construct families of matrices that allow us to achieve the LHV value
of Bell inequalities representing scenarios with arbitrary (unbounded)
number of measurement settings per party. Moreover, we can do this without applying any optimization procedure.
This connection might inspire researchers to develop more efficient methods to find the LHV value for other classes of Bell inequalities.
In the paper~$[\hyperlink{\paperslist}{\rm A5}]$ we summarize several known mathematical methods to calculate maximal excess
and we show how to construct some families of correlation matrices $M$ of order $n=(q-1)m$, which provide
classical value of BI for bipartite scenarios with unbounded number of measurement settings.
Meanwhile, let us present a sequence of additional
observations\footnote{Proofs of all observations and corollaries are provided in Appendix of~$[\hyperlink{\paperslist}{\rm A5}]$.} concerning the classical
value of a matrix $M$.

\begin{proposition}\label{Resupperr}
If a matrix $M$ of order $N=qm$ achieves the maximal excess with respect to $q$-equivalence,
then %the following upper bound holds for the induced Bell inequality
\begin{equation}\label{upperr}
\mathcal{C}(M)\leq Nr(M),
\end{equation}
where $r(M)$ is the numerical radius of the matrix $M$, defined as
\begin{equation}
r(M)=\max\big\{|\langle\psi|M|\psi\rangle| : |\psi\rangle\in\mathcal{H}_N, \langle\psi|\psi\rangle=1 \big\}. %=N\langle\phi|M|\phi\rangle}
\end{equation}
\end{proposition}
\noindent Note that the above upper bound for the classical value $\mathcal{C}$ is more restrictive than the upper
bound for the quantum value~\cite{EKB13,V15}, which is presented below.
\begin{observation}
\label{observation_Q}
The quantum value of a BI induced by $M$ satisfies
\begin{equation}\label{upperQ}
\mathcal{Q}(M)\leq N\sigma(M),
\end{equation}
where $\sigma(M)$ denotes the maximal singular value of $M$.
\end{observation}
\noindent In general, full classification of matrices $M$ that saturate the above bounds is an open problem.
Let us provide here a new
upper bound for bipartite scenario, for which we can fully characterize its tightness.
\begin{proposition}\label{resUpperC}
If the excess of a matrix $M$ of order $N=qm$ is maximal with respect to $q$-equivalence, then
\begin{equation}\label{uppernu}
\mathcal{C}(M)\leq \sqrt{N}\,\nu(M),
\end{equation}
where $\displaystyle{\nu(M)=\sqrt{\sum_{j=0}^{N-1}\Biggl|\sum_{k=0}^{N-1}M_{jk}\Biggr|^2}}$. Upper bound (\ref{uppernu})
is saturated iff the matrix $M$ has a constant row sum equal to $s$ and then $\mathcal{C}(M)=Ns$.\label{proposition_taxicab}
\end{proposition}
\noindent The constant row sum value $s$ presented above is always a positive real number provided that
$M$ has maximal excess and obeys the symmetry condition in~\eqref{symm}.

From Observation~\ref{observation_Q} and Proposition~\ref{proposition_taxicab} we conclude that for
a unitary matrix $M$ with constant row sum having maximal excess,
related BI does not have quantum advantage, $\mathcal{C}(M)=\mathcal{Q}(M)=N$.
For example, the three inequivalent not normalized Hadamard matrices of size $N=16$~\cite{H61} with constant row sum,
define three inequivalent bipartite BI with $m=16$ measurement settings and $q=2$ outcomes.
In this case, we have $\mathcal{C}(M)=\mathcal{Q}(M)=64=16\sqrt{16}$.

Non-unitary matrices with constant row sum can imply a correlation BI having quantum advantage. For instance,
a circulant matrix $M$ with
${\rm core(M)}={\rm circ}\left[0,-1, 1\right]$ of order three,\footnote{Note that in this case $q=2$ and $m=3$,
and the dimension $N=6$ is reduced to $n=3$, which is the size of the core of $M$. See Section~\ref{sec:BI}.}
satisfies the relation $\mathcal{C}(M)=4<3\sqrt{3}=\mathcal{Q}(M)$. The quantum value $\mathcal{Q}(M)$ is achieved
with assistance of the following  measurement settings
$A_j$ and $B_j$ (corresponding to parties $A$ and $B$, respectively) of the form of $X_j D X^{\dagger}_j$, for
\begin{equation}
X_j=\left[\begin{array}{rr}
\cos\varphi_j & \sin\varphi_j\\
-\sin\varphi_j  &  \cos\varphi_j
\end{array}\right],\quad
D=\left[\begin{array}{rr}
1 & 0\\
0 & -1
\end{array}\right],
\end{equation}
and $\varphi_j=0$, $\dfrac{2}{3}\pi$, $\dfrac{1}{3}\pi$ for $A_0$,
$A_1$, $A_2$ and $\varphi_j=\dfrac{1}{4}\pi$, $\dfrac{7}{12}\pi$, $\dfrac{11}{12}\pi$ for $B_0$, $B_1$, $B_2$,
respectively.

%%%%%%%%%%%%%%%%%%%%%%%%%%%%%%%%%%%%%%%%%%%%%%%%%%%%%%%%%%%%%%%%%%%%%%%%%%%%%%%%%%%%%%%%

\subsection{Family of Bell Inequalities with Quantum Advantage}
\label{sec:QA_example}

Further investigations led us to a more extended example and a tentative statement concerning
the existence of infinite family of constant row sum matrices with a quantum advantage.
We did not actually prove that this statement holds in every dimension,
but detailed numerical analysis suggests this might be true.\footnote{In this example we also consider the core of a real matrix
but we do not write it explicitly.}

Consider $n$ two-dimensional measurement operators
$A_j=X(\alpha_j)$ and $B_j=X(\beta_j)$, where
\begin{equation}
X(\varphi_j)=\left[
\begin{array}{rr}
\cos \dfrac{2\pi  \varphi_j}{n} &  -\sin \dfrac{2\pi \varphi_j}{n}\\
\\
-\sin \dfrac{2\pi \varphi_j}{n} & -\cos \dfrac{2\pi \varphi_j}{n}\\
\end{array}\right],
\end{equation}
so that each operator depends on a single phase $\varphi_j$, which corresponds to $n\alpha_j/2\pi$ or $n\beta_j/2\pi$ for $j\in\{1,2,..., n\}$.
For further brevity, we collectively write all the phases of $A_j$ and $B_j$ as vectors
\begin{equation}
\alpha^{(n)}=\left(\alpha_1,\alpha_2,...,\alpha_n\right) \qquad \text{and} \qquad
\beta^{(n)}=\left(\beta_1,\beta_2,...,\beta_n\right)\label{phase_vector},
\end{equation}
where $\alpha_j$, $\beta_j\in\mathbb{R}$.
%All phases are rescaled by $\pi/n$, which, in the following examples, will allows us to cast them onto the integer domain.
Hence, we have the following scenario: $m=n$ measurement settings and $q=2$ outputs per party.

Define a family of non-unitary real circulant matrices $M_n$ of size $n$ taking the form of
\begin{equation}
M_n=\texttt{circ}[c_1,c_2,...,c_n]=\texttt{circ}\Big[\underbrace{-1, ..., -1}_{\lfloor n/2\rfloor},\underbrace{1, ..., 1}_{\lceil n/2\rceil}\Big]\in\mathbb{R}^{n\times n}.\label{circulant_family}
\end{equation}
For example:
\begin{align}
M_3&=\texttt{circ}\Big[-1,1,1\Big]\label{circ_3}\\
M_4&=\texttt{circ}\Big[-1,-1,1,1\Big]\\
M_5&=\texttt{circ}\Big[-1,-1,1,1,1\Big]\\
M_6&=\texttt{circ}\Big[-1,-1,-1,1,1,1\Big]\label{circ_6}\\
&\vdots\nonumber
\end{align}
By definition a circulant matrix, $M_n$ has a constant row sum.
Using a well known formula for eigenvalues $\lambda_j$ of a circulant matrix~\cite{CGTV21},
\begin{equation}
\lambda_{j}=\sum_{k=1}^{n}c_k\omega_n^{(j-1)k} \quad : \quad \omega_N=\exp\Big\{\frac{2\pi i}{n}\Big\},\quad j\in\{1,...,n\},
\end{equation}
we can calculate the maximal singular value for even orders of $M_n$,
\begin{equation}
\sigma_n = |\lambda_n| = \frac{2}{\sin\frac{\pi}{n}},
\end{equation}
from which we get the limit
\begin{equation}
\lim_{n\to\infty} \frac{\sigma_n(M_n)}{n}=\frac{2}{\pi}\label{sigma_Mn}.
\end{equation}
Similar calculations can be done for odd values of $n$.

From numerical simulations, we obtain the following sequence of classical values of the Bell operator induced by a matrix $M_n$ of order $n$,
\begin{equation}
\begin{array}{r|c|c|c|c|c|c|c|c|c|c|c|c|c|c|c|c}
                           n & 3 & 4 &  5 &   6 &  7  &   8 &  9  & 10  & 11 & 12 & 13 & 14   & 15 & 16 & 17 & ...\\
\hline
\mathcal{C}(M_n) & 5 & 8 & 13 & 20 & 25 & 32 & 41 & 52 & 61 & 72 & 85 & 100 & 113     &    128  & 145 & ...
\end{array}
\end{equation}
One can see that numbers $\mathcal{C}(M_n)$ form a logical pattern\footnote{Actually, this is only an observation
and we cannot present yet a formal proof of this fact.}.
To see this better, we shall divide the sequence $\mathcal{C}(M_n)$ onto two subsequences for odd and even values of the parameter $n$,
\begin{align}
\mathcal{A}&=(5,13,25,41,61,85,113,145,...),\label{classic_A}\\
\mathcal{B}&=(8,20,32,52,72,100,128,...).\label{classic_B}
\end{align}
Such splitting is additionally justified by the natural asymmetric form of $M_n$ depending on the parity of dimension $n$,
or equivalently, the number of $-1$'s in $M_n$. See Eqs.~\eqref{circ_3}--\eqref{circ_6}.
Clearly, $\mathcal{A}$ is a progression with increasing difference $\delta=8+4k$ with $k\in\mathbb{N}_0$.

As for the second subsequence $\mathcal{B}$, one observes that every two consecutive numbers differ by $12,12,20,20,28,28,..., 12+8k,...$ for $k\in\mathbb{N}_0$, respectively.
One can split it further onto other two subsubsequences to see that they can be expressed as $\mathcal{B}_{\rm odd}=(8,32,72,128,...)$ so $\mathcal{B}_{\rm odd}(k)=8k^2$ and $\mathcal{B}_{\rm even} = (20,52,100,...)$ which implies $\mathcal{B}_{\rm even}(k)=4\mathcal{A}(k)$.

To calculate the quantum value $\mathcal{Q}$ of the Bell inequality induced by the matrix $M_n$, we should optimize the quantity
$\langle\psi|\mathbb{B}|\psi\rangle$, where $\mathbb{B}$ represents the left-hand side of~\eqref{BellM} and 
we choose the particular form of the state
$|\psi\rangle=|\psi(\gamma)\rangle=[\cos\gamma,0,0,\sin\gamma]^{\rm T}$,
which depends on a single parameter $\gamma$.
Clearly, the Bell operator depends on phases $\alpha^{(n)}$ and $\beta^{(n)}$, which were
introduced in~\eqref{phase_vector}.
It turns~out that we can find such (optimal) values of $\alpha^{(n)}$, $\beta^{(n)}$, and $\gamma$, that
$\langle\psi|\mathbb{B}|\psi\rangle=n\sigma_n$.
Moreover, we can simplify the problem fixing $\gamma=\pi/4$
and $\alpha^{(n)}=(0,1,...,n-1)$
since,
observing results from many simulations, such configuration provided exceptionally elegant outputs for any dimension $3\leqslant n\leqslant 14$.

Two classes of results presented in Table~\ref{tab:phases} can be distinguished.
For $n=3+4k$ or $n=4+4k$ and $k\in\mathbb{N}_0$, we see that
phases for $B_j$ always take half-integer values (highlighted rows),
while for complementary values of $n$, there are only pure integers for both $\alpha^{(n)}$ and $\beta^{(n)}$.
Perfect monotonic alignment in each numerical simulation
allows us to present the following compact formulas:
\begin{equation}
\begin{cases}
\alpha^{(n)} &= \ \ (0,1,2,..., n-1)\\
\beta^{(n)} &= \ \ \big(2k+3,(2k+3)\oplus_{2n}2,(2k+3)\oplus_{2n}4,...\big)\frac{1}{2}
\end{cases}
\end{equation}
for $n=3+4k$ or $n=4+4k$ and
$k\in\mathbb{N}_0$ where $\oplus_{2n}$ denotes addition modulo $2n$. Similarly,
\begin{equation}
\begin{cases}
\alpha^{(n)} &= \ \ (0,1,2,..., n-1)\\
\beta^{(n)} &= \ \ (2+k,(2+k)\oplus_n1,(2+k)\oplus_n2,...)
\end{cases}
\end{equation}
for $k\in\mathbb{N}_0$ and remaining values of $n$.

\begin{table}[ht!]
\center
\begin{tabular}{r|rrrrrrrrrrrrrr}
%\begin{array}{r|rrrrrrrrrrrrrr}
\hline
\rowcolor{gray!20}
  $\alpha^{(3)}=$ & 0 & 1 & 2\\
\rowcolor{gray!20}
 $2\beta^{(3)}=$ & 3 & 5 & 1\\
 \hline
\rowcolor{gray!20}
  $\alpha^{(4)}=$ & 0 & 1 & 2 & 3\\
\rowcolor{gray!20}
 $2\beta^{(4)}=$ & 3 & 5 & 7 & 1\\
\hline
  $\alpha^{(5)}=$ & 0 & 1 & 2 & 3 & 4\\
  $b^{(5)}=$ & 2 & 3 & 4 & 0 & 1\\
 \hline
  $\alpha^{(6)}=$ & 0 & 1 & 2 & 3 & 4 & 5\\
  $\beta^{(6)}=$ & 2 & 3 & 4 & 5 & 0 & 1\\
\hline
\rowcolor{gray!20}
  $\alpha^{(7)}=$ & 0 & 1 & 2 & 3 & 4 & 5 & 6\\
\rowcolor{gray!20}
 $2\beta^{(7)}=$ & 5 & 7 & 9 & 11& 13& 1 & 3\\ 
\hline
\rowcolor{gray!20}
  $\alpha^{(8)}=$ & 0 & 1 & 2 & 3 & 4 & 5 & 6 & 7\\
\rowcolor{gray!20}
 $2\beta^{(8)}=$ & 5 & 7 & 9 & 11& 13& 15& 1 & 3\\
\hline
  $\alpha^{(9)}=$ & 0 & 1 & 2 & 3 & 4 & 5 & 6 & 7 & 8\\
  $\beta^{(9)}=$ & 3 & 4 & 5 & 6 & 7 & 8 & 0 & 1 & 2\\ 
\hline
  $\alpha^{(10)}=$ & 0 & 1 & 2 & 3 & 4 & 5 & 6 & 7 & 8 & 9\\
  $b^{(10)}=$ & 3 & 4 & 5 & 6 & 7 & 8 & 9 & 0 & 1 & 2\\
\hline
\rowcolor{gray!20}
  $\alpha^{(11)}=$ & 0 & 1 & 2 & 3 & 4 & 5 & 6 & 7 & 8 & 9 &10\\
\rowcolor{gray!20}
 $2\beta^{(11)}=$ & 7 & 9 & 11 & 13 & 15 & 17 & 19 & 21 & 1 & 3 & 5\\
\hline
\rowcolor{gray!20}
  $\alpha^{(12)}=$ & 0 & 1 & 2 & 3 & 4 & 5 & 6 & 7 & 8 & 9 & 10& 11 \\
\rowcolor{gray!20}
 $2\beta^{(12)}=$ & 7 & 9 & 11 & 13 & 15 & 17 & 19 & 21 & 23 & 1 & 3 & 5\\
\hline
  $\alpha^{(13)}=$ & 0 & 1 & 2 & 3 & 4 & 5 & 6 & 7 & 8 & 9 & 10 & 11 & 12\\
  $\beta^{(13)}=$ & 4 & 5 & 6 & 7 & 8 & 9 & 10 & 11 & 12 & 0 & 1 & 2 & 3\\
\hline
  $\alpha^{(14)}=$ & 0 & 1 & 2 & 3 & 4 & 5 & 6 & 7 & 8 & 9 & 10 & 11 & 12 & 13  \\
  $\beta^{(14)}=$ & 4 & 5 & 6 & 7 & 8 & 9 & 10 & 11& 12 & 13 & 0 & 1 & 2 & 3\\
\hline
\vdots & \vdots
%\end{array}
\end{tabular}
\caption{Optimal vectors of phases $\alpha^{(n)}$ and $\beta^{(n)}$ introduced in~\eqref{phase_vector}
for measurement operators for dimension $3\leqslant n\leqslant 14$.
Each row should be understood as components of vectors $\alpha^{(n)}$ or $\beta^{(n)}$.
Phases for $B_k$ in every second pair of dimensions, $(3,4),(7,8),(11,12),...$, are characterized by half-integer values,
in contrary to other values of $n$.
%For other values of $N$, for both operators $A_j$ and $B_k$, only integer values emerge.
}
\label{tab:phases}
\end{table}

We claim that for any $n\geqslant 3$, matrices $M_n$~\eqref{circulant_family} give rise to a Bell inequality,
such that it provides a quantum advantage as the quantum value reads
\begin{equation}
\mathcal{C}(M_n) < \mathcal{Q}(M_n)=n\sigma_n.
\end{equation}
This fact can be easily inferred from the following observation.
In the bipartite scenario with $m=n$ and $q=2$, the Bell operator $\mathbb{B}$ takes a very simple form depending only on
a single real parameter~$c$, 
\begin{equation}
\mathbb{B}=\mathbb{B}(c)=\left[
\begin{array}{rrrr}
c & 0 & 0 & c\\
0 & -c & c & 0\\
0 & c & -c & 0\\
c & 0 & 0 & c
\end{array}\right]
\end{equation}
where
\begin{equation}
c=\sum_{j,k=1}^n \big[M_n\big]_{jk}\cos \frac{2\pi}{n} \alpha_j\cos \frac{2\pi}{n}\beta_k.
\end{equation}
%with $M_n$ denoting associated constant row sum matrix.
Simple, although technical proof of this fact requires elementary trigonometry and, 
provided that $M_n$ is of the form of~\eqref{circulant_family},
one can show that for even values of $n$,
the maximal singular value of $\mathbb{B}$ reads
\begin{equation}
\sigma_n(\mathbb{B})=2c=\frac{2}{\sin\frac{\pi}{n}}.
\end{equation}
Comparing this with classical sequences~\eqref{classic_A} and~\eqref{classic_B},
we observe strict inequality $\mathcal{C}<\mathcal{Q}$ for every $n\geqslant 3$.
Similar reasoning can be done for odd values of $n$.

All of this, together with~\eqref{sigma_Mn}, allows us to estimate the asymptotic behavior of the quantum advantage $Q$,
\begin{align}
Q=\lim_{n=2k+1\to\infty}\frac{\mathcal{Q}(M_n)}{\mathcal{A}(M_n)}&%=\lim_{k\to\infty}\frac{(2k+1)\sigma_{2k+1}}{2k(k+1)+1}
=\frac{2}{\pi}\lim_{k\to\infty}\frac{(2k+1)^2}{2k(k+1)+1}=\\
\lim_{n=4k\to\infty}\frac{\mathcal{Q}(M_n)}{\mathcal{B}(M_n)}&%=\lim_{k\to\infty}\frac{4k\sigma_{4k}}{8k^2}
=\frac{2}{\pi}\lim_{k\to\infty}\frac{(4k)^2}{8k^2}=\\
\lim_{n\to\infty}\frac{\mathcal{Q}(M_n)}{\mathcal{C}(M_n)}&=\frac{4}{\pi}.
\end{align}
Numerical value of the ratio $Q=\mathcal{Q}/\mathcal{C}$ can be compared with the maximal bound obtained in the CHSH scenario
\begin{equation}
\frac{4}{\pi}\approx 1.2732< 1.4142 = \sqrt{2}=\frac{\mathcal{Q}_{\rm CHSH}}{\mathcal{C}_{\rm CHSH}}.
\end{equation}

\section{Excess and Mutually Unbiased Bases}
In this section, we show that the classical value of the Bell inequality induced by a correlation
matrix $M$ with a core equal to a Hadamard matrix is closely related
to the notion of MUB (mutually unbiased bases)~\cite{DEBZ10}.
We start with the following observation based on Theorem 3 in~\cite{B77} and \cite{DBV17} (where a method
of reducing the number of dichotomic variables involved in calculations of the maximal excess of a Hadamard matrix was introduced).
\begin{observation}
\label{C_mub1}
The LHV value of a BI with two outcomes $\mp1$ induced by a correlation matrix $M$ with a core, written ${\rm core}(M)$, of order $n=(q-1)m$ is given by
\begin{equation}\label{optreduce}
\mathcal{C}(M)=\max_{x\in\{-1,1\}^n}\|{\rm core}(M)|x\rangle\|_1,
\end{equation}
where $\displaystyle{\Big\|\sum_{j=0}^{n-1} v_j|j\rangle\Big\|_1=\sum_{j=0}^{n-1}|v_j|}$ is the taxicab norm of a vector.
\end{observation}
\noindent From this observation we infer
\begin{proposition}
\label{C_mub2}
Let $M$ be a correlation matrix with a core being a Hadamard matrix of order $n=(q-1)m$. If there exists a vector $|x\rangle$ with entries
$\pm1$, which is unbiased to the rows of $M$, then the classical value
$\mathcal{C}(M)=n\sqrt{n}$. Conversely, if $\mathcal{C}(M)=n\sqrt{n}$, then there exists a vector $|x\rangle$ unbiased
to the rows of $M$. 
\end{proposition}
\noindent In particular, if $M$ is a correlation matrix, ${\rm core}(M)$ has constant row sum,
and a vector $|x\rangle\in\mathbb{R}^{n}$ has all its entries equal to unity,
then $\mathcal{C}(M)=\sqrt{n}\nu(M)$, in accordance with Observation~\ref{resUpperC}. %Note that we introduced $n$ for a size of the core of correlation matrix.

\begin{corollary}
If $\forall\,k : n\neq k^2$ (not a square number that corresponds to the size of core of $M$), then any pair of real mutually unbiased bases in $\mathbb{R}^n$
is strongly unextendible (there is no single vector unbiased to the pair of bases~\cite{HMBRBC21, MBGW14, Th16}).
\end{corollary}

\begin{corollary}
Suppose, there exist three real MUB $\{\mathbb{I},H_1,H_2\}$ in $\mathbb{R}^n$.
Then both $H_1$ and $H_2$ are $2$-equivalent to a Hadamard matrix with a constant row sum.
\end{corollary}

\begin{corollary}
For any pair of real MUB $\{\mathbb{I},H\}$, with $H$ being a Hadamard matrix with a constant row sum,
there exists at least a single vector mutually unbiased with respect to both bases.
\end{corollary}

One readily notices that in dimensions $n=2$, $8$, and $12$, which are not squared numbers,
no triplets of real MUB exist. There are no Hadamard matrices of this size with a constant row sum either.

%%%%%%%%%%%%%%%%%%%%%%%%%%%%%%%%%%%%%%%%%%%%%%%%%%%%%%%%%%%%%%%%%%%%%%%%%%%%%%%%%%%%%%%%%%%%%%%%%%%%%%%%%%%%%%%
\section{Tight Bell Inequalities}

{\sl Tight Bell inequalities} are defined as the facets (hyperplanes that completely characterize
correlations compatible with a LHV model) of the LHV polytope~\cite{AFKKPLA}.
In the quantum case, the space of correlations for a two-party scenario with $m$ measurement
settings and $q$ outcomes, is defined by all real vectors $v\in\mathbb{R}^{q^2m^2}$ with entries
of the form of $\langle A^s_x\otimes B^t_y\rangle$, for $x,y\in\{0,\dots,m-1\}$ and $s,t\in\{0,\dots,q-1\}$.
Unitary operators $A_x$ and $B_y$ have eigenvalues being $q^{\rm th}$ roots of the unity.
Taking into consideration the no-signaling conditions~\cite{PR94},
correlations are further restricted to a subspace of size $d=(q-1)^2m^2$.
Facets of the LHV polytope have dimension of $d-1$, so for $q=2$ (two outcomes scenario) there are 
$d-1=m^2-1$ linearly independent vectors ($v$), associated to LHV strategies $\langle A^s_x\otimes B^t_y\rangle=a^s_xb^t_y$, with $a_x,b_y=\pm1$.
We examined tightness of correlation BI induced by Hadamard matrices $H={\rm core}(M)$ of orders $2\leqslant n\leqslant 20$ (for some $M$),
and the numerical results, summarized in Table~\ref{tab:tight_BI}, suggest the following conjecture:
\begin{conjecture}
A Hadamard matrix of order $n=(q-1)m$ induces a tight correlation Bell inequality with $m=n$ settings and $q=2$, iff it is
not $2$-equivalent to matrix with a constant row sum.
\end{conjecture}

\begin{table}[ht!]
\begin{tabular}{l|l|l|l}
\# settings $(m)$ & \# vertices & \# affine independent vertices &  tightness\\
 \hline
 2 & 4 & 3 & tight\\ 
 4$^{\rm crs}$ & 4 & 3 & non-tight\\ 
 8 & 64 & 63&{\bf tight}$^{\star}$\\ 
 12 &2640&143&{\bf tight}$^{\star}$\\
 16$_{[1]}^{\rm crs}$ &896&105&non-tight\\
 16$_{[2]}^{\rm crs}$ &192&81&non-tight\\
 16$_{[3]}^{\rm crs}$ &64&45&non-tight\\
 16$_{[4]}$ &21504&255&{\bf tight}$^{\star}$\\
 16$_{[5]}$ &21504&255&{\bf tight}$^{\star}$\\
 20$_{[1]}$ &20064&399&{\bf tight}$^{\star}$\\
 20$_{[2]}$ &20064&399&{\bf tight}$^{\star}$\\
 20$_{[3]}$ &20064&399&{\bf tight}$^{\star}$\\
\end{tabular}
\caption{Study of tightness for bipartite correlation Bell inequalities with $m$ measurement settings and $q=2$ outcomes
of the form of $\displaystyle{\Big\langle\sum_{x,y=0}^{n-1}H_{xy}A_x\otimes B_y\Big\rangle\leq\mathcal{C}(H)}$,
where $H$ denotes a Hadamard matrix of order $n=(q-1)m=m$. Note that $H={\rm core}(M)$ and $M$ has dimension $N=qm$.
For a tight Bell inequality, the number of affine independent vertices must be equal to $N^2-1$.
Symbol $^{\rm crs}$ tells that a matrix $H$ has a constant row sum and subscripts in brackets
enumerate inequivalent Hadamard matrices.
All cases denoted by $^{\star}$ are believed to be new.}
\label{tab:tight_BI}
\end{table}

%%%%%%%%%%%%%%%%%%%%%%%%%%%%%%%%%%%%%%%%%%%%%%%%%%%%%%%%%%%%%%%%%%%%%%%%%%%%%%%%%%%%%%%%%%%%%%%%%%%%%%%%%%%
%%%%%%%%%%%%%%%%%%%%%%%%%%%%%%%%%%%%%%%%%%%%%%%%%%%%%%%%%%%%%%%%%%%%%%%%%%%%%%%%%%%%%%%%%%%%%%%%%%%%%%%%%%%
%%%%%%%%%%%%%%%%%%%%%%%%%%%%%%%%%%%%%%%%%%%%%%%%%%%%%%%%%%%%%%%%%%%%%%%%%%%%%%%%%%%%%%%%%%%%%%%%%%%%%%%%%%%

\newpage

\section{Summary}

In this chapter, we have established a relation between the classical (LHV) value of bipartite Bell inequalities
and the mathematical notion of the excess of a matrix. 
This one-to-one connection can be used to obtain the LHV value for infinitely many families of
Bell inequalities with an arbitrarily high number of measurement settings per party
without solving any optimization problem.
We derived upper bounds for the LHV value for a bipartite BI, which are stronger than the upper bounds for the quantum value.

Furthermore, we described a relation of some classes of BI to mutually unbiased bases.
Full set of optimal LHV strategies of BI, which are induced by a Hadamard matrix $H$
coincide with the entire set of vectors, which are mutually unbiased to the pair $\{\mathbb{I},H\}$.

Studying LHV strategies can identify tight BI.
We showed seven tight BI for the scenario with two outcomes and a number of measurement settings between $8$ and $20$.
This led us to a conjecture that %, which states that
every Hadamard matrix that is not equivalent to the one with a constant row sum, induces a tight Bell inequality.
To the best of our knowledge, presented observations and results are new and should provide new insights in the field of quantum nonlocality.

%\newpage
%\thispagestyle{empty}

\begin{figure}[ht!]
\center
\includegraphics[width=0.5in]{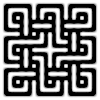}
\end{figure}

%%%%%%%%%%%%%%%%%%%%%%%%%%%%%%%%%%%%%%%%%%%%%%%%%%%%%%%%%%%%%%%%%%%%
\clearpage\null\thispagestyle{empty}
\chapter{Concluding Remarks}

Quantum entanglement, no matter how accessibly expressed, either by means of matrices with
a particularly designed structure or in any other mathematical language, still remains a
mysterious and elusive (although quite practical) property of the quantum world.
On the other hand, the analytical classification of certain classes of unitary matrices is far
beyond the reach of current theoretical and numerical tools due to their enormous complexity. 
The objective of this Thesis was twofold. 
Firstly, we wanted to broaden our knowledge of the analytical treatment of a set of unitary matrices
and their significant subsets. 
Secondly, we accepted the challenge of improving the description of particular aspects of quantum entanglement, especially
those concerning correlations among multipartite physical systems associated with states,
the existence of which was questioned by many experts.
We are convinced that in both cases, the goal was achieved.

\medskip

Let us recapitulate the main results presented in this work.
In the second chapter (\ref{chap:CHM}), we started with the problem of classification
of complex Hadamard matrices (CHM) which is a long-term project, launched in 2006~\cite{TZ06}.
The first unresolved case in the set of CHM concerns the six-dimensional case and all
consecutive dimensions are pending for a complete description too.
We provided two new important analytical representatives of eight- and nine-dimensional CHM.
Former one being a nontrivial one-parameter nonaffine family of inequivalent matrices,
while the latter one is isolated.
Based on the particularly interesting form of the nine-dimensional solution,
we proposed a general construction in any dimension $N\geqslant 8$.
This observation allowed us to combine the notion of Latin squares (or circulant structures)
with the set of CHM. Supported by the numerical evidence, we
drew two conjectures concerning the existence of infinitely many isolated CHM. %in any dimension greater than five.
We also resolved some other minor issues regarding the analytical description of CHM in low dimensions.

In Chapter~\ref{chap:DELTA}, we developed the idea of the restricted defect of a Hermitian unitary matrix,
which can be considered as an extension of the notion of the defect of an arbitrary unitary matrix, extensively used in work on
classification of CHM.
While the standard unitary defect can serve as a simple criterion to distinguish
isolated and potentially non-isolated (CHM) matrices, its Hermitian counterpart
is applicable to a subset of Hermitian unitary matrices having constant diagonal.
Having this, we were able to identify the internal structure of quantum measurement operators
in terms of the possibility of introduction of free parameters. 
We checked several classes of quantum measurements with prescribed symmetry: mutually unbiased bases,
equiangular tight frames and symmetric informationally complete positive-operator valued measures.
It turned out that in several cases such objects reflect isolated character, but it is also possible
to find examples for which the restricted defect does not vanish so that they
might be extended to multidimensional orbits of solutions labeled by free parameters.
Relatively large robustness of defect with respect to numerical inaccuracies
made it possible to calculate the value of the defect for a wide range of complex Hadamard matrices.
These results contribute to our understanding of the complicated structure of the set of complex
Hadamard matrices for intermediate dimensions $8\leqslant N\leqslant 16$.

The fourth chapter~(\ref{chap:AME46}) opens the second part of the Thesis where we directly
focused on (multipartite) quantum entanglement.
After many attempts, checking numerous possibilities, drawing a dozen of (often false) conjectures, we arrived
with an affirmative answer to the long-standing question concerning the existence of
absolutely maximally entangled states of four parties,
each having six degrees of freedom. In view of many facts pointing out that such a
state apparently does not exist, this unexpected result solved several problems at once.
This includes the existence of $2$-unitary matrices of size $36$ for which
very particular realignment of entries does not affect unitarity (these matrices maximize entangling power), perfect tensors of four indices,
a pair of orthogonal quantum Latin squares of size six (a quantum counterpart of the classical Euler problem of $36$ officers),
and special classes of pure quantum error correcting codes designed for the alphabet consisting of six letters.
Several other interesting properties of such state were investigated.

In the last chapter~(\ref{chap:EXCESS}), we briefly reviewed the concept
of combining two seemingly distinct areas: excess of a matrix and bipartite Bell inequalities.
Excess of a real matrix defined as the sum of all its elements
provides a way for introducing new families of the Bell inequalities.
This correspondence allows us to calculate the classical value for infinitely many Bell inequalities
without applying any optimizations.
In some cases we obtained stronger results than these offered by currently known bounds.
Relation to mutually unbiased bases was described and
most likely new results regarding tight Bell inequalities were presented.

\medskip

We hope that these modest results and propositions will find some place
in the contemporary theoretical and experimental fields of physics and will be used to develop even more interesting
theories in the vividly developing branches of quantum information, quantum combinatorics, quantum computation and nonlocality.
\newpage
Lastly, we would like to emphasize something.
When looking at the mentioned at the beginning two lists with ``the hottest'' open problems~\cite{KCIKProblems,TenAnnoyingProblems}
one recognizes a common thing. Almost every problem can be expressed in a simple language
which does not require extraordinary expert knowledge to comprehend the overall idea.
Some people say that all low hanging fruits have already been harvested. That might not necessarily be true.
Anyway, we believe, that even nowadays there is a niche that permits us to use some simplest methods in order
to accomplish long-awaited or just new results in many areas. We witnessed this phenomenon
mainly in Chapter~\ref{chap:AME46} but also in Chapters~\ref{chap:CHM} and~\ref{chap:EXCESS},
where by means of relatively easy tools or observations we arrived at moderately important conclusions.

All numerical problems presented in this Thesis were calculated using:
{\sl GNU Octave, version 5.2.0}, {\sl Matlab 9.1.0.441655 (R2016b)}
and occasionally
{\sl Mathematica 12.0.0.0}, which was mostly used to generate pictures and to confirm outputs from the previous software.
Most {\sl *.m} scripts and {\sl *.nb} notebooks are publicly available on the GitHub platform~\cite{GITHUB}.

\section{Open Questions and Future Work}

Many problems presented in this Thesis are far from being complete.
At the end of three first chapters we listed a sequence of open questions associated
with a given topic.
Let us recall the most important problems that we plan to continue to work on.

First of all, the paper concerning isolated complex Hadamard matrices $[\hyperlink{\paperslist}{\rm A2}]$, which is the scaffolding of the second chapter,
is currently being prepared in parallel with the Thesis. It should be released by the end of 2021
independently with another paper being a continuation of exploring mathematical aspects of the excess of non-Hadamard matrices, which is
in some sense a mathematical supplement to $[\hyperlink{\paperslist}{\rm A5}]$ containing physical aspects of excess.
Preprints~$[\hyperlink{\paperslist}{\rm A4}]$ and~$[\hyperlink{\paperslist}{\rm A5}]$ are waiting for the referees' decision.

Remaining problems, which are going to be considered in the next stage,
focus mainly on further development of the ideas described in the second part of the Thesis.
They are listed below:
\begin{itemize}
\item Does real AME$(4,6)$ state exist?
There are many clues that such state does not exist. But, a year and half ago we had twice as many clues that a complex AME$(4,6)$ state
should not exist and we were ready to publish a paper with very convincing arguments supporting such a statement!
\item It is roughly possible to introduce strictly analytical and deterministic algorithm to
construct an AME$(4,6)$ state, however at this stage it is
very cumbersome and suffers from many unclear statements.
How to fix and simplify this process and, moreover, how to adapt it for other dimensions,
for which other AME states are expected to be found?
\item Complex AME$(4,6)$ state can be realized experimentally or simulated on the quantum computer (suppose there exists such advanced
simulator). What is the most optimal decomposition of this state onto smaller/universal gates?
\item Do four or more mutually unbiased bases exist in dimension six? What about a complete set of seven MUB in $\mathbb{C}^6$?
Does the existence of AME$(4,6)$, related to dimension six, help us to make any progress in this direction?
\item What about new classes of Bell inequalities that can be constructed by the method we proposed in Chapter~\ref{chap:EXCESS}?
Will application of the excess of a matrix to quantum nonlocality reveal some previously unknown properties of Bell inequalities?
\end{itemize}

%%%%%%%%%%%%%%%%%%%%%%%%%%%%%%%%%%%%%%%%%%%%%%%%%%%%%%%%%%%%%%%%%%%%
%\clearpage\null\thispagestyle{empty}

%%%%%%%%%%%%%%%%%%%%%%%%%%%%%%%%%%%%%%%%%%%%%%%%%%%%%%%%%%%%%%%%%%%%

\begin{appendix}
\chapter{Papers}
\label{chap:papers_list}

     The papers do not belong to the main body of the Thesis. They were listed here at the convenience of the Reader
    who might wish to confront the content or quickly follow some additional information
    which was intentionally abandoned in the previous chapters.

\medskip

\noindent In order of appearance in the text:
\begin{enumerate}
\item[{\bf [A1]}] {\bf W. Bruzda};\\
{\sl Extension of the Set of Complex Hadamard Matrices of Size $8$},\\
\href{https://link.springer.com/article/10.1007/s11786-018-0379-8}{Math. Comput. Sci. {\bf 12}(4), 459--464} (2018).
\item[{\bf [A2]}] {\bf W. Bruzda};\\
{\sl Block-Circulant Complex Hadamard Matrices},\\
\href{https://arxiv.org/abs/2204.11727}{arXiv:2204.11727} (2022).
\item[{\bf [A3]}] {\bf W. Bruzda}, D. Goyeneche, K. \.{Z}yczkowski;\\
{\sl Quantum Measurements with Prescribed Symmetry},\\
\href{https://journals.aps.org/pra/abstract/10.1103/PhysRevA.96.022105}{Phys. Rev. A {\bf 96}, 022105} (2017).
\item[{\bf [A4]}] S. Ahmad Rather, A. Burchardt, {\bf W. Bruzda}, G. Rajchel-Mieldzio\'c, A. Lakshminarayan, K. \.{Z}yczkowski;\\
{\sl Thirty-six Entangled Officers of Euler},\\
%\href{https://arxiv.org/abs/2104.05122}{arXiv:2104.05122} (2021)\\
\href{https://doi.org/10.1103/PhysRevLett.128.080507}{Phys. Rev. Lett. {\bf 128}, 080507} (2022).
\item[{\bf [A5]}] D. Goyeneche, {\bf W. Bruzda}, O. Turek, D. Alsina, K. \.{Z}yczkowski;\\
{\sl Local Hidden Variable Value Without Optimization Procedures},\\
\href{https://arxiv.org/abs/2004.00695}{arXiv:2004.00695} (2021).
%\item[{\bf [\hyperlink{paper_A6.1}{A6}]}] D. Goyeneche, {\bf W. Bruzda}, O. Turek, K. \.{Z}yczkowski; {\sl Mathematical Properties of Excess}, {\sl in preparation} (2021).
\end{enumerate}

\end{appendix}

\clearpage\null\thispagestyle{empty}

\end{document}